\PassOptionsToPackage{unicode}{hyperref}
\PassOptionsToPackage{hyphens}{url}
\documentclass[
  12pt,
]{article}
\usepackage{xcolor}
\usepackage[a4paper,left=25mm,right=25mm,top=25mm,textheight=676pt]{geometry}
\usepackage{amsmath,amssymb}
\usepackage{iftex}
\ifPDFTeX
  \usepackage[T1]{fontenc}
  \usepackage[utf8]{inputenc}
  \usepackage{textcomp} % provide euro and other symbols
\else % if luatex or xetex
  \usepackage{unicode-math} % this also loads fontspec
  \defaultfontfeatures{Scale=MatchLowercase}
  \defaultfontfeatures[\rmfamily]{Ligatures=TeX,Scale=1}
\fi
\usepackage{lmodern}
\ifPDFTeX\else
\fi
\IfFileExists{upquote.sty}{\usepackage{upquote}}{}
\IfFileExists{microtype.sty}{% use microtype if available
  \usepackage[]{microtype}
  \UseMicrotypeSet[protrusion]{basicmath} % disable protrusion for tt fonts
}{}
\usepackage{setspace}
\makeatletter
\@ifundefined{KOMAClassName}{% if non-KOMA class
  \IfFileExists{parskip.sty}{%
    \usepackage{parskip}
  }{% else
    \setlength{\parindent}{0pt}
    \setlength{\parskip}{6pt plus 2pt minus 1pt}}
}{% if KOMA class
  \KOMAoptions{parskip=half}}
\makeatother
\makeatletter
\ifx\paragraph\undefined\else
  \let\oldparagraph\paragraph
  \renewcommand{\paragraph}{
    \@ifstar
      \xxxParagraphStar
      \xxxParagraphNoStar
  }
  \newcommand{\xxxParagraphStar}[1]{\oldparagraph*{#1}\mbox{}}
  \newcommand{\xxxParagraphNoStar}[1]{\oldparagraph{#1}\mbox{}}
\fi
\ifx\subparagraph\undefined\else
  \let\oldsubparagraph\subparagraph
  \renewcommand{\subparagraph}{
    \@ifstar
      \xxxSubParagraphStar
      \xxxSubParagraphNoStar
  }
  \newcommand{\xxxSubParagraphStar}[1]{\oldsubparagraph*{#1}\mbox{}}
  \newcommand{\xxxSubParagraphNoStar}[1]{\oldsubparagraph{#1}\mbox{}}
\fi
\makeatother

\usepackage{longtable,booktabs,array}
\usepackage{calc} % for calculating minipage widths
\usepackage{etoolbox}
\makeatletter
\patchcmd\longtable{\par}{\if@noskipsec\mbox{}\fi\par}{}{}
\makeatother
\IfFileExists{footnotehyper.sty}{\usepackage{footnotehyper}}{\usepackage{footnote}}
\makesavenoteenv{longtable}
\usepackage{graphicx}
\makeatletter
\newsavebox\pandoc@box
\newcommand*\pandocbounded[1]{% scales image to fit in text height/width
  \sbox\pandoc@box{#1}%
  \Gscale@div\@tempa{\textheight}{\dimexpr\ht\pandoc@box+\dp\pandoc@box\relax}%
  \Gscale@div\@tempb{\linewidth}{\wd\pandoc@box}%
  \ifdim\@tempb\p@<\@tempa\p@\let\@tempa\@tempb\fi% select the smaller of both
  \ifdim\@tempa\p@<\p@\scalebox{\@tempa}{\usebox\pandoc@box}%
  \else\usebox{\pandoc@box}%
  \fi%
}
\def\fps@figure{htbp}
\makeatother

\NewDocumentCommand\citeproctext{}{}

\makeatletter
 \let\@cite@ofmt\@firstofone
 \def\@biblabel#1{}
 \def\@cite#1#2{{#1\if@tempswa , #2\fi}}
\makeatother
\newlength{\cslhangindent}
\newlength{\csllabelwidth}
\newenvironment{CSLReferences}[2] % #1 hanging-indent, #2 entry-spacing
 {\begin{list}{}{%
  \setlength{\itemindent}{0pt}
  \setlength{\leftmargin}{0pt}
  \setlength{\parsep}{0pt}
  \ifodd #1
   \setlength{\leftmargin}{\cslhangindent}
   \setlength{\itemindent}{-1\cslhangindent}
  \fi
  \setlength{\itemsep}{#2\baselineskip}}}
 {\end{list}}
\usepackage{calc}

\usepackage{booktabs}
\usepackage{longtable}
\usepackage{array}
\usepackage{multirow}
\usepackage{wrapfig}
\usepackage{float}
\usepackage{colortbl}
\usepackage{pdflscape}
\usepackage{tabularx}
\usepackage{xltabular}
\usepackage{threeparttable}
\usepackage{threeparttablex}
\usepackage[normalem]{ulem}
\usepackage{makecell}
\usepackage{xcolor}
\usepackage{etoolbox}
\usepackage{pdflscape}
\AtBeginEnvironment{longtable}{\footnotesize\setlength{\tabcolsep}{4pt}\singlespacing}
\AtBeginEnvironment{thebibliography}{\singlespacing}
\AtBeginEnvironment{CSLReferences}{\singlespacing}
\makeatletter
\@ifpackageloaded{tcolorbox}{}{\usepackage[skins,breakable]{tcolorbox}}
\@ifpackageloaded{fontawesome5}{}{\usepackage{fontawesome5}}
\definecolor{quarto-callout-color}{HTML}{909090}
\definecolor{quarto-callout-note-color}{HTML}{0758E5}
\definecolor{quarto-callout-important-color}{HTML}{CC1914}
\definecolor{quarto-callout-warning-color}{HTML}{EB9113}
\definecolor{quarto-callout-tip-color}{HTML}{00A047}
\definecolor{quarto-callout-caution-color}{HTML}{FC5300}
\definecolor{quarto-callout-color-frame}{HTML}{acacac}
\definecolor{quarto-callout-note-color-frame}{HTML}{4582ec}
\definecolor{quarto-callout-important-color-frame}{HTML}{d9534f}
\definecolor{quarto-callout-warning-color-frame}{HTML}{f0ad4e}
\definecolor{quarto-callout-tip-color-frame}{HTML}{02b875}
\definecolor{quarto-callout-caution-color-frame}{HTML}{fd7e14}
\makeatother
\makeatletter
\@ifpackageloaded{caption}{}{\usepackage{caption}}
\AtBeginDocument{%
\ifdefined\contentsname
  \renewcommand*\contentsname{Table of contents}
\else
  \newcommand\contentsname{Table of contents}
\fi
\ifdefined\listfigurename
  \renewcommand*\listfigurename{List of Figures}
\else
  \newcommand\listfigurename{List of Figures}
\fi
\ifdefined\listtablename
  \renewcommand*\listtablename{List of Tables}
\else
  \newcommand\listtablename{List of Tables}
\fi
\ifdefined\figurename
  \renewcommand*\figurename{Figure}
\else
  \newcommand\figurename{Figure}
\fi
\ifdefined\tablename
  \renewcommand*\tablename{Table}
\else
  \newcommand\tablename{Table}
\fi
}
\@ifpackageloaded{float}{}{\usepackage{float}}
\floatstyle{ruled}
\@ifundefined{c@chapter}{\newfloat{codelisting}{h}{lop}}{\newfloat{codelisting}{h}{lop}[chapter]}
\floatname{codelisting}{Listing}

\makeatother
\makeatletter
\@ifpackageloaded{caption}{}{\usepackage{caption}}
\@ifpackageloaded{subcaption}{}{\usepackage{subcaption}}
\makeatother
\usepackage{bookmark}
\IfFileExists{xurl.sty}{\usepackage{xurl}}{} % add URL line breaks if available
\hypersetup{
  pdftitle={Inference for Standard Trial Estimands after Data-Driven Subgroup Discovery},
  pdfauthor={Larry F. León; Keaven M. Anderson},
  pdfkeywords={Infinitesimal jackknife, Multiplier
resampling, Post-selection inference, Subgroup identification, Winner's
curse},
  hidelinks,
  pdfcreator={LaTeX via pandoc}}

\title{Inference for Standard Trial Estimands after Data-Driven Subgroup
Discovery}
\author{Larry F. León \and Keaven M. Anderson}
\date{2026-08-20}
\begin{document}
% arXiv merge: article.cls's \maketitle discards the title machinery
% after use; keep it so the supplement's title block can be set below.
\makeatletter
\let\arxivSaveMaketitle\maketitle
\let\arxivSaveAtMaketitle\@maketitle
\let\arxivSaveTitle\title
\let\arxivSaveAuthor\author
\let\arxivSaveDate\date
\let\arxivSaveAnd\and
\let\arxivSaveThanks\thanks
\makeatother
\maketitle
\begin{abstract}
Data-driven subgroup identification is increasingly used in clinical
trials, yet the effect reported for a discovered subgroup is typically
the standard analysis: a Cox or generalized-linear-model treatment
coefficient fitted within that subgroup. Selection by an effect-driven
rule --- rewarding large estimated effects subject to screening and size
criteria --- inflates the reported coefficient: a winner's curse
invalidating naive intervals whether or not the identifier's internal
estimand was de-biased by cross-fitting. We develop a procedure-agnostic
framework for post-selection inference that separates discovery from
reporting: forest search, difference-in-natural-parameters estimators,
and causal forests generate interpretable candidate subgroups; selection
is aligned with the coefficient to be reported; and the target is the
standard-analysis effect of the selected subgroup, the candidate family
held fixed --- a conditional, data-adaptive estimand. Resampling the
search then reduces to jointly perturbing all candidate effects,
yielding refit-free multiplier resampling with an
infinitesimal-jackknife interval whose coverage we characterize under
standard regularity and an explicit condition on the limiting
competition. It matches the full bootstrap on a fixed family; on
model-generated families it remains valid for the same conditional
target, which the bootstrap reproduces only under further conditions.
Two trial applications and matched simulations show mitigated selection
bias and improved coverage.
\end{abstract}

% \setstretch{1.655} % arXiv: referee-mode line spacing removed

\par\bigskip

\section{Inference for the standard trial analysis}\label{sec-target}

\subsection{Identified subgroups and the winner's
curse}\label{sec-winners-curse}

Randomized trials are powered for an overall treatment effect, yet some
of their most consequential questions are subgroup questions: is there
an identifiable set of patients in whom the treatment is harmful, or in
whom an apparently inert treatment works? A subgroup \emph{identified
from the trial data} rather than prespecified carries the contamination
of its own discovery: chosen because its estimated effect is extreme,
its estimate is, by construction, biased toward extremity --- the
winner's curse of subgroup search --- and a confidence interval computed
as if the subgroup had been prespecified under-covers (León et al.
2024). The same phenomenon arises for any parameter selected by
optimization, for which conditional estimators and confidence sets have
been developed (Andrews et al. 2024). Post-selection inference is the
repair: a point estimate purged of the optimism selection injected, and
an interval carrying the variability of the selection itself.

The identification procedures we consider share a common form: each
scores a family of candidate subgroups by an estimated treatment effect,
retains those clearing a harm threshold, and reports the qualifying
candidate of largest effect --- the maximizer \(\widehat H\), the source
of the optimism. Three are treated: the forest search (FS) of León et
al. (2024), which enumerates the candidate family directly, and two
model-based identifiers adapted here to return interpretable,
threshold-defined subgroups --- the difference-in-natural-parameters
(DINA) estimator of Gao and Hastie (2025) and the causal forest, fit via
generalized random forests (GRF) (Wager and Athey 2018; Athey et al.
2019; Cui et al. 2023).

\subsection{The target of inference: the standard trial
analysis}\label{sec-standard-target}

We take the target of inference to be, \emph{for better or worse}, the
\textbf{standard analysis}: the treatment effect that would be reported
had the subgroup been prespecified --- for a time-to-event outcome the
partial-likelihood log hazard ratio of a within-subgroup Cox model (Cox
1972), and for binary, count, or continuous outcomes the within-subgroup
generalized linear model (GLM) treatment coefficient. Writing
\(\beta(g)\) for the population coefficient of the trial's working model
restricted to candidate \(g\), a candidate qualifies when
\(\hat\beta(g) > c\) (the harm threshold), \(\widehat{H}\) is the
maximizer, and the goal is a de-biased estimate of \(\beta(\widehat H)\)
with an interval valid under the selection that produced \(\widehat H\);
we report \(\theta(\widehat H) := \exp\{\beta(\widehat H)\}\), coverage
of the two coinciding by monotonicity. Here \(\beta(\widehat H)\) is the
\emph{least-false} within-subgroup coefficient --- the population root
of the working-model estimating equation, not assumed correctly
specified.

The candidate family is held \textbf{fixed at what the data produced},
so \(\beta(\widehat H)\) is a data-adaptive target parameter (Hubbard et
al. 2016) --- inference \emph{for that conditional estimand}, valid
under ordinary regularity with no assumption about how the cut locations
defining \(\widehat H\) were estimated (its relation to the
unconditional bootstrap is taken up in Supplementary Section S3). The
choice is institutional, not a claim of causal optimality: the standard
analysis is the scale on which trials are designed, monitored, and read
by regulators and clinicians, and a flagged subgroup must ultimately
support the claim \emph{``had this subgroup been prespecified, its
analysis would have shown this effect, with this precision.''}

The second design choice is \emph{interpretable membership}: subgroups
are defined by explicit covariate thresholds --- ``\(X_1 \le c_1\) and
\(X_2 > c_2\)'' --- evaluable from the measurements alone, whereas a
score-thresholded set \(\{x : \hat\tau(x) > t\}\) has the fitted model
as its membership rule. The reportable unit throughout is the triple
(interpretable definition, standard-analysis effect, valid
post-selection interval).

\subsection{What modern identifiers target
instead}\label{sec-ml-targets}

The machine-learning literature on heterogeneous treatment effects
supplies powerful subgroup identifiers built around different estimands
than the standard analysis --- a difference that determines their
selection statistics. GRF and causal forests estimate conditional
average treatment effects on the difference-in-means scale, their
survival extension differences in restricted mean survival time rather
than the Cox log hazard ratio (Wager and Athey 2018; Athey et al. 2019;
Cui et al. 2023); DINA targets the causal contrast of natural parameters
\(\tau(x) = \eta_1(x) - \eta_0(x)\), estimated by orthogonalized
cross-fitting in the R-learner tradition (Gao and Hastie 2025; Nie and
Wager 2021), and proposes subgroups by ranking the subgroup mean of the
fitted per-patient effects. The DINA scale coincides with the standard
analysis's --- a coincidence this paper exploits --- but the estimand
and the selection statistic do not: a covariate-conditional contrast
averaged over a subgroup is not the within-subgroup coefficient of the
trial's working model, as a population quantity or, what matters most,
as a random object under resampling.

None of this is a criticism --- these are strong discovery engines. The
difficulty is downstream: a subgroup \emph{selected} by ranking one
statistic but \emph{reported} through another lives on two different
probabilistic objects, and post-selection inference must either model
the selection performed or align it with the report. The resolution
(Section~\ref{sec-alignment}) is the second --- such procedures are
treated as \emph{candidate generators}, re-ranked on the inferential
effect --- but the framing itself is the point here: the paper's central
move is to demote discovery models from estimators of the reported
effect to generators of candidate regions, and to conduct post-selection
inference on the standard trial-analysis effect of the region ultimately
reported.

\subsection{What this paper contributes}\label{sec-this-paper}

This paper develops an inferential framework for a reporting problem
that data-driven subgroup identification leaves open: a subgroup is
discovered by an adaptive procedure, yet the effect reported for it is
the standard trial analysis --- the within-subgroup Cox or GLM
coefficient that would have been used had the region been prespecified.
The winner's curse of Section~\ref{sec-winners-curse} is created by
searching many candidate regions under an effect-driven selection rule
--- screens and size preferences included --- and it arises whether or
not the identifier's internal estimand was de-biased by cross-fitting,
because the optimism is in the maximization over the search, not in the
fit. The first contribution is to separate candidate generation from
inference: forest search, DINA, and the causal forest are treated as
ways of proposing interpretable candidate regions, and once a family of
candidates is on the table, the inferential problem is defined on the
per-candidate standard-analysis coefficient --- a common selection map
acting on candidate effects (Section~\ref{sec-fs-family}) under which
apparently different identifiers become instances of one reporting
problem.

The second contribution is an alignment principle. A post-selection
correction for the reported effect is faithful only when the selection
rule is applied to the statistic that will be reported. If an identifier
ranks regions on a fitted effect surface or a doubly robust score while
the trial reports a within-subgroup coefficient, the resampling law of
the ranking statistic is not the resampling law of the reported
coefficient; model-based identifiers therefore enter as candidate
generators, their qualifying regions re-ranked on the inferential effect
(Section~\ref{sec-alignment}). This is not a computational convenience
--- it is what makes the selection event correspond to the estimand
being reported. A byproduct of practical interest is that model-based
identifiers thereby acquire what their fitted surfaces do not natively
provide: an interpretable covariate-cut subgroup, together with
post-selection inference for the effect the trial reports; the
constructions are stated as explicit algorithms in Supplementary
Sections S1 and S2.

The third contribution is the target of inference and its scope. The
guaranteed target is the conditional estimand \(\beta(\widehat H)\) of
Equation~\ref{eq-conditional-target} --- the standard-analysis effect in
the subgroup the search actually drew, with the realized candidate
family held fixed --- a data-adaptive quantity distinct from the effect
in a hypothetical true subgroup. Whether the family is fixed or
regenerated under resampling is a primary inferential distinction, not a
technical proviso: on a fully enumerated family a bootstrap re-runs the
same competition and the conditional and full-search views coincide to
first order, whereas a model-generated family that changes under
resampling makes the regenerated-family bootstrap answer a different
question (Section~\ref{sec-scope}).

The fourth contribution is an efficient implementation. On a fixed
family, re-running the search under resampling reduces, to first order,
to perturbing all candidate effects jointly and re-applying the
selection rule (Section~\ref{sec-fs-family},
Section~\ref{sec-correction}). The resulting refit-free
multiplier-resampling procedure subtracts two resampling discrepancies
--- the optimism at the \emph{re-selected} subgroup in the lineage of
Harrell et al. (1996), and the estimator's own finite-sample discrepancy
at the \emph{observed} one, a term the Harrell correction omits --- and
forms a selection-aware infinitesimal-jackknife (IJ) interval (Efron
2014; Wager et al. 2014) from the same draws. The formulation here
extends the inferential target, the alignment condition, and the
fixed-versus-regenerated distinction to identifiers whose selection is,
or is re-ranked to be, a functional of the per-candidate
standard-analysis effects, and to the GLM family on the
natural-parameter scale.

\section{Forest search as a fixed candidate family}\label{sec-fs-family}

We begin with forest search because it is the fixed-family prototype of
the framework of Section~\ref{sec-this-paper}: the family is enumerated
before effects are ranked, and the same standard-analysis coefficient
both selects and reports the subgroup. Forest search (León et al. 2024)
is an exhaustive identification procedure: enumerate an interpretable
family of candidate subgroups, evaluate each with the trial's own
analysis, screen on effect size and internal replicability, and select
among the remaining candidates. We present it here in a non-trivial
instance --- two continuous baseline covariates --- because this
instance isolates, without clutter, the structural property the rest of
the paper rides on: the candidate family is \textbf{fixed}, and
consequently a bootstrap of the \emph{entire search} collapses, to first
order (under regularity conditions), into a re-ranking of jointly
perturbed standard-analysis coefficients.

\subsection{The candidate family from two
covariates}\label{sec-family-construction}

We observe \(\{(Y_i,A_i,X_i)\}_{i=1}^{n}\), with \(A \in \{0,1\}\) the
randomized treatment, \(X\) baseline covariates (binary, categorical, or
continuous) from which to form candidate subgroups, and \(Y\) the
outcome (GLM or time-to-event with event indicator); forest search can
additionally include pre-specified cuts, such as biomarker levels
(\(\leq 0\%, \leq 5\%, \leq 10\%\)).

For illustration take two continuous covariates \(X_1, X_2\) and place
cut grids \(\mathcal C_j = \{c_{j,1} < \cdots < c_{j,k_j}\}\) at
prespecified locations of the observed distribution --- by default the
mean, median, and quartiles (\(k_j = 4\)), with a \(J\)-quantile grid as
the general alternative. Each cut point \(c\) --- a candidate subgroup
\emph{boundary} --- defines the two one-sided conditions it can
generate, so the base set of subgroup-defining conditions is
\begin{equation}\protect\phantomsection\label{eq-base}{
\mathcal B \;=\; \big\{\, \mathbf 1\{X_j \le c\},\; \mathbf 1\{X_j > c\}
\;:\; c \in \mathcal C_j,\; j = 1,2 \,\big\},
\qquad |\mathcal B| \;=\; L \;=\; 2\,(k_1 + k_2),
}\end{equation} and the \textbf{candidate family} is every conjunction
of at most two distinct conditions,
\begin{equation}\protect\phantomsection\label{eq-family}{
\mathcal F \;=\;
\big\{\, \varphi \;:\; \varphi \in \mathcal B \,\big\}
\;\cup\;
\big\{\, \varphi \wedge \varphi' \;:\; \varphi, \varphi' \in \mathcal B,\; \varphi \neq \varphi' \,\big\},
\qquad
|\mathcal F| \;\le\; L + \binom{L}{2}.
}\end{equation} Its members read as clinical criteria --- one-sided
sets, corners, and intervals from two cuts on one covariate. Empty
intersections and candidates failing minimum-size and minimum-event
guards are dropped (default four cuts per covariate: \(L = 16\),
\(|\mathcal F| \le 136\)); the construction generalizes to conjunctions
of up to \(d_{\max}\) conditions, with \(d_{\max} = 2\) suggested for
interpretability.

The property to register is what \(\mathcal F\) does \emph{not} depend
on: determined by the covariate cut grids alone --- before any outcome
model is fit or treatment assignment consulted --- it depends on no
outcome, fitted effect, or learned surface. Throughout, \emph{fixed
family} means precisely this: the list of subgroup definitions is one
fixed object, and any resampling re-evaluates the \emph{same} list
rather than generating a new one (cut locations from observed quantiles
are held fixed under all resampling).

\subsection{The per-candidate standard analysis}\label{sec-percand}

Each candidate receives the trial's own analysis. For
\(g \in \mathcal F\), fit the working model to the members of \(g\):
under proportional hazards,
\(\lambda_i(t) = \lambda_{0,g}(t)\,\exp\{\beta(g)\,A_i\}\) for
\(i \in g\), with \(A_i \in \{0,1\}\) the randomized treatment; for a
GLM outcome,
\(g_{\mathrm{link}}\big(\mathbb E[Y_i \mid A_i]\big) = \alpha_g + \beta(g)\,A_i\)
--- optionally covariate-adjusted in either case. The estimate
\(\hat\beta(g)\) solves the within-subgroup score equation
\(\sum_{i \in g} \psi_i(\beta; g) = 0\), and the per-subject empirical
influence is the dfbeta
\begin{equation}\protect\phantomsection\label{eq-dfbeta}{
\mathrm{db}_{g,i} \;=\; \mathcal I_g^{-1}\,\psi_i\big(\hat\beta(g);\, g\big)
\;\approx\; \hat\beta(g) - \hat\beta_{(-i)}(g),
\qquad
\sum_{i \in g} \mathrm{db}_{g,i} = 0,
}\end{equation} with \(\mathcal I_g\) the within-subgroup observed
information. The sum of squared influences,
\(\sigma_{D,g}^2 = \sum_{i \in g} \mathrm{db}_{g,i}^2\), is the robust
(sandwich / infinitesimal-jackknife) variance of \(\hat\beta(g)\) (Lin
and Wei 1989; White 1980; Efron 1982). Orientation is fixed once: harm
corresponds to \(\beta > c\) on the model's own scale, with \(c\) the
log of a ratio threshold for ratio measures and the threshold itself for
a mean difference.

\subsection{The splitting-consistency screen}\label{sec-consistency}

A candidate is admitted not on the size of its estimate alone but on the
\emph{internal replicability} of that estimate. The criterion of León et
al. (2024) is a repeated random split: divide \(g\) into two halves by
independent fair coins, refit the working model on each half, and call
the split \textbf{consistent} if both half-estimates exceed the
consistency threshold \(c_{\mathrm{cons}}\); over many splits, the
consistency rate \(p_{\mathrm{cons}}(g)\) is the fraction of consistent
splits, and \(g\) is admitted if \(p_{\mathrm{cons}}(g) \ge p^\star\)
(e.g., \(0.90\)) alongside the screening requirement
\(\hat\beta(g) \ge c_{\mathrm{screen}}\) on the full fit (e.g.,
\(c_{\mathrm{screen}} = \log 1.25\), \(c_{\mathrm{cons}} = \log 1.0\)).

To first order the split is a Rademacher multiplier bootstrap of the
subgroup estimating function: writing \(G_i = 2Z_i - 1\), the two
half-estimates form a mirror pair \(\hat\beta(g) \pm D_g\) with
\(D_g = \sum_{i \in g} G_i\,\mathrm{db}_{g,i}\), whose conditional
variance \(\sigma_{D,g}^{2}\) is the robust variance of
\(\hat\beta(g)\). Supplementary Section S3 derives this identity and
both readings of the screen --- a closed form for the consistency rate,
computed without splitting, and a calibration equating admission at
level \(p^{\star}\) with a one-sided robust lower confidence bound
clearing the threshold --- and places the fair-coin split alongside its
score- and bootstrap-resampling cousins.

What carries forward to the bridge of Section~\ref{sec-bridge} is the
structural reading: per candidate, the estimate and its consistency
screen are functionals of one fitted model --- its coefficient
\(\hat\beta(g)\) and its influence values \(\{\mathrm{db}_{g,i}\}\).

\subsection{Selection}\label{sec-selection}

Admission is per candidate; identification \emph{selects}. Collecting
the screen and the calibrated consistency requirement into per-candidate
thresholds \(t_g = \max\big(c_{\mathrm{screen}},\; c_{\mathrm{cons}} +
z_{(1+p^\star)/2}\,\sigma_{D,g}\big)\), the procedure applies a
deterministic rule to the candidate effects --- among the admitted
\(\{g : \hat\beta(g) \ge t_g\}\), take the largest effect, or the most
consistent candidate, with subgroup size breaking ties --- which we
write as a \textbf{selection map} acting on the vector of candidate
effects: \begin{equation}\protect\phantomsection\label{eq-selmap}{
\widehat H \;=\; \mathcal S\big(\{\hat\beta(g) : g \in \mathcal F\}\big).
}\end{equation} The particular rule matters less than its two structural
properties: \(\mathcal
S\) is deterministic given the candidate effects (with the
\(\sigma_{D,g}\) entering only through the fixed thresholds \(t_g\)),
and it \textbf{ranks on the inferential effect} --- the same
\(\hat\beta(g)\) the report is about.

\subsection{The bridge: a fixed family connects re-selection and
resampling}\label{sec-bridge}

The de-biasing of León et al. (2024) is a bootstrap of the
\emph{search}: each resample re-runs enumerate--screen--select, carrying
not just re-estimation of a fixed subgroup but \textbf{re-selection} ---
a possibly different winner each resample. Four steps show how the
multiplier resampling already introduced captures that re-selection
through the fixed family.

\textbf{(i) One resample is one weight vector.} A bootstrap resample
assigns subject \(i\) a multiplicity \(K^{*}_{bi}\)
(\(\sum_i K^{*}_{bi} = n\)); refitting candidate \(g\) weights each
member's score contribution by \(K^{*}_{bi}\), and since the unweighted
scores vanish at \(\hat\beta(g)\), only the \emph{centred} weights
\(G_i^{(b)} = K^{*}_{bi} - 1\) perturb the solution --- to first order a
dot product with the influences:
\begin{equation}\protect\phantomsection\label{eq-boot}{
\hat\beta^{*}_b(g) - \hat\beta(g)
\;=\; \sum_{i \in g} \big(K^{*}_{bi} - 1\big)\,\mathrm{db}_{g,i} + o_p\!\big(n_g^{-1/2}\big)
\;=:\; D_g(b).
}\end{equation} This is the fair-coin split perturbation of
Section~\ref{sec-consistency} under a different mean-zero, unit-variance
weight law, and with centred-Poisson weights it is the
infinitesimal-jackknife linearization of the nonparametric bootstrap
(Efron 2014): the consistency screen and the bootstrap of a Cox/GLM
coefficient are one resampling object under two multiplier laws.

\textbf{(ii) With the family fixed, re-running the search is re-applying
the rule.} Bootstrapping the search, the family cannot change ---
\(\mathcal F\) is the same fixed list of definitions
(Section~\ref{sec-family-construction}) --- so the per-candidate
analyses change only through the weights (Equation~\ref{eq-boot}) while
the rule \(\mathcal S\) stays fixed. Thus ``re-run the entire search on
resample \(b\)'' collapses, to first order, into ``re-apply the rule to
the perturbed effects'':
\begin{equation}\protect\phantomsection\label{eq-reselect}{
\widehat H^{*}_b \;=\;
\mathcal S\big(\{\hat\beta(g) + D_g(b) \;:\; g \in \mathcal F\}\big),
}\end{equation} with the thresholds \(t_g\) held at their observed
values. Re-selection is thereby a deterministic functional of exactly
the quantities the single-subgroup bootstrap already controls; the only
expansion anywhere is Equation~\ref{eq-boot}, the dfbeta linearization
already underwriting the consistency screen (Supplementary Equation S4).

\textbf{(iii) One shared vector couples the candidates.} Selection is a
competition decided by \emph{differences} \(D_g(b) - D_{g'}(b)\), never
by one candidate in isolation, so the arithmetic broadcasts a
\textbf{single} multiplier vector to all candidates on each draw,
endowing the perturbations with the candidates' true overlap covariance,
\begin{equation}\protect\phantomsection\label{eq-overlap}{
\operatorname{Cov}\big(D_g(b),\, D_{g'}(b) \,\big|\, \text{data}\big)
\;=\; \sum_i \mathrm{db}_{g,i}\,\mathrm{db}_{g',i}.
}\end{equation} Two near-identical candidates are pushed alike on their
shared members, which cancel in the difference, so the winner flips
between them only on draws that genuinely favour one --- exactly the
bootstrap's behaviour (independent per-candidate perturbations would
manufacture selection variability the bootstrap never produces).
Computationally it is one matrix product
\(\mathbf P = \mathbf B_{\mathrm{eff}}^{\top}\boldsymbol\Xi\): the data
enter only through the influence matrix \(\mathbf B_{\mathrm{eff}}\),
the randomness only through the multiplier draws \(\boldsymbol\Xi\).

\textbf{(iv) What this buys.} On each draw the re-selected winner
carries its own perturbation \(D_{\widehat H^{*}_b}(b)\); because the
rule rewards large perturbed effects it is positive on average, and that
average \emph{is} the selection optimism --- subtracting it, together
with the same-draws term at the observed \(\widehat H\), is the two-term
León correction developed with its infinitesimal-jackknife interval in
Section~\ref{sec-correction}. The resampling is
\emph{procedure-agnostic}: steps (i)--(iii) used only a \textbf{fixed}
family carrying each candidate's standard analysis
\(\big(\hat\beta(g), \{\mathrm{db}_{g,i}\}\big)\) and a selection that
is a \textbf{functional of those candidate effects}
(Equation~\ref{eq-selmap}). Forest search satisfies both natively;
identifiers ranking on a \emph{different} statistic (a fitted
per-patient effect, a doubly-robust score) or whose family is
\emph{regenerated} by each resample do not, and bringing them under the
template --- by re-ranking their qualifying candidates on the
inferential effect, conditional on the proposed family --- is the
subject of the alignment and scope sections
(Section~\ref{sec-alignment}, Section~\ref{sec-scope}).

\section{Correcting the selected-subgroup effect}\label{sec-correction}

The bridge (Section~\ref{sec-bridge}) reduced re-selection over a fixed
family to one pass of multiplier arithmetic: a single shared draw
perturbs every candidate at once (Equation~\ref{eq-boot}), re-selection
is the deterministic rule applied to the perturbed effects
(Equation~\ref{eq-reselect}), and the entire competition is read from
the one matrix
\(\mathbf P=\mathbf B_{\mathrm{eff}}^{\top}\boldsymbol\Xi\). This
section uses that arithmetic twice. First it removes the leading
selection bias from the point estimate; then, from the \emph{same} draws
and at no extra cost, it attaches an interval that carries the
uncertainty of the selection itself. Neither step is tied to a
particular discovery engine: the section requires only candidate
definitions, their standard within-subgroup fits, and a deterministic
selection map acting on those effects. In the fixed-family forest-search
setting the construction reproduces, to first order, the full-search
bootstrap correction of León et al. (2024), with each resample refit
replaced by its influence approximation Equation~\ref{eq-boot}.

\subsection{The two-term de-biased estimate}\label{sec-debias-point}

On each draw the re-selected winner \(\widehat H^{*}_b\) carries its own
perturbation \(D_{\widehat H^{*}_b}(b)=\mathbf P[\widehat H^{*}_b,b]\).
Because the rule rewards a large perturbed effect
\(\hat\beta(g)+D_g(b)\), conditioning on the winner draws its
perturbation from the \emph{upper} part of the candidates' distribution
--- a maximum-type order statistic over the competition --- so it is
positive on average. That average is the optimism of selection;
alongside it sits a same-draws term evaluated at the fixed observed
subgroup, \begin{equation}\protect\phantomsection\label{eq-biasterms}{
\widehat{\mathrm{bias}}_{\mathrm{sel}}=\frac1B\sum_{b=1}^{B}D_{\widehat H^{*}_b}(b),
\qquad
\widehat{\mathrm{bias}}_{\mathrm{fix}}=\frac1B\sum_{b=1}^{B}D_{\widehat H}(b),
}\end{equation} and the \textbf{de-biased estimate} of the selected
effect subtracts both,
\begin{equation}\protect\phantomsection\label{eq-debiased}{
\tilde\beta(\widehat H)=\hat\beta(\widehat H)
-\widehat{\mathrm{bias}}_{\mathrm{sel}}
-\widehat{\mathrm{bias}}_{\mathrm{fix}}.
}\end{equation} The first term is the winner's curse --- the expected
perturbation of whatever wins. The second is taken at the \emph{fixed}
\(\widehat H\) and so has expectation zero under centred multipliers
(\(\mathbb E\,D_{\widehat H}=0\)); León et al. (2024) nonetheless keep
it. It is precisely the term that Harrell et al. (1996) drops as noise,
and dropping it is harmless for the point estimate but not for its
variance: mean-zero though it is, the same-draws discrepancy supplies
the per-draw residual the interval below is built from. The complement
subgroup \(\widehat H^{c}\) is de-biased identically, from
\(D_{\widehat H^{c*}_b}(b)\) and \(D_{\widehat H^{c}}(b)\).

Expressed as the bootstrap-minus-observed average of León et al. (2024)
(their Eq. 7), \begin{equation}\protect\phantomsection\label{eq-leon7}{
\hat\beta^{*}(\widehat H)=\hat\beta(\widehat H)
-\frac1B\sum_{b=1}^{B}\big(\eta^{*}_b(\widehat H^{*}_b)+\eta^{*}_b(\widehat H)\big),
\qquad \eta^{*}_b(g)=\hat\beta^{*}_b(g)-\hat\beta(g),
}\end{equation} every \(\eta^{*}_b(g)\) is a same-subgroup discrepancy
at one fixed \(g\), so the dfbeta linearization Equation~\ref{eq-boot}
applies term by term, \(\eta^{*}_b(g) = D_g(b)\), and turns
Equation~\ref{eq-leon7} into Equation~\ref{eq-debiased} identically: the
point estimator is that correction, linearized. The only approximation
anywhere is Equation~\ref{eq-boot} --- exact for linear and
mean-difference outcomes, loosest in the small-\(n_{\widehat H}\) tail,
where the minimum-event guard keeps the working model stable --- and the
selection map enters both routes through the same re-selected
\(\widehat H^{*}_b\), so ties in the rule perturb both equally and the
gap to the full bootstrap is governed solely by Equation~\ref{eq-boot}.

\subsection{The interval: an infinitesimal jackknife on the same
draws}\label{sec-ij-interval}

The de-biased estimate Equation~\ref{eq-debiased} is a smooth function
of the resampling weights --- an average, over draws, of quantities read
from \(\mathbf P\). León et al. (2024) estimate its variance by viewing
it as a \emph{bagged} estimator and applying the infinitesimal jackknife
(Efron 2014; Wager et al. 2014): an observation that, when resampled
more often, swings the estimate more is more influential, and the
variance is the sum over observations of those squared influences. The
influence of observation \(i\) is estimated by how the per-draw
correction co-varies with \(i\)'s resampling weight --- and that weight
is the centred multiplicity \(K^{*}_{bi}-1=G_i^{(b)}\) that the draws
\(\boldsymbol\Xi\) already carry.

Write the per-draw residual as the centred total bias,
\begin{equation}\protect\phantomsection\label{eq-residual}{
r_b=\big(\widehat{\mathrm{bias}}_{\mathrm{sel}}+\widehat{\mathrm{bias}}_{\mathrm{fix}}\big)
-D_{\widehat H^{*}_b}(b)-D_{\widehat H}(b)
}\end{equation} --- the total bias removed in Equation~\ref{eq-debiased}
minus the two perturbations that draw \(b\) contributed to it. The
infinitesimal-jackknife variance and its finite-\(B\) correction are
then \begin{equation}\protect\phantomsection\label{eq-ij}{
\tilde V=\sum_{i=1}^{n}\widetilde{\mathrm{cov}}_i^{\,2},\quad
\widetilde{\mathrm{cov}}_i=\frac1B\sum_{b=1}^{B}\big(K^{*}_{bi}-\bar K^{*}_i\big)\,r_b,
\qquad
\hat V=\tilde V-\frac nB\,\overline{r^2},
}\end{equation} where \(\bar K^{*}_i=B^{-1}\sum_b K^{*}_{bi}\) is
subject \(i\)'s mean multiplicity across draws and
\(\overline{r^2}=B^{-1}\sum_b r_b^{2}\); the subtracted term removes the
upward Monte-Carlo bias of \(\tilde V\) at finite \(B\) (Wager et al.
2014).

Stacking the residuals into \(\mathbf r=(r_1,\dots,r_B)^{\top}\) and
row-centring the draws,
\(\tilde{\boldsymbol\Xi}=\boldsymbol\Xi-\operatorname{rowmean}\boldsymbol\Xi\),
the same matrix delivers this with no second pass,
\begin{equation}\protect\phantomsection\label{eq-ij-matrix}{
\widetilde{\mathrm{cov}}=\tfrac1B\,\tilde{\boldsymbol\Xi}\,\mathbf r,\qquad
\tilde V=\lVert\widetilde{\mathrm{cov}}\rVert^{2},\qquad
\hat V=\tilde V-\tfrac nB\,\overline{r^2},
}\end{equation} with \(\mathbf r\) read from the same
\(\mathbf P=\mathbf B_{\mathrm{eff}}^{\top}
\boldsymbol\Xi\) that produced the point estimate. The de-biased
interval is
\begin{equation}\protect\phantomsection\label{eq-ij-interval}{
\tilde\beta(\widehat H)\;\pm\;z_{1-\alpha/2}\sqrt{\hat V},
}\end{equation} exponentiated on the hazard- or odds-ratio scale, with
\(\sqrt{\tilde V}\) and finally the fixed-subgroup
\(\sigma_{D,\widehat H}\) as fallbacks should \(\hat V\le 0\) at small
\(B\).

The interval is \emph{selection-aware} by the same mechanism as the
point correction: the naive fixed-subgroup variance
\(\sigma^{2}_{D,\widehat H}\) --- the diagonal of
Equation~\ref{eq-overlap} --- treats \(\widehat H\) as prespecified,
whereas the residual \(r_b\) (Equation~\ref{eq-residual}) moves with the
\emph{re-selected} winner, so \(\hat V\) additionally carries the
variability of which subgroup was selected; how the two compare is not a
universal ordering, a finite-sample question measured in
Section~\ref{sec-simulation}. Nothing here used how \(\mathcal F\) was
built or why \(\mathcal S\) ranks, so any identifier expressible in the
Section~\ref{sec-bridge} template inherits this interval, the one
proviso being a \emph{fixed} family (Section~\ref{sec-scope}); its
contrast with the sharp resampling bound of Supplementary Section S3.5
is taken up there.

Collected into one procedure, the candidate construction of
Section~\ref{sec-fs-family} and the correction of this section run as
follows.

\begin{tcolorbox}[enhanced jigsaw, arc=.35mm, bottomrule=.15mm, breakable, colback=white, colframe=quarto-callout-note-color-frame, left=2mm, leftrule=.75mm, opacityback=0, rightrule=.15mm, toprule=.15mm]

\textbf{Algorithm --- post-selection inference for the selected
subgroup.}

\emph{Input.} Data \(\{(Y_i,A_i,X_i)\}_{i=1}^{n}\); candidate family
\(\mathcal F\); standard within-subgroup working model (Cox for
survival, a GLM otherwise); selection map \(\mathcal S\); number of
multiplier draws \(B\).

\begin{enumerate}
\def\labelenumi{\arabic{enumi}.}
\item
  \textbf{Per-candidate fits.} For each \(g\in\mathcal F\), fit the
  working model within \(g\) and record the treatment effect
  \(\hat\beta(g)\) and the observation-level influences
  \(\mathrm{db}_{g,i}\); collect them into the effect matrix
  \(\mathbf B_{\mathrm{eff}}\).
\item
  \textbf{Select.} Apply the selection map to the observed effects,
  \(\widehat H=\mathcal S(\{\hat\beta(g)\}_{g\in\mathcal F})\). For a
  model-based identifier, \(\mathcal F\) is its qualifying candidates
  and \(\mathcal S\) re-ranks them on \(\hat\beta\)
  (Section~\ref{sec-alignment}).
\item
  \textbf{Perturb all candidates jointly.} Draw centred, unit-variance
  multipliers \(\boldsymbol\Xi\) (\(n\times B\)) and form
  \(\mathbf P=\mathbf B_{\mathrm{eff}}^{\top}\boldsymbol\Xi\), so one
  shared draw perturbs every candidate at once: \(\hat\beta(g)+D_g(b)\)
  with \(D_g(b)=\mathbf P[g,b]\) (Equation~\ref{eq-boot}).
\item
  \textbf{Re-select under each draw.} For \(b=1,\dots,B\), re-apply the
  rule to the perturbed effects,
  \(\widehat H^{*}_b=\mathcal S(\{\hat\beta(g)+D_g(b)\})\)
  (Equation~\ref{eq-reselect}).
\item
  \textbf{De-bias.} Subtract the winner's-curse and same-draws terms
  (Equation~\ref{eq-biasterms}),
  \(\tilde\beta(\widehat H)=\hat\beta(\widehat H)
  -\widehat{\mathrm{bias}}_{\mathrm{sel}}-\widehat{\mathrm{bias}}_{\mathrm{fix}}\)
  (Equation~\ref{eq-debiased}).
\item
  \textbf{Variance on the same draws.} Form the per-draw residuals
  \(r_b\) (Equation~\ref{eq-residual}) and the infinitesimal-jackknife
  variance \(\hat V\) from the same \(\mathbf P\)
  (Equation~\ref{eq-ij-matrix}).
\item
  \textbf{Interval.} Report
  \(\tilde\beta(\widehat H)\pm z_{1-\alpha/2}\sqrt{\hat V}\),
  exponentiated to the hazard- or odds-ratio scale
  (Equation~\ref{eq-ij-interval}). The complement \(\widehat H^{c}\) is
  treated identically.
\end{enumerate}

\emph{Output.} A de-biased estimate \(\tilde\beta(\widehat H)\) of the
selected-subgroup effect and a selection-aware confidence interval. On a
fixed family the procedure is first-order exact against the full
bootstrap that re-runs \(\mathcal S\) on each resample; on a data-driven
family it is exact conditional on the proposed family
(Section~\ref{sec-scope}).

\end{tcolorbox}

\section{Aligning the selection criterion}\label{sec-alignment}

The issue with a model-based identifier is not that it is a biased
estimator of its own target --- cross-fitting sees to that --- but that
its target is not the coefficient the trial will report. DINA and the
causal forest estimate a per-patient causal contrast --- a log odds
ratio, log hazard ratio, or restricted-mean difference --- with
cross-fitted, orthogonalized machinery (Gao and Hastie 2025; Wager and
Athey 2018) that removes the nuisance and regularization bias of that
contrast --- for the causal survival forest, the asymptotic normality of
its restricted-mean contrast under cross-fitting (Cui et al. 2023) is
exactly the premise this paper takes as given --- and that contrast is
what they \emph{rank} on. But ranking is not reporting. To deliver an
\emph{interpretable} subgroup, both search the space forest search
searches --- conjunctions of covariate-pair cuts --- and take the one
with the largest effect, the same enumerate-screen-argmax selection of
Section~\ref{sec-fs-family}, reporting the standard within-subgroup
analysis there. Cross-fitting de-biased the ranking contrast; it did
nothing about that argmax, nor about the within-subgroup effect finally
reported. Selecting the maximum of many searched effects inflates the
reported effect at the winner (Section~\ref{sec-winners-curse}) whether
or not the statistic that ranked them was de-biased. That selection
optimism, not the contrast's nuisance bias, is the residual this paper
removes; it is why the correction of Section~\ref{sec-correction} is
required for a DINA- or forest-identified subgroup alike, and why
everything below concerns the \emph{selection}, not the fit. The bias
that remains after cross-fitting is the bias of the search.

\subsection{The interpretable-subgroup
construction}\label{sec-align-construct}

A model-based identifier turns its fitted surface into an interpretable
subgroup by searching the same covariate-pair family as forest search
(Section~\ref{sec-fs-family}) and reporting the standard within-subgroup
analysis at the selected region; the region-mean of the fitted
per-patient effect is a linear functional of the global fit. The
construction is stated as an explicit algorithm, for the DINA and
causal-forest surfaces, in Supplementary Section S2.

\subsection{The alignment condition}\label{sec-align-condition}

The bridge (Section~\ref{sec-bridge}) requires two things of an
identifier: a fixed family of subgroup definitions, and a selection that
is a functional of the per-candidate effects \(\hat\beta(g)\). Forest
search supplies both; a model-based identifier supplies neither for
free. As Section~\ref{sec-ml-targets} noted, procedures such as DINA
(Gao and Hastie 2025), causal forests, and virtual twins (Foster et al.
2011) rank candidates by a \emph{fitted} statistic --- a
covariate-conditional effect averaged over the subgroup --- rather than
by the within-subgroup coefficient the trial reports. This section makes
precise why that gap breaks the de-biasing of
Section~\ref{sec-correction} if left alone, and why a single change ---
re-ranking on the effect --- repairs it. We work it through DINA, whose
natural-parameter scale already coincides with the standard analysis's
(Section~\ref{sec-ml-targets}); the causal forest and the penalized rule
differ only in the native statistic and are deferred to the appendix.

\textbf{The condition.} Every term León subtracts is an \emph{effect}
discrepancy: in Equation~\ref{eq-leon7} each
\(\eta^{*}_b(g)=\hat\beta^{*}_b(g)-\hat\beta(g)\) perturbs the
inferential coefficient, so the optimism removed in
Equation~\ref{eq-debiased} is the optimism of a rule that ranks on
\(\hat\beta\). Writing \(S_g\) for the statistic the identifier actually
ranks on, the procedure's re-selection
\(\widehat H^{*}_b=\mathcal S(\{\hat\beta(g)+D_g(b)\})\)
(Equation~\ref{eq-reselect}) is the faithful linearization of the
identifier's selection exactly when \(S_g=\hat\beta(g)\).

\textbf{Alignment.} The de-biased estimate Equation~\ref{eq-debiased}
and its interval (Section~\ref{sec-ij-interval}) reproduce León's
bootstrap to first order if and only if the identifier ranks on the
inferential effect, \(S_g=\hat\beta(g)\). Re-ranking a model-based
identifier's \emph{qualifying} candidates on \(\hat\beta(g)\) enforces
this by construction --- at the price of demoting the model from the
arbiter of the winner to a generator of the family.

\subsection{Why a fitted-effect ranking
mis-corrects}\label{sec-align-dina}

Ranking on the fitted-effect mean rather than the reported coefficient
mis-corrects: under resampling the region-mean effect perturbs through
the \emph{global} fit, the reported coefficient through the
\emph{within-subgroup} dfbeta (Equation~\ref{eq-boot}); the influence
functions differ, so alignment must be enforced on the statistic the
rule ranks --- the perturbation algebra is given in Supplementary
Section S2.

\subsection{Re-ranking restores the template}\label{sec-align-fix}

The repair makes the selection a functional of the effects after all:
among the candidates DINA \emph{qualifies}, re-rank on the inferential
effect \(\hat\beta(g)\) and let the rule choose from that ordering. DINA
is thereby a \emph{generator} of the candidate family, not the arbiter
of the winner; the reported subgroup is the effect-defined one, the
procedure's re-selection is the faithful linearization of that
selection, and the de-biased estimate (Equation~\ref{eq-debiased}) and
IJ interval (Section~\ref{sec-ij-interval}) are first-order exact.
Re-ranking corrects the reported subgroup with its own effect optimism,
removing the mismatch. The difference is visible in the GBSG analysis:
standalone DINA returns a subgroup on its own scale with no inferential
coefficient to report; DINA-as-generator returns one carrying the
de-biased hazard ratios tabulated there.

One caveat remains and heads the scope section: re-ranking fixes the
statistic, not the family. DINA's candidates are read off a cross-fit
surface a genuine bootstrap would regenerate, so the de-biased estimate
and interval are exact for the estimand \emph{conditional on the
proposed family}, the gap to the unconditional target being the omitted
family-generation component --- vanishing as \(\mathcal F\) approaches
the full enumerated space, growing when it is a tight surface-selected
handful --- quantified against a family-regenerating bootstrap in
Section~\ref{sec-scope}.

The causal forest and the penalized rule tell the same story under a
different native statistic --- a subgroup mean of doubly-robust scores
for the forest, a regularization-path ordering for the lasso --- and
admit the same repair; Supplementary Section S1 gives their influence
forms and confirms the re-ranking carries over unchanged, and
Supplementary Section S2 casts the forest into the membership mapping
just as it does DINA.

\section{Scope: the fixed-family condition}\label{sec-scope}

The alignment section (Section~\ref{sec-alignment}) settled the first
requirement the bridge (Section~\ref{sec-bridge}) placed on an
identifier --- selection ranks on the inferential effect. The second ---
the framework's second inferential distinction --- is that the family
the re-selection ranges over be \textbf{fixed}.

\textbf{Two sources of optimism.} Selection optimism has two parts ---
the \emph{screen} admits candidates whose effect crossed the threshold,
and the \emph{rule} chooses among the admitted --- and a full bootstrap
captures both by re-running enumerate→screen→select. To match it, the
re-selection Equation~\ref{eq-reselect} must range over the \textbf{full
enumerated family}, re-applying the screen on each draw: restricting the
competition to the post-screen survivors would capture the rule's
optimism but omit the screen's. The family for
Equation~\ref{eq-debiased} is the one the search enumerated, not the one
it returned.

\textbf{A fixed family makes the procedure exact.} When that family is
the fixed enumerated space of Section~\ref{sec-family-construction}, a
genuine bootstrap re-selects over the same candidate set on every
resample: the full bootstrap (León et al. 2024) and the
multiplier-resampling procedure range over identical families, the
re-selection laws coincide to first order (Equation~\ref{eq-reselect}),
and the de-biased estimate and interval are exact for the estimand the
bootstrap targets. Nothing about the data chooses the family, so there
is no regeneration to account for --- the fully enumerated forest
search, the clean case.

\textbf{A data-driven family makes it conditional.} The procedure also
admits a model-based front-end --- a prognostic lasso prefilter, a GRF
or DINA candidate selector --- that shapes which cuts enter the
enumeration. A full bootstrap accommodates such a front end by mimicking
the entire algorithm on each resample (León et al. 2024), proposing a
possibly different family \(\mathcal F^{*}_b\) and so capturing the
variability of which conjunctions are offered. The multiplier-resampling
procedure cannot --- it enumerates the observed-data family once and
holds it fixed --- so it captures the within-family effect-selection
optimism but not the family regeneration. The target it computes is
therefore conditional on the proposed family,
\begin{equation}\protect\phantomsection\label{eq-conditional-target}{
\beta(\widehat H),\qquad
\widehat H=\arg\max_{g\in\mathcal F}\hat\beta(g)\ \text{(per the rule)},\
\mathcal F\ \text{held fixed at the realized family}.
}\end{equation} and the gap to the unconditional target is the omitted
family-generation component of both the bias and the variance
(Section~\ref{sec-align-fix}). Its sign is not settled a priori: the
fixed-family infinitesimal-jackknife variance runs mildly \emph{wide} of
the bootstrap while the unaccounted regeneration term pushes toward
under-coverage, so the net is a quantitative question, measured against
a family-regenerating full bootstrap (Section~\ref{sec-simulation}).

\textbf{The practical reading.} Because exactness rests on a fixed
family, the recommended default --- assumed throughout --- enumerates
fully and lets no fitted model decide the family. A model-based front
end stays admissible (Section~\ref{sec-alignment}) but trades exactness
for the conditional-on-family target, the full bootstrap as reference: a
prognostic (main-effect) lasso prefilter (León et al. 2024) bears on the
family, not the ranking, so removing it restores exactness, whereas a
predictive (interaction) lasso (Tian et al. 2014) is genuine subgroup
selection and enters as the GRF and DINA front ends do. Supplementary
Section S3 places the conditional target within the broader inference
literature.

\subsection{First-order validity, formally}\label{sec-formal-validity}

The principal contribution of this paper is the inferential framework of
the preceding sections --- candidate generation separated from
inference, selection aligned with the reported coefficient
(Section~\ref{sec-alignment}), and the conditional data-adaptive target
(Equation~\ref{eq-conditional-target}). The theorems below establish
first-order validity for its multiplier-resampling implementation on a
fixed family, and delimit precisely what is guaranteed when the family
is model-generated. We record the regularity conditions in interpretable
form and give the two first-order results; their precise statements, two
supporting lemmas, and all proofs are deferred to Supplementary Section
S4. The headline is the \emph{local} regime, in which the candidate
effects are comparable in size and selection stays genuinely stochastic
--- the regime in which the correction is first-order material; the
well-separated regime, where one subgroup dominates, follows as a
corollary.

Throughout, \(\mathcal F=\{g_1,\dots,g_M\}\) (\(M=|\mathcal F|\)) is the
realized candidate family,
\(\widehat H=\arg\max_{g\in\mathcal F}\hat\beta(g)\) the selected
subgroup, and \(\beta(\widehat H)\) the conditional estimand of
Equation~\ref{eq-conditional-target}. The conditions are stated
precisely, with interpretation, in Supplementary Section S4: randomized
i.i.d. sampling with independent censoring \textbf{(A1)}; a fixed,
finite candidate family whose cuts --- prespecified or data-computed ---
are held fixed under resampling \textbf{(A2)}; regular asymptotically
linear within-subgroup estimators with dfbeta influence
(Equation~\ref{eq-dfbeta}) --- the standard Cox (Lin and Wei 1989) and
GLM (White 1980) sandwich conditions \textbf{(A3)}; exchangeable
mean-zero, unit-variance multiplier weights (Præstgaard and Wellner
1993) \textbf{(A4)}; a regular limiting competition with a unique
maximum, realized in a \emph{separated} or a \emph{local} regime
\textbf{(A5)}; and, needed only for Theorem 2's conservativeness, a sign
control on the selection-weighted overlap covariance \textbf{(A6)}.

\textbf{Theorem 1 (the multiplier reproduces the bootstrap on a fixed
family).} \emph{Under (A1)--(A5), conditionally on the data and writing
\(P^{*}\) and \(P^{*}_{\mathrm{boot}}\) for the multiplier and
full-bootstrap re-selection laws: the re-selection laws agree,
\(\sup_{g\in\mathcal F}\lvert P^{*}(\widehat H^{*}_b=g)-P^{*}_{\mathrm{boot}}(\widehat H^{*}_b=g)\rvert\xrightarrow{p}0\);
the de-biased estimate agrees with the León et al. (2024) full-bootstrap
correction Equation~\ref{eq-leon7},
\(\tilde\beta(\widehat H)-\hat\beta^{*}(\widehat H)=o_p(n^{-1/2})\); and
the infinitesimal-jackknife variances agree to first order.}

On the enumerated family the refit-free correction is therefore not an
approximation to the full bootstrap but its first-order image --- the
conclusion the simulations of Section~\ref{sec-simulation} bear out,
where the two routes agree up to a finite-sample discrepancy.

\textbf{Theorem 2 (the conditional estimand: limiting law, bias, and
sufficient conditions for conservative coverage).} \emph{Under
(A1)--(A6) with centred-Poisson weights, in the two regimes of (A5):}

\emph{(i) (Consistency.) \(\tilde\beta(\widehat H)\) is consistent for
\(\beta(\widehat H)\) in both regimes. Under separation (A5-i),
\(P(\widehat H=g^\star)\to1\) and
\(\sqrt n\,\{\tilde\beta(\widehat H)-\beta(\widehat H)\}\Rightarrow
N(0,\sigma^{2}_{D,g^\star})\) with bias \(o(n^{-1/2})\).}

\emph{(ii) (Local limit law.) Under (A5-ii), with competition field
\(W=\mu+\zeta\), \(\zeta\sim N(0,\Sigma)\), and \(G\) the limiting
selection map of Equation~\ref{eq-selmap},}
\[\sqrt n\,\{\tilde\beta(\widehat H)-\beta(\widehat H)\}\;\Rightarrow\;
\Lambda(\zeta)\;=\;\zeta_{G}-m(\mu+\zeta),\qquad
m(w)=\mathbb E_{\zeta^{*}}\!\bigl[\zeta^{*}_{G(w+\zeta^{*})}\bigr],\]
\emph{\(\zeta^{*}\) an independent copy of \(\zeta\). \(\Lambda\) is
non-Gaussian and the error is \(O_p(n^{-1/2})\); asymptotic normality is
specific to (i).}

\emph{(iii) (Local bias.) Writing \(m^{(S)}\) for the optimism under
noise covariance \(S\),}
\[\mathbb E[\Lambda]\;=\;m(\mu)-\mathbb E\bigl[m(\mu+\zeta)\bigr]
\;=\;m^{(\Sigma)}(\mu)-\tfrac12\,m^{(2\Sigma)}(\mu),\] \emph{positive at
ties and in a neighbourhood of them, negative under sufficient
separation; where it is nonzero the correction leaves a residual of
exact order \(n^{-1/2}\). Under (A5-i) both the optimism and the
residual are \(o(n^{-1/2})\), recovering (i).}

\emph{(iv) (Variance and coverage.) \(\hat V\) of Equation~\ref{eq-ij}
is conservative for \(\operatorname{Var}\tilde\beta(\widehat H)\) ---
under (A6) in the local regime, unconditionally under separation ---
retaining \(\sigma^2_{D,\widehat H}\) through the same-draws term. Under
(A5-i) the interval Equation~\ref{eq-ij-interval} is asymptotically
conservative for \(\beta(\widehat H)\). Under (A5-ii) its asymptotic
coverage is \(P(|\Lambda|\le z_{1-\alpha/2}\sqrt V)\), at least
\(1-\alpha\) whenever}
\[q_{1-\alpha}\bigl(\max\nolimits_g|\zeta_g|\bigr)+m^{(\Sigma)}(0)
\;\le\;z_{1-\alpha/2}\sqrt{V_{\star}},\qquad V_{\star}=\inf\nolimits_w V(w).\]

Read plainly: the correction removes most of the winner's curse but not
all of it; what remains is of the same order as the standard error and
changes sign as the competition separates; and the interval covers the
selected subgroup's effect provided that remainder is small against the
interval's own width. In the local regime \(\Lambda\) is the noise
realized at the winner minus the resampling estimate of the optimism
evaluated at the observed field, so the residual of (iii) is a genuine
\(n^{-1/2}\) term rather than a remainder: the correction removes the
leading optimism \(m(\mu)\) --- the subtracted winner's-curse term of
Equation~\ref{eq-debiased} --- up to a curvature term of the same order.
Its tie value, the sign reversal, and the form of \(m\) under the
implemented selection map are given in Supplementary Section S4, with
the condition in (iv) and the sense in which it, unlike (ii)--(iii), is
uniform in \(\mu\). As in Theorem 1 the point-estimate de-biasing holds
for any (A4) weights, while the variance and coverage statements are
specific to the centred-Poisson weights the procedure draws, for which
\(\hat V\) is Efron's infinitesimal jackknife (Efron 2014).

The guarantee is for the conditional estimand \(\beta(\widehat H)\), and
only for it: Theorem 2 does not assert coverage of the marginal effect
\(\theta^{\dagger}(H)\) in a prespecified true subgroup --- a distinct
functional, and the clinically salient target when the discovered region
is close to a true one. Both coverages are examined empirically in
Section~\ref{sec-simulation}, under a genuine-harm and a borderline-null
design.

The interval is thus selection-aware by construction, valid for the
data-adaptive target \(\beta(\widehat H)\) in the sense of Hubbard et
al. (2016).

These two regimes --- separated (a finite-sample refinement) and local
(first-order material) --- and the no-tie boundary between them are
interpreted in subgroup terms in Supplementary Section S5.

\section{Applications}\label{sec-applications}

We apply the proposals to two trials: the GBSG tumour-recurrence data on
the hazard-ratio scale, and the AIDS Clinical Trials Group 175 (ACTG175)
CD4 non-response data on the odds-ratio scale.

All three identifiers search the same candidate family of
Section~\ref{sec-fs-family} --- conjunctions of at most \(d_{\max}=2\)
covariate-pair cuts --- and report the standard within-subgroup analysis
of Section~\ref{sec-percand} at the selected region; a candidate is
eligible only if it holds at least \(60\) subjects with \(10\) events in
each arm. Forest search admits candidates clearing the screening
threshold with splitting-consistency rate
\(p_{\mathrm{cons}}(g) \ge 0.9\) (screening and consistency thresholds
both \(\log 1.0\)) and returns the largest-effect region whose size is
maximal within a \(10\%\) neighbourhood of that maximum, with
per-candidate threshold \(t_g\) as in Equation~\ref{eq-selmap}. DINA and
GRF enter as candidate \emph{generators} (Section~\ref{sec-align-fix}),
their regions re-ranked on the inferential coefficient \(\hat\beta(g)\):
the GRF forest is grown to depth \(2\) and screens its restricted-mean,
doubly robust (DR-score) contrast over a horizon at \(0.80\) of the
follow-up support, DINA on its mean-effect scale, with DR-score and
mean-effect floors of \(0\) and \(\log 1.0\) (mirroring FS). Because the
DINA and GRF families regenerate under resampling, their results here
and in Section~\ref{sec-simulation} are empirical evaluations of how the
correction behaves on such families, not instances of Theorem 2, whose
scope is the fixed enumerated family. Supplementary Sections S6 (GBSG)
and S7 (ACTG175) report the harm subgroup \(\widehat{H}\) and its
complement \(\widehat{H}^{c}\) with a naive estimate and two de-biased
estimates: the full nonparametric bootstrap (FB) and the refit-free
multiplier resampling (MR), based on \(1{,}000\) and \(5{,}000\) draws
respectively.

\subsection{GBSG and ACTG175 analyses}\label{sec-application}

The GBSG trial (via the \texttt{survival} package (Therneau 2026))
records baseline age, menopausal status (meno), tumour size (size), a
grade-3 indicator (grade3), positive lymph nodes (nodes), and
progesterone (pgr) and oestrogen (er) receptor concentrations; for pgr
and er we use \(J = 10\) cuts (Section~\ref{sec-family-construction}).
Forest search additionally includes pre-specified candidates evaluating
positive receptor-defined subgroups directly via \(\{er \le 0\}\) and
\(\{pgr \le 0\}\).

The subgroups identified (full de-biased hazard-ratio table in the
supplement) are: The GRF identifier selected \(\{er \le 0\}\) and the FS
identifier \(\{er \le 0\} \cap \{size \le 35\}\); DINA selected
\(\{grade3 \ge 1\} \cap \{pgr \le 10\}\). The corrections attenuate all
three identifiers' estimates, to differing degrees: forest search
remains the most elevated (1.94 FB, 1.44 MR); DINA's two routes fall on
opposite sides of the null (0.89 FB, 1.11 MR); and GRF corrects to
approximately the null (1.04, 1.05). The two de-biasing routes differ
sharply in cost: at \(B =\) 1,000 FB ran from about 72 seconds to about
10.1 minutes across the three identifiers, whereas for MR (5,000 draws)
in 0.11 seconds to 0.38 seconds --- roughly 671 to 1,710 times faster, a
margin that widens with \(B\).

The odds-ratio companion analysis of the ACTG175 trial exercises the
same machinery on the GLM path: forest search and GRF identify closely
overlapping high-weight, high-CD4 regions. The corrections again
attenuate all three: forest search remains clearly elevated (2.43 FB,
1.79 MR); GRF corrects to marginally above one (1.09, 1.10); and DINA
corrects to below one (0.85, 0.95).

\section{Simulation evaluation of resampling de-biasing
procedures}\label{sec-simulation}

We evaluate the proposals in simulations built from the GBSG data,
following the data-generating model of León et al. (2024) as implemented
in the R package \texttt{forestsearch}: an accelerated-failure-time
super-population with a true harm subgroup of prevalence approximately
\(12.5\%\), at a borderline (null) design with
\(\theta^{\dagger}(H) = 1.00\) and \(\theta^{\dagger}(H^{c}) = 0.66\).
The design stresses false identification --- screening thresholds are
deliberately permissive, and no subgroup is supplied, so each identifier
must rediscover it from the raw covariates. On each replicate we compare
four estimates of the within-subgroup hazard ratio on the \emph{same}
data and selected subgroup: the \textbf{oracle} refit on the true \(H\),
the \textbf{naive} plugin on \(\widehat H\), \textbf{FB}, and
\textbf{MR}. Bias is reported against the marginal hazard ratio
\(\theta^{\dagger}(\cdot)\), controlled direct effects
\(\theta^{\ddagger}(\cdot)\) (León et al. 2024), and the conditional
estimand \(\theta(\cdot) = \exp\{\beta(\cdot)\}\) characterized by
Theorem 2. Full design details, per-identifier estimation and coverage
tables, and the classification summaries are in Supplementary Section
S8.

Detection rates are properties of the identification procedures, not of
the correction: under the null a detection is a false harm declaration,
so forest search, DINA, and GRF declare false harm in 66\%, 70\%, and
98\% of replicates respectively. The correction addresses the bias and
coverage of the reported effect \emph{conditional on} a declaration; it
neither targets nor controls the false-declaration rate itself, which is
governed by the screening thresholds (Section~\ref{sec-consistency}) and
the discovery procedure's tuning.

The winner's curse dominates the naive analysis and the correction
removes most of it: on the harm subgroup the naive plugin overstates the
marginal hazard ratio by 63\% (forest search), 58\% (DINA), and 57\%
(GRF), while multiplier resampling brings each close to the marginal
effect (-6.6\%, +6.5\%, and -9.8\%, against oracle residuals +13.2\%,
+9.4\%, +4.7\%). The full bootstrap, the unconditional comparator,
tracks the multiplier on the fixed forest-search family but
\emph{undershoots} on the regenerating DINA and GRF families (-25.9\%
and -24.1\%) --- the conditional-versus-unconditional distinction,
visible in the estimates. For coverage of the marginal hazard ratio
\(\theta^{\dagger}(H)\), the multiplier intervals reach at or above
nominal in the complement for all three identifiers and at (or above)
nominal in the harm subgroup; the wider full bootstrap covers
throughout. Coverage of the conditional estimand \(\theta(\widehat H)\)
that Theorem 2 characterizes is above-nominal for forest search (0.98),
whose candidate family is fixed, as the theorem requires; at this
borderline-null design the DINA and GRF families, though not strictly
fixed, are also above-nominal (0.99, 0.98). Under the genuine-harm
design, by contrast, conditional coverage is near-nominal only on the
fixed forest-search family, as Theorem 2's scope predicts: a strong
signal approaches the separated regime of (A5), where the theorem's
prediction is exact rather than conservative coverage, so a modest
finite-sample shortfall is consistent with it. Per-analysis bias and
coverage figures, with the classification (sensitivity/specificity)
summaries, are in Supplementary Section S8.

The simulations here use the survival hazard-ratio scale; the binary,
odds-ratio path appears in Supplementary Section S8, on the ACTG175
covariate distribution, where the same protective-region reversal and
its repair recur across forest search, DINA, and GRF.

\section{Discussion}\label{sec-discussion}

Subgroup discovery and subgroup reporting are different statistical
problems. The effect a trial reports for a discovered subgroup is the
standard within-subgroup analysis, evaluated at a subgroup selected
because its estimate was extreme, so the bias to remove is that of the
\emph{search}, not the nuisance bias of the fitted contrast that
cross-fitting already handles. This paper corrects on the reporting
scale, treating any identifier as a generator of candidates re-ranked on
the inferential effect: on a fixed enumerated family the correction is a
refit-free multiplier resampling --- one shared draw perturbs every
candidate, the rule is re-applied, and a two-term de-biased estimate
with an infinitesimal-jackknife interval is read from a single matrix
--- reproducing the full bootstrap to first order at a fraction of its
cost (Section~\ref{sec-simulation}).

Two conditions delimit the correspondence with the full bootstrap:
\emph{alignment} (the selecting statistic must be the reported one),
enforced by re-ranking a model-based identifier's qualifying regions on
the reported coefficient, and \emph{fixedness} (an enumerated family
makes the correction first-order exact against the full bootstrap, a
regenerated family leaves it exact only conditional on the proposed
family). These are also the method's limits. The reported estimand is a
working-model coefficient --- hazard and odds ratios are
non-collapsible, and the corrected estimate inherits the interpretive
limits of the standard analysis --- and on a data-driven family the
procedure does not itself price the family regeneration, so the cleanest
application remains the fully enumerated forest search, with model-based
front ends admissible as conditional extensions checked against a full
bootstrap where unconditional inference is required.

Several supporting developments are deferred to the supplement: the GRF
(Supplementary Section S1) and DINA (Supplementary Section S2)
interpretable-subgroup constructions; the multiplier-resampling form of
the consistency criterion (Supplementary Section S3) and its closest
prior work (Supplementary Section S3.5); the regularity conditions,
lemmas, and proofs (Supplementary Section S4); the ACTG175 odds-ratio
application; and additional simulation design points (Supplementary
Section S8). The tables there isolate the most consequential error ---
declaring a protective subgroup harmful: even with a protective truth
(\(\theta^{\dagger}(H) = 0.75\)), selecting on apparent excess hazard
drives the naive hazard ratio above one, and multiplier resampling
restores coverage toward or above nominal where the naive interval
essentially never covers the protective target; the binary ACTG175 study
shows the same reversal and repair on the odds-ratio scale.

\textbf{Reporting guidance.} Because a subgroup's declaration rate ---
governed by the identifier and its thresholds, not by the correction
(Section~\ref{sec-simulation}) --- cannot be read from the observed
data, we recommend estimating it by simulation calibrated to the
analyst's own trial, using the \texttt{forestsearch} data-generating
mechanisms, and reporting it alongside, but distinctly from, the
de-biased effect.

\textbf{Why the conditional estimand.} Reporting \(\beta(\widehat H)\)
rather than the marginal \(\theta^{\dagger}(H)\) of a hypothetical true
subgroup is a choice of question: \(\beta(\widehat H)\) is the number
actually reported and inflated by the winner's curse
(Section~\ref{sec-scope}, Section~\ref{sec-simulation}). Because the
correction conditions on the realized region, the covariate naming a
discovered subgroup may be a correlate of the modifier rather than the
modifier itself --- the classification-accuracy caveat of Supplementary
Sections S5 and S8.

The broader point is a single principle for subgroup reporting in
trials: discovery may be flexible, but valid inference requires
correcting the search that produced the subgroup, on the same scale the
trial will report.

\section*{Data availability
statement}\label{data-availability-statement}
\addcontentsline{toc}{section}{Data availability statement}

The GBSG (Schumacher et al. 1994) and ACTG175 (Hammer et al. 1996) trial
data are publicly available in the R packages \texttt{survival}
(Therneau 2026) and \texttt{speff2trial} (Juraska et al. 2022),
respectively. Methods and code reproducing all tables, figures, and
applications are in the R package \texttt{forestsearch} (León 2026)
(\texttt{https://github.com/larry-leon/forestsearch}); computational
details are given in Supplementary Section S9.

\section*{Generative AI Disclosure}\label{generative-ai-disclosure}
\addcontentsline{toc}{section}{Generative AI Disclosure}

During the preparation of this work the authors used Claude (Anthropic;
Opus 4 series) for editorial assistance --- language refinement,
drafting and revision of exposition, coding support, and consistency
checking of the manuscript. All methodology, theory, simulations, and
analyses were developed by the authors, who reviewed and verified all
AI-assisted output and take full responsibility for the content of the
article, including the accuracy of its results and references.

\section*{Disclosure Statement}\label{disclosure-statement}
\addcontentsline{toc}{section}{Disclosure Statement}

Both authors are employees of Merck \& Co., Inc., Rahway, NJ, USA, and
hold stock in the company. The authors report no other competing
interests.

\protect\phantomsection\label{refs}

% ---- arXiv merge: supplementary material ---------------------------
% Title block as in the supplement's preamble. The supplement set its
% S-prefixed numbering in \AtBeginDocument; the same redefinitions are
% applied here, with every counter they number reset.
\clearpage
\makeatletter
\let\maketitle\arxivSaveMaketitle
\let\@maketitle\arxivSaveAtMaketitle
\let\title\arxivSaveTitle
\let\author\arxivSaveAuthor
\let\date\arxivSaveDate
\let\and\arxivSaveAnd
\let\thanks\arxivSaveThanks
\makeatother
\title{Web-based supporting materials for `Inference for Standard Trial
Estimands after Data-Driven Subgroup Discovery' by L. F. León and K. M.
Anderson}
\author{Larry F. León \and Keaven M. Anderson}
\date{2026-08-20}
\setcounter{section}{0}
\setcounter{subsection}{0}
\setcounter{subsubsection}{0}
\setcounter{figure}{0}
\setcounter{table}{0}
\setcounter{equation}{0}
\setcounter{footnote}{0}
\renewcommand{\thesection}{S\arabic{section}}
\renewcommand{\thesubsection}{\thesection.\arabic{subsection}}
\renewcommand{\thesubsubsection}{\thesubsection.\arabic{subsubsection}}
\renewcommand{\thefigure}{S\arabic{figure}}
\renewcommand{\thetable}{S\arabic{table}}
\renewcommand{\theequation}{S\arabic{equation}}
% hyperref anchors, so S-numbered targets do not collide with the main
% paper's (link targets only; printed numbers are set above).
\renewcommand{\theHsection}{S\arabic{section}}
\renewcommand{\theHfigure}{S\arabic{figure}}
\renewcommand{\theHtable}{S\arabic{table}}
\renewcommand{\theHequation}{S\arabic{equation}}
\maketitle
\begin{abstract}
This supplement collects the auxiliary constructions, proofs, and
additional simulation results supporting the main paper.
\end{abstract}

% \setstretch{1.655} % arXiv: referee-mode line spacing removed
This supplement gathers five kinds of material kept out of the main
text. First, the constructions that cast the two model-based front ends
into the interpretable-subgroup mapping the correction operates on ---
the causal forest (Section~\ref{sec-appendix-grf}) and the DINA surface
(Section~\ref{sec-appendix-dina}). Second, the forest search consistency
criterion implemented via multiplier resampling, with its closed form
and its relation to the closest prior work
(Section~\ref{sec-appendix-multiplier}, Section~\ref{sec-prior-guo}).
Third, the formal development behind Section 5.1 of the main paper: the
regularity conditions, two supporting lemmas, and the proofs of both
theorems (Section~\ref{sec-appendix-proofs}). Fourth, the GBSG and
ACTG175 application tables; and fifth, the additional simulation design
points --- other harm strengths and sample sizes --- that complement the
main-text study of Section 7 of the main paper. The organizing
distinction throughout is whether inference conditions on the realized
candidate family or regenerates it under resampling.

\begin{figure}

\centering{

\pandocbounded{\includegraphics[keepaspectratio]{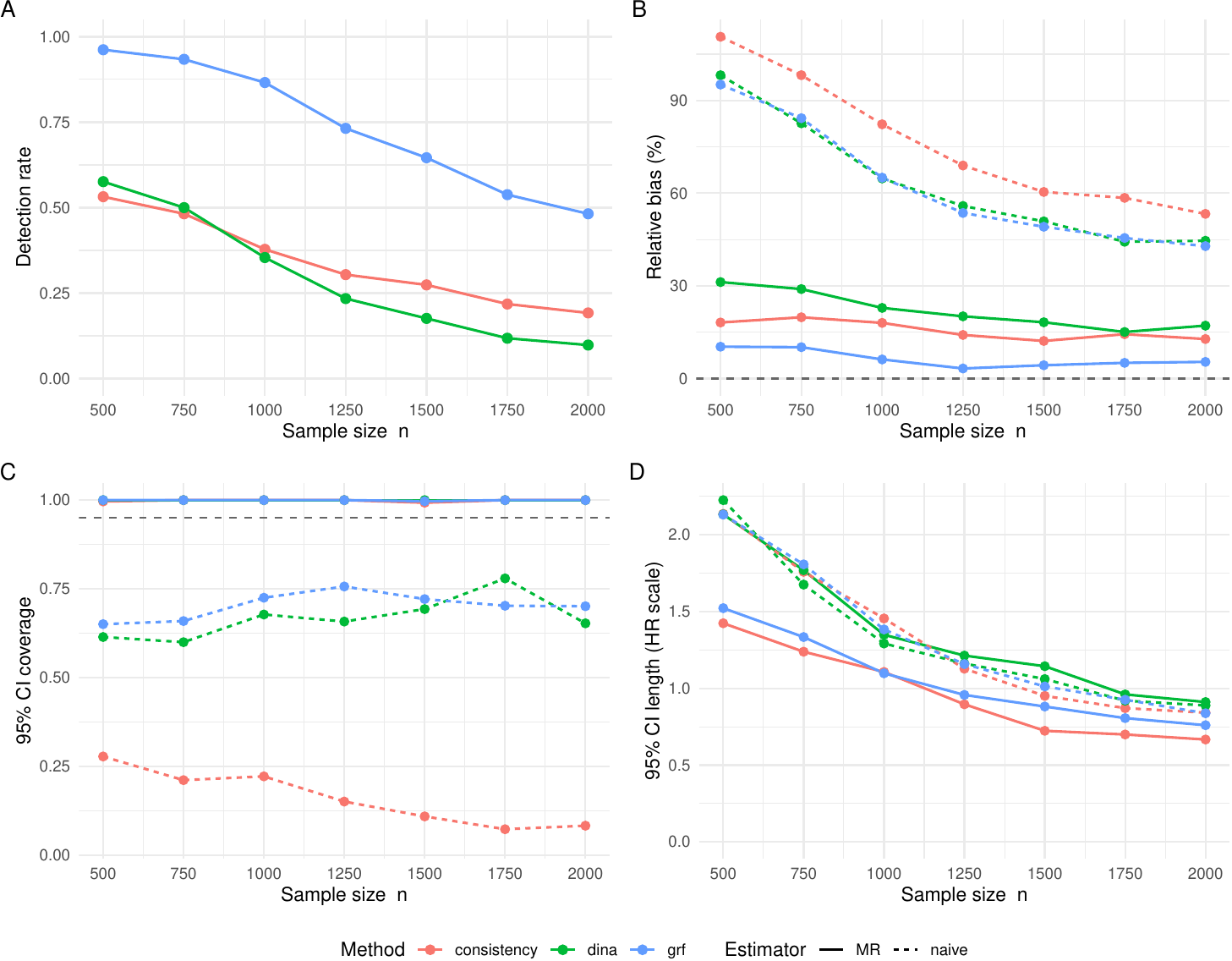}}

}

\caption{\label{fig-mr-coverage-h075m}(A) Subgroup-detection rate vs n
by detector, 500 simulations per cell. (B) Bias of the MR \& naive
estimator (line type) relative to the marginal subgroup hazard ratio
(C†), as a percentage of the target, on the harm subgroup Ĥ, vs n; the
dashed line at 0 marks no bias. (C) Empirical 95\% CI coverage of the
marginal subgroup hazard ratio (C†) for the MR \& naive estimator (line
type), on the harm subgroup Ĥ, vs per-trial sample size n (true
\(\theta^{\dagger}(H) = 0.75\), \(\theta^{\dagger}(H^{c}) = 0.66\)
(marginal HR); \(\theta^{\ddagger}(H) = 0.70\),
\(\theta^{\ddagger}(H^{c}) = 0.58\) (CDE); 500 simulations per cell,
NULL MR draws; detectors: consistency, dina, grf). Dashed line = 0.95
nominal. Uncertainty bands omitted for legibility (select a single
estimator to show them). All rates conditional on identifying a
subgroup. (D) Median 95\% CI width for the MR \& naive estimator (line
type) on the harm subgroup Ĥ, vs n (HR scale).}

\end{figure}

These additional design points evaluate the multiplier-resampling (MR)
correction against the naive analysis across the three identifiers ---
forest search, DINA, and GRF --- at three values of
\(\theta^{\dagger}(H)\) (0.75, 1.00, and 1.51) and per-trial sample
sizes \(n = 500\)--\(2000\), each scored against both the marginal
hazard ratio and the conditional estimand \(\theta(\widehat H)\)
(Figure~\ref{fig-mr-coverage-h075c} through
Figure~\ref{fig-mr-coverage-h15m}); the full bootstrap (FB), which
re-runs the entire candidate search on every resample, is
computationally prohibitive across all these parameter combinations and
is not swept. The below-null case \(\theta^{\dagger}(H) = 0.75\)
(Figure~\ref{fig-mr-coverage-h075m}) is the most cautionary: the
subgroup is in truth protective, yet because a region is flagged only on
apparent excess hazard, the naive plug-in is severely upward-biased ---
for forest search at \(n = 500\) it reports a hazard ratio of 1.59, a
111\% overstatement of the protective truth, and its interval
essentially never covers \(\theta^{\dagger}(H)\), so an uncorrected
analysis would indicate harm where the truth is benefit. Multiplier
resampling reduces this bias to about 18\% and restores coverage to 1.00
--- substantial, if for this identifier incomplete, mitigation --- while
the false-harm detections that drive the problem grow rarer as the
sample grows (from 53\% to 19\% for forest search as \(n\) rises from
500 to 2000). These design points all stay on the survival hazard-ratio
scale. A parallel binary study, built on the ACTG175 covariate
distribution and reported on the odds-ratio scale, carries the same
correction onto the GLM path: with a \emph{protective} subgroup of
marginal odds ratio \(\theta^{\dagger}(H) = 0.75\) against a fixed
complement \(\theta^{\dagger}(H^{c}) \approx 0.66\) --- the binary
analogue of the below-null survival point --- the naive within-subgroup
odds ratio is again driven the wrong way, and multiplier resampling
restores coverage of \(\theta^{\dagger}(H)\) to near nominal across
forest search, DINA, and GRF (Figure~\ref{fig-mr-coverage-or075m}).

\section{Appendix: aligning the causal forest}\label{sec-appendix-grf}

The alignment of Section 4 of the main paper was worked through DINA;
the causal forest is the same argument under a different native
statistic, recorded here for completeness. A generalized random / causal
forest (Wager and Athey 2018; Athey et al. 2019) ranks a region by a
subgroup mean of doubly-robust scores,
\begin{equation}\protect\phantomsection\label{eq-grf-stat}{
S_g=\frac1{|g|}\sum_{i\in g}\hat\Gamma_i,
}\end{equation} where \(\hat\Gamma_i\) is the augmented
inverse-probability-weighted pseudo-outcome assembled from the forest's
nuisance estimates. Under a resample \(S_g\) perturbs through those
nuisance fits, so its influence --- the influence of \(\hat\Gamma_i\)
--- is again a \emph{global}-fit object, distinct from the \emph{local}
within-subgroup dfbeta \(\mathrm{db}_{g,i}\) of Equation 5 of the main
paper. As with DINA, then, \(S_g\neq\hat\beta(g)\) and the perturbations
differ, so an effect-ranked procedure re-selects different winners than
the forest's own bootstrap and mis-corrects (the alignment proposition
of Section 4 of the main paper). The remedy is identical: score the
forest's \emph{qualifying} candidates by the inferential effect
\(\hat\beta(g)\), demoting the forest to a candidate generator; the
de-biased estimate (Equation 10 of the main paper) and interval (Section
3.2 of the main paper) are then first-order exact, conditional on the
proposed family.

The forest is in fact the sharpest case of the family caveat of Section
5 of the main paper. Its native estimand is a restricted-mean or
survival-probability contrast, not the Cox or GLM coefficient the trial
reports (Section 1.3 of the main paper), and --- more to the point here
--- the \emph{forest itself} determines which conjunctions qualify, so a
literal bootstrap that re-fits the forest proposes a different family on
each resample. Conditional on the proposed family (Equation 16 of the
main paper) the procedure is exact; the unconditional target
additionally bootstraps the forest, the regime in which the omitted
family-generation term of Section 5 of the main paper is largest.

\textbf{A note on the lasso.} The lasso does not appear above as a
subgroup selector, and deliberately so. In León et al. (2024) and in
routine use it is a prognostic prefilter --- it reduces the covariate
set before enumeration --- so it shapes the family rather than the
ranking, is governed by the conditional-versus-unconditional distinction
of Section 5 of the main paper rather than by alignment, and is removed
under the recommended fully-enumerated default. There is a genuine lasso
\emph{selector} distinct from this use: a penalized
treatment-by-covariate interaction model whose non-zero coefficients
define the effect-modification rule (Tian et al. 2014; Zhao et al.
2022). That selector ranks on the penalized-path statistic, not the
within-subgroup effect, and so falls under the alignment proposition of
Section 4 of the main paper exactly as DINA and the forest do ---
re-ranked on \(\hat\beta(g)\), conditional on the family it proposes.

\section{Appendix: casting a fitted surface into the subgroup
mapping}\label{sec-appendix-dina}

The alignment section (Section 4 of the main paper) described in prose
how a model-based identifier produces an interpretable subgroup: fit a
per-patient effect surface, search axis-aligned signatures for the
region of largest effect, and report the standard within-subgroup
analysis. This appendix states that construction as an algorithm, for
the DINA surface specifically, and records the membership mapping that
lets the resulting subgroup enter the correction of Section 3 of the
main paper unchanged. It is the model-based analog of the
enumerate-screen-select procedure of Section 2 of the main paper,
differing only in how candidates are generated and scored.

\begin{tcolorbox}[enhanced jigsaw, arc=.35mm, bottomrule=.15mm, breakable, colback=white, colframe=quarto-callout-note-color-frame, left=2mm, leftrule=.75mm, opacityback=0, rightrule=.15mm, toprule=.15mm]

\textbf{Algorithm --- DINA signature search.}

\emph{Input.} Cross-fitted DINA fit
\(\hat\tau(x)=\hat\beta_0+x^{\top}\hat\beta\) with coefficient
covariance \(\widehat{\operatorname{Var}}(\hat\beta)\); covariates
\(x_1,\dots,x_K\); harm threshold \(m\); minimum size \(n_{\min}\)
(default \(60\)); maximum depth \(\in\{1,2\}\); depth-2 quantile grid
(default deciles); focus rule (default \emph{largest qualifying
region}).

\begin{enumerate}
\def\labelenumi{\arabic{enumi}.}
\item
  \textbf{Per-patient effects.} Compute
  \(\hat\tau_i=\hat\beta_0+x_i^{\top}\hat\beta\) for all \(i\). (The
  contrast is already de-biased by cross-fitting; the steps below do not
  touch it.)
\item
  \textbf{Depth-1 candidates.} For each covariate \(x_j\), each
  direction \(\in\{\le,\ge\}\), and each unique value \(q\) of \(x_j\),
  form the signature \(s=(j,\text{dir},q)\) with membership mask
  \(M_s=\mathbf 1\{x_j\le q\}\) or \(\mathbf 1\{x_j\ge q\}\).
\item
  \textbf{Depth-2 candidates} (if depth \(=2\)). For each pair of
  distinct covariates and each (direction, threshold) drawn from the
  per-covariate quantile grid, form the AND-conjunction with mask
  \(M_{s}=M_{s_1}\!\cdot M_{s_2}\) (the intersection). Low-cardinality
  covariates are searched at their unique values.
\item
  \textbf{Qualify.} Retain a candidate iff
  \(\sum_i M_{s,i}\ge n_{\min}\) \textbf{and} \(\bar\tau_s\ge m\), where
  \(\bar\tau_s=\bigl(\sum_i M_{s,i}\hat\tau_i\bigr)/
  \sum_i M_{s,i}\) is the region-mean effect.
\item
  \textbf{Select.} Among qualifying candidates (depth-1 and depth-2
  ranked jointly), return the one chosen by the focus rule --- by
  default the largest --- so a conjunction is returned only when it
  outranks every singleton. Let \(\widehat S\) denote the selected
  signature, \(M_{\widehat S}\) its mask.
\end{enumerate}

\emph{Output.} The signature \(\widehat S\) (a
covariate--direction--threshold triple, or a pair), its membership mask
\(M_{\widehat S}\), region size, and region-mean effect
\(\bar\tau_{\widehat S}=\bar a_{\widehat S}^{\top}\hat\beta\) with
\(\bar a_{\widehat S}=(1,\bar x_{\widehat S})\), whose variance
\(\bar a_{\widehat S}^{\top}\widehat{\operatorname{Var}}(\hat\beta)\,
\bar a_{\widehat S}\) is conditional on \(\widehat S\) --- \emph{not}
selection-adjusted.

\end{tcolorbox}

Two features of the output matter for the rest of the paper. First, the
\textbf{membership mapping} is the same indicator representation forest
search uses (Equation 2 of the main paper): a signature is a conjunction
of covariate-threshold indicators, so \(M_{\widehat S}\) defines a
subgroup the standard analysis can be fit on directly. The selected
signature is therefore handed to the within-signature standard model ---
a Cox model for survival, a GLM otherwise, comparing arms among the
patients in \(M_{\widehat S}\) --- and \emph{that} coefficient, not the
region-mean \(\bar\tau\), is the effect the trial reports and the input
to Section 3 of the main paper. Second, the \textbf{ranking statistic is
the global-fit functional} \(\bar a_S^{\top}\hat\beta\): its dependence
on the fold-fitted \(\hat\beta\) (not on a within-subgroup
re-estimation) is precisely why an effect-ranked procedure must re-rank
these candidates rather than inherit DINA's ordering, the point of
Section 4 of the main paper. The search itself --- steps 2--5 --- is an
argmax over many candidates; its optimism is the selection bias the
correction removes, and cross-fitting in step 1 does nothing to it.

The same template casts the \textbf{causal forest} into the mapping; its
native statistic and the alignment it requires are the subject of
Section~\ref{sec-appendix-grf}. The forest forms its interpretable
region one of two ways: as the policy tree's root-to-leaf path(s) ---
each leaf a conjunction of split indicators (left child
\(\Rightarrow\{x\le c\}\), right child \(\Rightarrow\{x>c\}\)), a
subgroup spanning several leaves is their disjunction, a union of
axis-aligned boxes --- or, exactly as DINA does, by enumerating the
forest's candidate cuts and their depth-two conjunctions and selecting
one by effect and size, yielding a single conjunction. Either way the
membership rule is axis-aligned and built directly from the cuts, after
which the within-signature standard analysis and the correction proceed
as above. Whatever the generator, the cast is the same: a fitted surface
yields an axis-aligned membership rule, the rule selects patients, and
the standard within-rule effect --- selection-corrected per Section 3 of
the main paper --- is what gets reported.

\subsection{Why a fitted-effect ranking mis-corrects: the perturbation
algebra}\label{sec-supp-dina-pert}

The construction ranks each candidate on its fitted-effect mean
\(\bar{\hat\tau}_g=\bar a_g^{\top}\hat\beta_{\mathrm D}\), a functional
of the \emph{global} DINA fit (Section 1 of the main paper). Write
\(\hat\tau_i=a_i^{\top}\hat\beta_{\mathrm D}\) for the per-patient
effect with \(a_i\) the contrast row, so
\(\bar a_g=\frac1{|g|}\sum_{i\in g}a_i\), and
\(\mathrm{IF}_{\mathrm D,i}\) for the influence of that fit at patient
\(i\). Under a resample the statistic therefore perturbs through the
global fit, \begin{equation}\protect\phantomsection\label{eq-dina-pert}{
E_g(b)=\bar a_g^{\top}\big(\hat\beta^{*}_{\mathrm D,b}-\hat\beta_{\mathrm D}\big)
=\sum_i\big(K^{*}_{bi}-1\big)\,
\underbrace{\bar a_g^{\top}\,\mathrm{IF}_{\mathrm D,i}}_{\textstyle \mathrm{dt}_{g,i}},
}\end{equation} This is a different object from the effect perturbation
\(D_g(b)\) of Equation 5 of the main paper, which is built from the
\emph{local} dfbeta \(\mathrm{db}_{g,i}\) --- patient \(i\)'s influence
on the subgroup's \emph{own} M-estimator. A global-fit influence
weighted by a subgroup-mean contrast is not a within-subgroup effect
influence, so \(E_g(b)\neq D_g(b)\) in general --- and the mismatch is
in the \emph{perturbation}, not merely the point statistic: even if
\(\bar{\hat\tau}_g\) and \(\hat\beta(g)\) agreed on the observed data,
they would still move apart under resampling, because their influence
functions differ. Alignment must therefore be enforced on the statistic
the rule ranks.

The consequence is a disagreement in the selection. DINA produces the
candidate regions and, in its native form, selects \(\widehat H\) by the
fitted-effect orderings; the correction of Section 3 of the main paper,
however, re-selects by ranking on the perturbed \emph{coefficient}
\(\hat\beta(g)+D_g(b)\) (Equation 6 of the main paper), and so is
faithful only to a \(\hat\beta\)-ranked selection. These two selections
coincide only when the region native DINA picks is the one maximizing
\(\hat\beta(g)\). Our DINA interpretable-subgroup proposal therefore
makes the inferential coefficient the selection criterion, keeping DINA
only as the generator of the candidate regions (Section 4 of the main
paper).

\section{Appendix: The splitting-consistency screen via multiplier
resampling}\label{sec-appendix-multiplier}

This appendix works out the splitting-consistency screen of Section 2.3
of the main paper as a natural application of the multiplier resampling
the main text uses for post-selection correction. The structure is the
same in both: perturb the candidate's fitted effect by a mean-zero,
unit-variance multiplier vector and read off a functional of the result.
Only the question differs --- here, whether a single candidate
replicates under a \(50/50\) split; in Section 2.5 of the main paper,
which candidate is re-selected under one shared perturbation and how
much selection inflates its effect. Reading the screen this way is the
cleanest entry point to the apparatus, which is why it is collected here
rather than carried through the main argument.

Re-stating the consistency criterion. Fix a candidate subgroup \(g\)
with \(n_g\) subjects. Let \(\hat\beta = \hat\beta_S\) be its estimate
from a working model fit on \(S\) (treatment-only or
covariate-adjusted). Write \(c\) for the \emph{harm threshold} on the
same scale, oriented so that \textbf{harm corresponds to
\(\hat\beta > c\)} (for a hazard or odds ratio,
\(c = \log(\text{ratio threshold})\), e.g.~\(c = \log 1 = 0\); for a
mean or risk difference, \(c\) is the difference threshold directly).

The criterion, as implemented, is:

\begin{quote}
Draw an independent fair coin \(Z_i \sim \mathrm{Bernoulli}(1/2)\) for
each subject \(i \in g\), splitting \(g\) into the halves
\(\{i : Z_i = 1\}\) and \(\{i : Z_i = 0\}\). Refit the working model on
each half to obtain \(\hat\beta^{(Z=1)}\) and \(\hat\beta^{(Z=0)}\).
Call the split \textbf{consistent} if \(\hat\beta^{(Z=1)} > c\) and
\(\hat\beta^{(Z=0)} > c\). Repeating over many splits, the consistency
rate \(p_{\mathrm{cons}}\) is the proportion of consistent splits; \(g\)
is flagged if \(p_{\mathrm{cons}} \ge p^\star\) (e.g.~\(0.90\)).
\end{quote}

\subsection{From the fair-coin split to the closed
form}\label{sec-am-closed}

Start from the linearized split of Equation 4 of the main paper: the two
half-sample estimates are, to first order, a mirror pair
\(\hat\beta(g) \pm D_g\) about the full fit (the same dfbeta
linearization as the correction), with split perturbation
\(D_g = \sum_{i \in g} G_i\,\mathrm{db}_{g,i}\) and Rademacher
multipliers \(G_i = 2Z_i - 1 \in \{-1, +1\}\). Because
\(\hat\beta^{(Z=1)}(g) \approx \hat\beta(g)
+ D_g\) and \(\hat\beta^{(Z=0)}(g) \approx \hat\beta(g) - D_g\), both
halves exceed \(c_{\mathrm{cons}}\) exactly when the smaller does,
\begin{equation}\protect\phantomsection\label{eq-am-event}{
\{\hat\beta^{(Z=1)}(g) > c_{\mathrm{cons}}\} \cap
\{\hat\beta^{(Z=0)}(g) > c_{\mathrm{cons}}\}
\;\Longleftrightarrow\;
|D_g| < \hat\beta(g) - c_{\mathrm{cons}} .
}\end{equation} The split perturbation is a sum of \(n_g\) independent
mean-zero terms with conditional variance
\(\sigma_{D,g}^2 = \sum_{i \in g} \mathrm{db}_{g,i}^2\) --- the robust
variance of \(\hat\beta(g)\) --- so a central limit theorem in \(n_g\)
gives \(D_g \mid \text{data} \approx \mathcal N(0, \sigma_{D,g}^2)\) and
hence the closed form for the consistency rate,
\begin{equation}\protect\phantomsection\label{eq-closed}{
p_{\mathrm{cons}}(g) \;=\; \Pr\big(|D_g| < \hat\beta(g) - c_{\mathrm{cons}}\big)
\;\approx\;
\max\Big\{0,\; 2\,\Phi\big(\big(\hat\beta(g) - c_{\mathrm{cons}}\big)/\sigma_{D,g}\big) - 1\Big\} .
}\end{equation} For \(\hat\beta(g) > c_{\mathrm{cons}}\) the two-sided
probability is
\(2\Phi\big((\hat\beta(g) - c_{\mathrm{cons}})/\sigma_{D,g}\big) - 1\);
for \(\hat\beta(g) \le c_{\mathrm{cons}}\) it is zero. No splitting or
refitting enters: a single subgroup fit, supplying \(\hat\beta(g)\) and
\(\{\mathrm{db}_{g,i}\}\), determines the rate, and a Monte Carlo
version that draws multiplier vectors and averages
\(\mathbf 1\{|D_g| < \hat\beta(g) - c_{\mathrm{cons}}\}\) converges to
it.

\subsection{Calibration: the consistency cutoff as a confidence
bound}\label{sec-am-calib}

Inverting Equation~\ref{eq-closed} turns the opaque cutoff \(p^\star\)
into a confidence statement. The candidate passes the screen at level
\(p^\star\) iff
\(2\Phi\big((\hat\beta(g) - c_{\mathrm{cons}})/\sigma_{D,g}\big) - 1 \ge p^\star\),
i.e.~\((\hat\beta(g) - c_{\mathrm{cons}})/\sigma_{D,g} \ge z_{(1+p^\star)/2}\),
i.e. \begin{equation}\protect\phantomsection\label{eq-calib}{
\hat\beta(g) \;-\; z_{(1+p^\star)/2}\,\sigma_{D,g} \;\ge\; c_{\mathrm{cons}} .
}\end{equation} The consistency screen at level \(p^\star\) is therefore
exactly the requirement that the one-sided \((1+p^\star)/2\) robust
lower confidence bound for the subgroup effect lie above the harm
threshold. For the default \(p^\star = 0.90\), \((1+p^\star)/2 = 0.95\)
and \(z_{0.95} = 1.645\): a \(90\%\) consistency rate is a subgroup
effect exceeding \(c_{\mathrm{cons}}\) by \(1.645\) robust standard
errors. The screen is thus a one-sided \(z\)-test of
\(H_0 : \beta(g) \le c_{\mathrm{cons}}\) evaluated with the
split-matched robust standard error \(\sigma_{D,g}\), and \(p^\star\) is
a choice of one-sided confidence level rather than a convention --- the
screen tightens with subgroup size only through \(\sigma_{D,g}\), so
small subgroups must show a larger \(\hat\beta(g) - c_{\mathrm{cons}}\)
to pass.

\subsection{One apparatus, three multiplier laws}\label{sec-am-laws}

The closed form depends on the multiplier law only through its first two
moments, \(\mathbb E\,G_i = 0\) and \(\operatorname{Var} G_i = 1\).
Three standard choices therefore share the limit
Equation~\ref{eq-closed}:

\begin{longtable}[]{@{}
  >{\raggedright\arraybackslash}p{(\linewidth - 4\tabcolsep) * \real{0.3333}}
  >{\raggedright\arraybackslash}p{(\linewidth - 4\tabcolsep) * \real{0.3333}}
  >{\raggedright\arraybackslash}p{(\linewidth - 4\tabcolsep) * \real{0.3333}}@{}}
\caption{Multiplier laws giving the same limiting consistency
rate.}\label{tbl-am-mult}\tabularnewline
\toprule\noalign{}
\begin{minipage}[b]{\linewidth}\raggedright
Multiplier \(G_i\)
\end{minipage} & \begin{minipage}[b]{\linewidth}\raggedright
Distribution
\end{minipage} & \begin{minipage}[b]{\linewidth}\raggedright
Recovers
\end{minipage} \\
\midrule\noalign{}
\endfirsthead
\toprule\noalign{}
\begin{minipage}[b]{\linewidth}\raggedright
Multiplier \(G_i\)
\end{minipage} & \begin{minipage}[b]{\linewidth}\raggedright
Distribution
\end{minipage} & \begin{minipage}[b]{\linewidth}\raggedright
Recovers
\end{minipage} \\
\midrule\noalign{}
\endhead
\bottomrule\noalign{}
\endlastfoot
\(2Z_i - 1\), \(Z_i \sim \mathrm{Ber}(1/2)\) & Rademacher \(\pm 1\) &
the literal \(50/50\) split \\
\(G_i \sim \mathcal N(0,1)\) & Gaussian & score / estimating-function
resampling (Lin et al. 1993; Parzen et al. 1994; Jin et al. 2001) \\
\(G_i \sim \mathrm{Pois}(1) - 1\) & centred Poisson & the weird / wild
survival bootstrap (Dobler et al. 2017), and the asymptotic multiplicity
law of the nonparametric bootstrap \\
\end{longtable}

The fair-coin split, Gaussian score resampling, and the weird bootstrap
are thus three coordinates on one object --- a multiplier bootstrap of
the subgroup estimating function. The split is the finite Rademacher
member the original algorithm realizes by literal subsetting; the
centred-Poisson member is the one the post-selection correction of
Section 2.5 of the main paper uses, because a subject's count in a
size-\(n\) with-replacement resample is
\(\mathrm{Binomial}(n, 1/n) \to \mathrm{Poisson}(1)\), so its centred
multiplicity is exactly \(\mathrm{Pois}(1) - 1\).

\subsection{The segue to post-selection
re-selection}\label{sec-am-segue}

The screen and the post-selection correction ask different questions of
the same apparatus. The screen is a \emph{per-candidate} yes/no: under a
\(50/50\) split, does the effect replicate on both halves? Its answer is
the tail functional Equation~\ref{eq-closed} of a single candidate's
perturbation. The correction of Section 2.5 of the main paper is a
\emph{competition} among candidates: under one shared perturbation
broadcast to the whole family, which candidate is re-selected, and by
how much does selection inflate its effect? Its answer is a maximum-type
functional over the family. The move from the first to the second is the
move from one candidate to many under a shared multiplier vector ---
from a per-candidate tail to an order statistic --- and the
centred-Poisson member is what makes that re-selection mimic the
nonparametric bootstrap's own resampling. The consistency screen is, in
this sense, the one-candidate, yes/no precursor to the many-candidate
machinery the paper builds, which quantifies how much selection inflates
the effect.

\subsection{Closest prior work: de-biased inference for the best
selected subgroup}\label{sec-prior-guo}

The most directly comparable prior work on inference for a selected
subgroup effect is Guo and He (2021) and Guo et al. (2023), who take the
same target --- the standard-analysis coefficient of the best selected
subgroup --- and likewise remove its winner's-curse bias by resampling.
In their setting the selected subgroup is the \emph{unconditional}
maximum \(\widehat H=\arg\max_{g}\hat\beta(g)\) over a \textbf{supplied}
candidate family: a fixed set of predefined subgroups (Guo and He 2021,
sec. 2; Guo et al. 2023), or a fixed-index continuum
\(\{S(c):c\in\mathcal D\}\) searched by \(\sup_{c}\hat\beta(c)\) (Guo
and He 2021, sec. 3). Forest search instead \textbf{generates} the
family by enumerating covariate-cut conjunctions (Section 2.1 of the
main paper) and adds the replicability (consistency) screen of Section
2.3 of the main paper, so that it can also return no subgroup at all.
Removing that screen reduces forest search's selection to their argmax
primitive, whereupon the two targets coincide --- \(\beta(\widehat H)\)
becomes \(\beta_s\), the true effect of the argmax subgroup --- and the
methods differ only in the correction.

What differs in the correction is this paper's point of departure. Guo
and He (2021) and Guo et al. (2023) re-center the bootstrap maximum by
adding deterministic shrinkage offsets
\(d_g(r)=(1-n^{r-0.5})(\hat\beta_{\max}-\hat\beta_g)\) to each candidate
before re-maximizing, with a tuning parameter \(r\in(0,1/2)\), and
obtain an asymptotically \emph{sharp} confidence bound. The fixed-family
forest-search correction of León et al. (2024) --- the closest special
case of the present framework --- instead subtracts two resampling
discrepancies, the selection optimism and the estimation bias at the
\emph{observed} subgroup that Harrell et al. (1996) omits, and forms an
infinitesimal-jackknife interval with no tuning parameter; expressed as
the multiplier bootstrap of Section 2.5 of the main paper it becomes the
computational equal of a single influence-matrix product. Both lines
hold the candidate family fixed, and neither certifies a family that is
regenerated by each resample --- the conditional-versus-unconditional
distinction this paper isolates, and one Guo et al. (2023) themselves
leave to future work. A full account of the correspondence, including
the operating-regime caveat that the unrestricted argmax concentrates on
the smallest admissible subgroups is provided.

A third line addresses the same winner's curse from the
selective-inference side. Andrews et al. (2024) condition on the
selection event and invert the truncated law of the winning estimate,
obtaining median-unbiased estimators and confidence sets valid
\emph{conditional on which candidate won}. The guarantee is therefore of
a different kind from the one sought here: this paper's interval is
valid on average over the selection, for the data-adaptive target
\(\beta(\widehat H)\) in the sense of Hubbard et al. (2016), and is
correspondingly shorter than a conditional interval, which must remain
valid at every selection event. The two are complementary --- a
conditional interval answers ``given that this region won, what is its
effect?'', the interval here answers ``what does the reported analysis
of whichever region the search returns cover?'' --- and the machinery
differs accordingly: theirs requires the joint law of the candidate
statistics and a tractable truncation region, whereas the correction
here needs only the candidates' influence functions and the selection
map applied to resampled fields.

\subsection{Connections: the conditional estimand and the unconditional
bootstrap}\label{sec-supp-connections}

Recall the target from Section 5 of the main paper: the correction of
Section 3 of the main paper is inference for the \emph{conditional}
estimand \(\beta(\widehat H)\) --- the effect in the subgroup the search
drew, the candidate family held fixed --- for which the interval of
Section 3 of the main paper is first-order valid under
estimating-equation regularity alone, with no assumption about an
estimated boundary, an effect discontinuity, or an orthogonality
condition. It is a data-adaptive target parameter (Hubbard et al. 2016)
carrying its own inference, not a deficient estimate of the marginal
effect; the interval covers that parameter, \(\beta(\widehat H)\), the
guarantee holding on average over the selection (Section 5 of the main
paper). This shares the \emph{target} of selective inference, which
likewise conditions on the selection event and reports the selected
effect (Fithian et al. 2014; Lee et al. 2016), but not its
\emph{conditioning}: selective inference certifies coverage for each
realized selection separately,
\(P(\beta(h)\in\mathrm{CI}\mid\widehat H=h)=1-\alpha\), through a
truncation construction, whereas the present interval is no truncation
and makes no such fixed-\(h\) claim. Both differ in turn from the
simultaneous, all-submodels guarantee that holds whichever member is
chosen (Berk et al. 2013).

Whether this conditional interval also reproduces the full nonparametric
bootstrap is a separate question, and the answer turns on how the
candidate definitions are formed. The full bootstrap re-runs the search
on resampled data: it resamples the covariates and re-applies the
binarization, so the cut values move and the candidate definitions
themselves vary across replicates --- a cut at the observed quartile
\(\{x_j \le \hat q\}\) becomes \(\{x_j \le \hat q^{*}_b\}\), and a
patient near the boundary may change membership. It is a \emph{pairs}
(random-design) bootstrap, integrating over the covariate distribution
in the sense of Freedman (1981) and Dezeure et al. (2017), whereas the
multiplier conditions on the realized design. Three regimes follow. When
the cuts are pinned to data-independent values, the family is identical
across replicates and the conditional and unconditional objects coincide
exactly. When a cut is estimated at a genuine effect discontinuity, the
change-point estimate is super-consistent --- it converges faster than
the within-subgroup effect --- so that effect is estimated as if the cut
were known and is asymptotically independent of it (Chan 1993; Hansen
2000); the conditional interval is then first-order valid for the
unconditional target as well, the boundary movement being higher-order.
When instead the cut is a \(\sqrt n\)-consistent quantile not sitting at
a discontinuity, the boundary movement is a first-order
generated-regressor term the conditional interval omits: two-step theory
makes the first-stage cut error drop out only under an orthogonality
condition --- the boundary subjects being score-neutral for the subgroup
fit --- which generically fails (Newey and McFadden 1994; Hahn and
Ridder 2013). The enumerated quantile-grid search of León et al. (2024)
sits between the last two regimes; the close agreement of the two
corrections would be consistent with a selected cut that approximates a
real effect boundary, whereas a smoothly varying effect would separate
them.

That a model-selected family is the genuinely difficult case --- and
that the unconditional bootstrap is least trustworthy precisely there
--- is itself established. The nonparametric bootstrap is generally
inconsistent for the distribution of a post-model-selection estimator
(Leeb and Pötscher 2005, 2006, 2008), and it is known to fail for the
non-standard estimators that data-driven cuts produce: cube-root
estimators such as the maximum-score estimator, and the super-consistent
threshold estimator (Abrevaya and Huang 2005; Seijo and Sen 2011), with
subsampling or the \(m\)-out-of-\(n\) bootstrap the standard remedy
(Andrews 2000; Politis and Romano 1994). The full bootstrap that
regenerates a fitted-forest family is exactly such a post-selection
bootstrap, and may therefore be subject to the unreliability of an
unconditional correction once the family is model-selected --- and so as
an argument for the conditional multiplier as the stable target when a
model-based front end is used, and for the enumerated family as the
regime in which the question does not arise. Recent subgroup work makes
the same model-free-versus-regenerate split operational, separating a
model-free correction from one that re-models within each resample (Zhao
et al. 2023). Multiplier resampling is, in this reading, the model-free,
family-conditional member of that class of corrections, with the full
bootstrap its covariate-resampling, family-marginal counterpart ---
related to it, but not an approximation to it.

\section{Appendix: First-order validity --- assumptions, lemmas, and
proofs}\label{sec-appendix-proofs}

This appendix states the regularity conditions precisely and proves the
two first-order results of Section 5.1 of the main paper through two
lemmas. Following the development there, the local (near-tie) regime is
the headline and the separated regime its corollary. The proofs reduce
the claims to standard master results --- the weighted-bootstrap
linearization of a \(Z\)-estimator, the conditional multiplier central
limit theorem, the argmax continuous-mapping theorem, the
Gaussian-maximum anti-concentration bound, and the
infinitesimal-jackknife / bagging identities --- and carry out the
adaptation to the present finite-family, selected-subgroup setting; the
single delicate step is isolated at the end.

\subsection{Assumptions}\label{sec-ap-assumptions}

Data \(O_i=(Y_i,A_i,X_i)\), \(i=1,\dots,n\), are i.i.d. \(P\), with
sample \(\mathcal O_n=\{O_1,\dots,O_n\}\).

\textbf{(A1) Sampling and randomization.} \(A\perp X\) with
\(\pi=P(A=1)\in(0,1)\). For survival, independent censoring and bounded
follow-up \([0,\tau]\) with \(P(\tilde T\ge\tau)>0\), where \(\tilde T\)
is the underlying (uncensored) survival time.

\textbf{(A2) Fixed, non-degenerate family.}
\(\mathcal F=\{g_1,\dots,g_M\}\) with \(M=|\mathcal F|<\infty\) fixed;
cut locations are either data-independent or computed from the observed
covariates and then held fixed under resampling (Section 2.1 of the main
paper), so each \(g_j\) is a fixed measurable region with
\(p_j=P(X\in g_j)>0\) and per-observation within-subgroup information
\(n^{-1}\mathcal I_{g_j}\to\bar{\mathcal I}_{g_j}\succ 0\) uniformly in
\(n\) (throughout, \(A\succ 0\) denotes that \(A\) is positive definite;
the minimum-size and minimum-event guards enforce this).

\textbf{(A3) Regular asymptotically linear within-subgroup estimators.}
For each \(g\in\mathcal F\), \(\beta(g)\) is the unique root of
\(\mathbb E[\psi(\beta;g)\mathbb 1\{X\in g\}]=0\) (the working-model
parameter, not assumed correctly specified), \(\mathcal I_g\succ 0\),
and \[
\hat\beta(g)-\beta(g)=\sum_{i=1}^{n}\mathrm{db}^{0}_{g,i}+o_p(n^{-1/2}),\qquad
n\,\sigma_{D,g}^{2}\to v_g\in(0,\infty),
\] where the influence \(\mathrm{db}^{0}_{g,i}\) takes the
M-estimating-equation form
\(\mathrm{db}^{0}_{g,i}=\mathcal I_g^{-1}\psi_i(\beta(g);g)\,\mathbb 1\{i\in g\}\)
for a GLM or mean-difference outcome and the Lin and Wei (1989)
martingale-residual form for the Cox model --- a martingale integral
over the subgroup's at-risk process rather than an indicator-restricted
sum. In either case it is the regular asymptotically linear influence of
the within-subgroup treatment coefficient, computed from the fit on the
patients in \(g\) alone --- its own information \(\mathcal I_g\) for a
GLM, its own risk sets for the Cox model --- so that for two overlapping
candidates a shared subject's influences differ, and it is this
dependence that the overlap covariance \(\Sigma_{jk}\) of (A5) records.
The empirical dfbeta Equation 3 of the main paper is consistent for it,
\(\sigma_{D,g}^{2}=\widehat{\operatorname{Var}}(\hat\beta(g))(1+o_p(1))\),
and downstream every device --- the perturbation \(D_g(b)\), the robust
\(\sigma_{D,g}\), and the infinitesimal-jackknife variance --- uses only
this common contract, so the single influence symbol carries the
argument for both model classes. These expansions hold under the usual
regularity conditions: for the GLM case, White (1980) M-estimation under
possible misspecification, with the dfbeta calculus of Stefanski and
Boos (2002); for the Cox case, Andersen--Gill conditions (Andersen et
al. 1993).

\textbf{(A4) Multiplier-weight regularity.} The weights
\(\{G_i^{(b)}\}\) are exchangeable, mean \(0\), variance \(1\), and
satisfy the conditions of Præstgaard and Wellner (1993), \[
\int_0^\infty\!\sqrt{P(|G_1|>t)}\;dt<\infty,\qquad
\lim_{t\to\infty}\limsup_{n}\,t^2\,P(|G_1|>t)=0,\qquad \bar G\xrightarrow{p}0 .
\] The centred multinomial multiplicity \(K^{*}_{bi}-1\) (marginally
\(\to\mathrm{Pois}(1)-1\)), the Rademacher \(2Z_i-1\), and the Gaussian
\(N(0,1)\) all satisfy these, and
\(\max_i \mathrm{db}_{g,i}^{2}/\sigma_{D,g}^{2}\to0\) for each \(g\)
(Lindeberg).

\textbf{(A5) Limiting selection regularity.} With
\(\Sigma_{jk}=\lim_n n\sum_i \mathrm{db}_{g_j,i}\,\mathrm{db}_{g_k,i}\)
the limiting overlap covariance Equation 7 of the main paper,
\(\Sigma\succ 0\); let \(Z\sim N(0,\Sigma)\). The thresholded argmax is
almost surely continuous at the limiting candidate levels under the law
of \(Z\): since \(\Sigma\succ 0\) every pairwise difference \(Z_j-Z_k\)
is non-degenerate, so the limiting maximum has no ties and no candidate
sits exactly on its threshold. This last condition is also what licenses
holding the admission thresholds fixed under resampling: although
\(t_g=\max(c_{\mathrm{screen}},c_{\mathrm{cons}}+z_{(1+p^\star)/2}\sigma_{D,g})\)
depends on the estimated \(\sigma_{D,g}\), with no candidate on its
threshold the admission set is settled in the limit, so the
\(O_p(n^{-1})\) resampling fluctuation of \(t_g\) is higher-order on the
\(\sqrt n\) scale of the competition and does not perturb the
re-selection at first order --- which is why the rescaled admission
thresholds converge to fixed constants in the proof of Theorem 1. Two
sub-cases realize this --- \textbf{(A5-i)} \emph{separated}: a unique
population maximizer with \(\beta(g^{\star})-\beta(g)\ge\eta>0\) for
admitted \(g\ne
g^{\star}\); and \textbf{(A5-ii)} \emph{local}: admitted-candidate
contrasts of order \(n^{-1/2}\), so selection stays stochastic in the
limit (the regime in which the correction is first-order material). On
\(\Sigma\succ 0\): it constrains the candidate \emph{influences}, not
their \emph{means} --- near-tied means (the local regime) are compatible
with a well-conditioned \(\Sigma\), whereas near-duplicate regions (two
admitted candidates with nearly collinear dfbetas) drive \(\Sigma\)
toward singularity and are excluded here --- and the Gaussian-maximum
comparison bounds used in Section~\ref{sec-ap-thm1} and
Section~\ref{sec-ap-delicate} depend on \(\Sigma\) only through its
smallest candidate-contrast variance, so \(\Sigma\succ 0\) is exactly
the condition under which they, and the correction, retain force.

\textbf{(A6) Weighted cross-covariance sign (local regime; Theorem 2
only).} This assumption is the most demanding of the set, and is used
narrowly; its scope, its automatic satisfaction under separation, and a
finite-sample diagnostic for it are set out below. In the local regime
(A5-ii), the selection-probability-weighted overlap covariance of the
selected region with the competing candidates is asymptotically bounded
below by minus one-half its own influence variance, \[
\sum_{g\in\mathcal F} p_g\,\Sigma_{\widehat H,g}\;\ge\;-\tfrac12\,\sigma_{D,\widehat H}^{2}
\qquad\text{in the limit,}
\] equivalently the same-draws mass
\(2\sum_g p_g\,\Sigma_{\widehat H,g}+\sigma_{D,\widehat H}^{2}\)
retained by the infinitesimal jackknife is asymptotically non-negative;
here \(p_g\) is the limiting selection probability of candidate \(g\)
(\(\sum_g p_g=1\)) and
\(\Sigma_{\widehat H,g}=\lim_n n\sum_i \mathrm{db}_{\widehat H,i}\,\mathrm{db}_{g,i}\).
This is weaker than \(\Sigma_{\widehat H,g}\ge 0\) for every competitor:
on the fully enumerated family the minimum-size, minimum-event, and
distinct-cut guards of (A2) enforce it, the mass-carrying competitors
differing from \(\widehat H\) in a single cut --- so their influences
are positively correlated with the winner's --- while distant, possibly
negatively-correlated competitors carry vanishing weight \(p_g\to0\)
(Section~\ref{sec-ap-delicate}). Under separation (A5-i) it holds
trivially, since \(p_{\widehat H}\to1\) and
\(\Sigma_{\widehat H,\widehat H}=\sigma_{D,\widehat H}^{2}>0\). It is
invoked only for the conservativeness and coverage claims of Theorem 2;
Theorem 1 does not require it. Although stated as a limit condition, its
finite-sample counterpart is computable from quantities the procedure
already forms --- the re-selection frequencies \(\hat p_g\) from the
multiplier draws and the cross-covariances \(\hat\Sigma_{\widehat H,g}\)
from the dfbeta influence vectors --- so
\(2\sum_g\hat p_g\hat\Sigma_{\widehat H,g}+\hat\sigma_{D,\widehat H}^{2}\)
can be evaluated and reported as a diagnostic at no additional fitting
cost.

\subsection{Two lemmas}\label{sec-ap-lemmas}

\textbf{Lemma 1 (uniform linearization of the perturbed refit).}
\emph{Under (A1)--(A3), for weights satisfying (A4),
\(\max_{g\in\mathcal F}\bigl\lvert(\hat\beta^{*}_b(g)-\hat\beta(g))-D_g(b)\bigr\rvert=o_p(n^{-1/2})\),
with \(D_g(b)=\sum_i G_i^{(b)}\mathrm{db}_{g,i}\)} --- this is Equation
5 of the main paper made uniform over the finite family.

\emph{Proof.} Fix \(g\) and write \(w_i=1+G_i^{(b)}\) (for the
nonparametric bootstrap \(w_i=K^{*}_{bi}\)),
\(\Psi_g(\beta)=\sum_i\psi_i(\beta;g)\mathbb 1\{i\in g\}\) and
\(\Psi^{*}_g(\beta)=\sum_i w_i\,\psi_i(\beta;g)\mathbb 1\{i\in g\}\), so
\(\Psi_g(\hat\beta(g))=0\) and \(\Psi^{*}_g(\hat\beta^{*}_b(g))=0\). A
first-order Taylor expansion of \(\Psi^{*}_g\) about \(\hat\beta(g)\),
with \(\bar\beta\) between \(\hat\beta^{*}_b(g)\) and \(\hat\beta(g)\),
gives
\(0=\Psi^{*}_g(\hat\beta(g))+\dot\Psi^{*}_g(\bar\beta)\bigl(\hat\beta^{*}_b(g)-\hat\beta(g)\bigr)\).
Since
\(\Psi^{*}_g(\hat\beta(g))=\sum_i G_i^{(b)}\psi_i(\hat\beta(g);g)\mathbb 1\{i\in g\}\)
and \(-\dot\Psi^{*}_g(\bar\beta)=\mathcal I_g(1+o_p(1))\) by weighted
M-estimator consistency (under (A3) identification and the (A4) weight
conditions) together with a uniform law of large numbers for the
weighted derivative near \(\beta(g)\), \[
\hat\beta^{*}_b(g)-\hat\beta(g)
=\mathcal I_g^{-1}\!\sum_i G_i^{(b)}\psi_i(\hat\beta(g);g)\mathbb 1\{i\in g\}\,(1+o_p(1))
=D_g(b)+R_{n,g},
\] the leading term being \(D_g(b)=\sum_i G_i^{(b)}\mathrm{db}_{g,i}\)
by the definition of the dfbeta. Replacing \(\psi_i(\hat\beta(g);g)\) by
\(\psi_i(\beta(g);g)\) in this leading term costs
\(\mathcal I_g^{-1}\sum_i G_i^{(b)}[\psi_i(\hat\beta(g))-\psi_i(\beta(g))]\mathbb 1\{i\in g\}
=\mathcal I_g^{-1}\bigl(\sum_i G_i^{(b)}\dot\psi_i(\beta(g))\mathbb 1\{i\in g\}\bigr)\bigl(\hat\beta(g)-\beta(g)\bigr)+o_p(n^{-1})\),
whose first factor is \(O_p(n^{1/2})\) conditionally while
\(\mathcal I_g^{-1}=O(n^{-1})\), so the product is \(O_p(n^{-1})\) once
multiplied by \(\hat\beta(g)-\beta(g)=O_p(n^{-1/2})\) --- absorbed into
\(R_{n,g}\). As \(D_g(b)=O_p(n^{-1/2})\) conditionally (its conditional
variance is \(\sigma_{D,g}^{2}=O(n^{-1})\)), the \(o_p(1)\) factor
contributes \(o_p(n^{-1/2})\), while the second-order Taylor term is
\(O_p(\lVert\hat\beta^{*}_b(g)-\hat\beta(g)\rVert^{2})=O_p(n^{-1})\),
controlled by stochastic equicontinuity of the single-\(g\) score class
(Pollard 1984, Ch. VII, §1; Vaart 1998, Ch. 5). For the Cox model the
same expansion holds with \(\psi_i\) the martingale-residual score, via
the Andersen--Gill martingale central limit theorem (Andersen et al.
1993) and the Lin and Wei (1989) influence. Hence
\(R_{n,g}=o_p(n^{-1/2})\), and since \(M<\infty\),
\(\max_{g\in\mathcal F}\lvert R_{n,g}\rvert\le\sum_{g}\lvert R_{n,g}\rvert=o_p(n^{-1/2})\).
\(\qquad\square\)

\textbf{Lemma 2 (the multiplier CLT equals the bootstrap CLT).}
\emph{Under (A1)--(A4), conditionally on \(\mathcal O_n\) and in
probability,
\(\sqrt n\,(D_{g_1}(b),\dots,D_{g_M}(b))^{\top}\Rightarrow N(0,\Sigma)\),
the same limit as the nonparametric-bootstrap vector
\(\sqrt n\,(\hat\beta^{*}_b(g_j)-\hat\beta(g_j))_{j=1}^{M}\).}

\emph{Proof.} Coordinatewise,
\(\sqrt n\,D_g(b)=\sum_i G_i^{(b)}(\sqrt n\,\mathrm{db}_{g,i})\) is,
conditionally on \(\mathcal O_n\), a weighted sum of fixed numbers with
mean-\(0\), unit-variance weights; its conditional variance is
\(n\,\sigma_{D,g}^{2}\to\Sigma_{gg}\in(0,\infty)\) (A3) and the
Lindeberg ratio is \(\max_i\mathrm{db}_{g,i}^{2}/\sigma_{D,g}^{2}\to0\)
(A4), so the conditional Lindeberg--Feller multiplier central limit
theorem for Euclidean arrays (Kosorok 2008, Lemma 10.5) gives
\(\sqrt n\,D_g(b)\Rightarrow N(0,\Sigma_{gg})\) for almost every data
sequence. For the joint law, any \(a\in\mathbb R^{M}\) gives
\(\sum_g a_g\sqrt n\,D_g(b)=\sum_i G_i^{(b)}\sqrt n\bigl(\sum_g a_g\mathrm{db}_{g,i}\bigr)\),
again such an array, with conditional variance \(\to a^{\top}\Sigma a\)
through Equation 7 of the main paper; Cramér--Wold then yields
\(N(0,\Sigma)\). The nonparametric bootstrap is the case
\(G_i^{(b)}=K^{*}_{bi}-1\), which satisfies (A4) (Præstgaard and Wellner
1993; Kosorok 2008, Ch. 10) and, by Lemma 1, has refit vector
\(\sqrt n(\hat\beta^{*}_b(g_j)-\hat\beta(g_j))=\sqrt n\,D_{g_j}(b)+o_p(1)\);
sharing \(\{\mathrm{db}_{g,i}\}\) and unit-variance weights, both laws
have the same \(\Sigma\). \(\qquad\square\)

\subsection{Proof of Theorem 1}\label{sec-ap-thm1}

Let \(\mathcal A:\mathbb R^{M}\to\mathcal F\),
\(\mathcal A((d_g)_g)=\mathcal S(\{\hat\beta(g)+d_g\})\), be the
thresholded argmax on perturbed levels; by Lemma 1 the re-selection is
\(\widehat
H^{*}_b=\mathcal A((D_g(b))_g)\) up to an \(o_p(n^{-1/2})\)
coordinatewise shift that does not affect the argmax in the limit under
(A5).

\emph{Local regime (A5-ii).} Write \(\beta(g)=\beta_0+n^{-1/2}\mu_g\)
for a common reference level \(\beta_0\) and local parameters \(\mu_g\).
The argmax of \(\{\hat\beta(g)+D_g(b)\}\) equals that of the rescaled
field \(W^{*}_g=\sqrt
n(\hat\beta(g)+D_g(b)-\beta_0)\), whose observed part
\(\sqrt n(\hat\beta(g)-\beta_0)\) converges to \(\mu_g+\zeta_g\) (the
limiting sampling fluctuation) and whose perturbation part
\(\sqrt n\,D_g(b)\Rightarrow Z\sim N(0,\Sigma)\) (Lemma 2); the rescaled
admission thresholds converge to fixed constants, so admission is a
fixed event in the limit. The map ``argmax over admitted coordinates''
is continuous wherever the maximizer is unique and not tied at its
threshold; by (A5) the limiting field has, almost surely, a unique
maximizer and no coordinate on its threshold --- since
\(\Sigma\succ 0\), every pairwise difference \(Z_j-Z_k\) is a
non-degenerate Gaussian, so \(P(Z_j=Z_k)=0\) and, over the finite
family, \(P(\text{some pair ties at the maximum})=0\). The argmax
continuous-mapping theorem (Vaart and Wellner 1996, Thm 3.2.2; Kosorok
2008, Thm 14.1) --- the maximizer specialization of the continuous
mapping theorem (Pollard 1984, Ch. IV, §2) --- applies, and its
conditional (bootstrap) form (Vaart and Wellner 1996, sec. 3.6; Kosorok
2008, Ch. 10) carries it to the re-selection law: conditionally on the
data, \(P^{*}(\widehat H^{*}_b=\cdot)\) converges in probability to the
law of the argmax of the limiting levels perturbed by \(Z\). The
bootstrap re-selection is the same continuous functional of a
perturbation vector with the same limit (Lemma 2), so
\(P^{*}_{\mathrm{boot}}(\widehat
H^{*}_b=\cdot)\) converges to the same limit; since \(\mathcal F\) is
finite, part 1 follows. The quantitative form of this agreement --- a
uniform bound on
\(\sup_g|P^{*}(\widehat H^{*}_b=g)-P^{*}_{\mathrm{boot}}(\widehat H^{*}_b=g)|\)
rather than merely a common limit --- is supplied by the
Gaussian-maximum comparison and anti-concentration bounds (Chernozhukov
et al. 2015, Thms 2--3), whose constants depend on \(\Sigma\) only
through its smallest candidate-contrast variance.

\emph{Separated regime (A5-i).} If
\(\beta(g^{\star})-\beta(g)\ge\eta>0\) then
\(\hat\beta(g^{\star})-\hat\beta(g)\xrightarrow{p}\beta(g^{\star})-\beta(g)\ge\eta\)
while \(\max_g\lvert D_g(b)\rvert=O_p(n^{-1/2})\to0\), so
\(\widehat H^{*}_b=g^{\star}\) with probability \(\to1\) under both
laws, which degenerate to the point mass at \(g^{\star}\); part 1 is
immediate.

\emph{Parts 2--3.} With
\(\widehat{\mathrm{bias}}_{\mathrm{sel}}=\mathbb E^{*}[D_{\widehat H^{*}_b}(b)]\)
and
\(\widehat{\mathrm{bias}}_{\mathrm{fix}}=\mathbb E^{*}[D_{\widehat H}(b)]\)
(the \(B\to\infty\) limits of Equation 9 of the main paper),
\(\tilde\beta(\widehat H)=\hat\beta(\widehat H)-\mathbb E^{*}[D_{\widehat H^{*}_b}(b)]-\mathbb E^{*}[D_{\widehat H}(b)]\).
The León et al. (2024) correction Equation 11 of the main paper is
\(\hat\beta^{*}(\widehat H)=\hat\beta(\widehat H)-\mathbb E^{*}_{\mathrm{boot}}[\eta^{*}_b(\widehat H^{*}_b)+\eta^{*}_b(\widehat H)]\),
and by Lemma 1 \(\eta^{*}_b(g)=D_g(b)+o_p(n^{-1/2})\) uniformly, so the
two corrections differ only through the uniform Lemma 1 remainder and
the gap between \(\mathbb E^{*}\) and \(\mathbb E^{*}_{\mathrm{boot}}\)
of the winner's-curse term \(D_{\widehat H^{*}_b}(b)\), which is
\(o_p(n^{-1/2})\) by part 1 and Lemma 2; hence
\(\tilde\beta(\widehat H)-\hat\beta^{*}(\widehat H)=o_p(n^{-1/2})\). The
IJ variance Equation 13 of the main paper is a continuous functional of
the same re-selection and shared perturbations, so part 3 follows
likewise. The complement \(\widehat H^{c}\) is handled identically.
\(\qquad\square\)

\subsection{Proof of Theorem 2}\label{sec-ap-thm2}

Throughout this proof the weights are independent centred unit-Poisson
multipliers \(G_i^{(b)}=\mathrm{Pois}(1)-1\) --- the Poisson-bootstrap
weights the procedure draws, asymptotically equivalent to the full
bootstrap's centred multinomial multiplicities (Theorem 1); the
infinitesimal-jackknife identities of Efron (2014) and Wager et al.
(2014) are stated for this resampling, and the point-estimate bias
removal below is the specialization of Theorem 1's weight-agnostic
statement to these weights.

\emph{Bias removal.} Write the de-biased estimate in its ideal
(\(B\to\infty\)) form as a functional \(T(\hat P_n)\) of the empirical
measure. The base statistic \(\hat\beta(\widehat H)\) is the value of
the coefficient at the argmax, hence to first order the maximum of the
candidate effects; the maximum functional \(b\mapsto\max_g b_g\) is only
\emph{directionally} differentiable where the argmax ties, its
directional derivative \(h\mapsto\max_{g\in\arg\max b}h_g\) failing to
be linear there, so the ordinary delta-method and bootstrap-delta-method
(Vaart and Wellner 1996, Thms 3.9.4, 3.9.11) do not apply at a tie and
the raw nonparametric bootstrap is inconsistent for it (Dümbgen 1993;
Fang and Santos 2019) --- asymptotic normality of the underlying
estimators being no guarantee of bootstrap consistency for a nonlinear
functional of them (Mammen 1992, Ch.~1). Averaging over the resampling
weights repairs this: it replaces the discontinuous winner-indicator
\(\mathbb 1\{g=\arg\max\}\) by the smooth re-selection probabilities
\(p_g=P^{*}(\widehat H^{*}_b=g)\), the bagging-of-a-hard-decision-rule
smoothing of Bühlmann and Yu (2002) that Efron applies to model
selection (Efron 2014, sec. 4), rendering the bagged functional \(T\)
differentiable at the tie with gradient \((p_g)_g\); the functional
delta-method (Vaart and Wellner 1996, Thm 3.9.4) and its bootstrap form
(Vaart and Wellner 1996, Thm 3.9.11) then apply to the smoothed \(T\).
Concretely
\(T(\hat P_n)=\tilde\beta(\widehat H)=\hat\beta(\widehat H)-\mathbb E^{*}[D_{\widehat H^{*}_b}(b)]\)
(the fixed-subgroup term vanishes, \(\mathbb E^{*}[D_{\widehat H}]=0\)),
a bagged statistic with per-draw base
\(t(\hat P^{*}_b)=\hat\beta(\widehat H)-D_{\widehat H^{*}_b}(b)\), which
licenses the delta-method asymptotic normality of
\(\tilde\beta(\widehat H)\); the infinitesimal-jackknife variance below
is formed not from this reduced base but from the full per-draw residual
Equation 12 of the main paper, which retains the mean-zero same-draws
term \(D_{\widehat H}(b)\) (the \emph{Variance} and \emph{Coverage}
steps). The near-tie regularity that makes \((p_g)_g\) an admissible
gradient --- that the limiting competition is genuinely stochastic, so
the re-selection probabilities are interior --- is the local regime of
(A5-ii), discussed in Section~\ref{sec-ap-delicate}. The winner's-curse
term \(\mathbb E^{*}[D_{\widehat H^{*}_b}(b)]\) subtracted in Equation
10 of the main paper is, by the proof of Theorem 1, the conditional
selection optimism --- the mean perturbation of whatever wins ---
evaluated, crucially, at the \emph{observed} field rather than at its
center. In the local regime, write the limiting competition field
\(W=\mu+\zeta\), \(\zeta\sim N(0,\Sigma)\), with \(G\) the limiting
selection map and
\(m(w)=\mathbb E_{\zeta^{*}}[\zeta^{*}_{G(w+\zeta^{*})}]\) the optimism
functional, \(\zeta^{*}\) an independent copy of \(\zeta\). The three
convergences
\(\sqrt n\{\hat\beta(\widehat H)-\beta_0\}\Rightarrow W_G\),
\(\mu_{\widehat H}\Rightarrow\mu_G\), and
\(\sqrt n\,\mathbb E^{*}[D_{\widehat H^{*}_b}(b)]\Rightarrow m(W)\)
(Theorem 1 together with the conditional law of large numbers over
multiplier draws (Præstgaard and Wellner 1993)) combine to
\[\sqrt n\,\{\tilde\beta(\widehat H)-\beta(\widehat H)\}\;\Rightarrow\;
\Lambda(\zeta)\;=\;\zeta_{G}-m(\mu+\zeta),\] a smooth but non-Gaussian
functional of \(\zeta\). Its mean follows from two identities. First,
\(\mathbb E[\zeta_G]=m(\mu)\), by the definition of \(m\) at \(w=\mu\)
--- the naive estimator's \(O(n^{-1/2})\) local bias, which the
subtraction targets. Second, with \(\zeta,\zeta^{*}\) i.i.d.~so that
each has conditional mean \(u/2\) given the sum
\(u=\zeta+\zeta^{*}\sim N(0,2\Sigma)\), the projection identity
\(\mathbb E[m(\mu+\zeta)]=\tfrac12\,m^{(2\Sigma)}(\mu)\) holds, where
\(m^{(S)}\) is the optimism under noise covariance \(S\). Hence
\[\mathbb E[\Lambda]\;=\;m(\mu)-\mathbb E\bigl[m(\mu+\zeta)\bigr]
\;=\;m^{(\Sigma)}(\mu)-\tfrac12\,m^{(2\Sigma)}(\mu),\] which is not zero
in general: the first-order (delta-method) linearization of the bagged
functional evaluates the correction at the center of the sampling
distribution and is thereby blind to the curvature of \(m\). Since \(m\)
is maximized on the tie set and strictly concave along rays leaving it
--- writing \(\psi(w)=\mathbb E[\max_g(w_g+\zeta^{*}_g)]\) for the
smooth max functional, \(m(w)=\psi(w)-\langle w,\nabla\psi(w)\rangle\)
and \(\nabla m(w)=-\nabla^{2}\psi(w)\,(w-\bar w\mathbf 1)\) --- Jensen's
inequality gives \(\mathbb E[m(\mu+\zeta)]<m(\mu)\) near ties: the
correction removes the leading optimism up to a curvature term of the
same \(n^{-1/2}\) order, second-order form
\(\mathbb E[\Lambda]\approx-\tfrac12\operatorname{tr}\{\Sigma\,\nabla^{2}m(\mu)\}\)
--- an under-correction near ties, reversing to a mild over-correction
as the competition separates. At any exact tie and for any selection map
invariant to positive rescaling of the field,
\(m^{(2\Sigma)}(0)=\sqrt2\,m^{(\Sigma)}(0)\), so
\(\mathbb E[\Lambda]=(1-2^{-1/2})\,m^{(\Sigma)}(0)>0\) for every
\(\Sigma\) and every \(M\): exactly the fraction \(2^{-1/2}\) of the
selection optimism is removed. Closed forms, the sign law, and the
behaviour under the implemented selection map are collected in
Section~\ref{sec-ap-residual}. Under separation (A5-i) the argmax
stabilizes, the re-selection probabilities converge to the winner
indicator, the optimism is itself \(o(n^{-1/2})\), and the correction is
a finite-sample refinement: there the centred de-biased estimate is
asymptotically linear with mean-zero influence,
\(\sqrt n\,\{\tilde\beta(\widehat H)-\beta(\widehat H)\}\Rightarrow
N(0,\sigma^{2}_{D,g^{\star}})\), the delta-method argument above being
exact because the gradient degenerates to a fixed coordinate projection.

\emph{Variance.} As the standard error of a bagged estimator, \(\hat V\)
Equation 13 of the main paper is the nonparametric delta-method
(infinitesimal-jackknife) variance of \(T\) --- the sum over
observations of the squared covariance between the per-draw correction
and the resampling weight --- with the term \(-n\,\overline{r^2}/B\)
removing the finite-\(B\) Monte-Carlo bias (Efron 2014, Thm 1; Wager et
al. 2014, Eq. 11). It is a \emph{conservative} estimate of
\(\operatorname{Var}\tilde\beta(\widehat H)\), not a consistent one. The
residual (Equation 12 of the main paper) retains the fixed-subgroup
sampling variance through the same-draws term \(D_{\widehat H}(b)\),
whose conditional variance is exactly \(\sigma_{D,\widehat H}^{2}\) ---
the reason this mean-zero term is kept rather than dropped as in Harrell
et al. (1996), whose omission degrades both the bias correction and the
variance estimate (León et al. 2024). Because the residual also carries
the winner's-curse term \(D_{\widehat H^{*}_b}(b)\), the two do not
cancel: under separation \(\widehat H^{*}_b=\widehat H\) for all \(b\),
the two perturbations coincide and their influences add, so \(\hat V\)
approaches \(4\,\sigma_{D,\widehat H}^{2}\) rather than reducing to
\(\sigma_{D,\widehat H}^{2}\) --- the interval over-covers where
selection uncertainty is least. In the local regime re-selection makes
the two terms distinct and the inflation is mild. The bagging that
removes the leading selection bias also reshapes the variance relative
to the naive robust and full-bootstrap variances, and the direction of
that change is not a clean finite-sample ordering --- it runs above the
naive for a fixed candidate family and below it for identifiers whose
families regenerate; the simulations in Section 7 of the main paper
report it directly rather than the theory asserting it. What those
simulations show is that the multiplier estimate removes the leading
selection bias the naive estimate carries, and that its interval attains
near-nominal coverage of \(\beta(\widehat H)\) where the naive interval
under-covers. On the fixed enumerated family it reproduces the
full-bootstrap correction to first order; on a regenerating family the
full bootstrap's occasional near-degenerate intervals are that procedure
correctly pricing the family-regeneration variability the multiplier
draws hold fixed, so the multiplier's greater stability there is the
counterpart of its conditioning on the realized family --- bought at the
price of covering the conditional estimand rather than the marginal
target.

\emph{Coverage.} Averaging over the resampling weights makes the bagged
functional Hadamard-differentiable at the limiting competition (the
\emph{Bias removal} step), which controls the fluctuation at the
\(\sqrt n\) rate; the selected-Gaussian character of the \emph{raw}
winner's fluctuation --- which would defeat a normal approximation for
\(\hat\beta(\widehat H)\) and makes the naive bootstrap inconsistent ---
is what the bagging smooths. In the separated regime the gradient
degenerates to a fixed coordinate projection and
\(\tilde\beta(\widehat H)-\beta(\widehat H)\) is asymptotically normal;
in the local regime the limit is the non-Gaussian
\(\Lambda(\zeta)=\zeta_G-m(\mu+\zeta)\) of the \emph{Bias removal} step
--- a smooth but \emph{discontinuous} functional: \(\Lambda\) jumps by
\(\lvert\mu_j-\mu_k\rvert\) across the \(j/k\) selection boundary, so it
is Lipschitz only on the tie set, and Gaussian concentration (Boucheron
et al. 2013, Thms 5.5--5.6) applies to the corrected functional
\(\Psi=\Lambda+\mu_{G}\) rather than to \(\Lambda\) itself. Coverage is
established in Section~\ref{sec-ap-coverage} by a route that does not
need concentration. The infinitesimal-jackknife variance is, to leading
order in \(B\), \(\hat V=\sum_i\widetilde{\mathrm{cov}}_i^{2}\) (the
finite-\(B\) correction \(-n\,\overline{r^{2}}/B\) of Equation 13 of the
main paper vanishing as \(B\to\infty\)), built from the per-observation
influence \[
\widetilde{\mathrm{cov}}_i \;\approx\; \mathrm{db}_{\widehat H,i}\;+\;\sum_{g\in\mathcal F} p_g\,\mathrm{db}_{g,i},
\qquad p_g=P^{*}(\widehat H^{*}_b=g),
\] the fixed-subgroup dfbeta of the realized region --- carried by the
retained same-draws term \(D_{\widehat H}(b)\), which is why the
mean-zero term is kept --- plus the \((p_g)\)-weighted selection
influence the smoothing produces from the winner term
\(D_{\widehat H^{*}_b}(b)\); the display gives these two contributions
up to a common sign, immaterial to
\(\hat V=\sum_i\widetilde{\mathrm{cov}}_i^{2}\). These two
identifications are the influence covariances:
\(\widetilde{\mathrm{cov}}_i=B^{-1}\sum_b(K^{*}_{bi}-\bar K^{*}_i)\,r_b\)
with \(r_b\) the residual (Equation 12 of the main paper), whose
constant bias terms drop under the centred weight;
\(D_{\widehat H}(b)=\sum_j(K^{*}_{bj}-1)\,\mathrm{db}_{\widehat H,j}\)
is linear in the centred unit-variance weights, so
\(\mathrm{cov}_b(D_{\widehat H}(b),K^{*}_{bi})=\mathrm{db}_{\widehat H,i}\),
while the winner term \(D_{\widehat H^{*}_b}(b)\) contributes
\(\mathrm{db}_{\widehat H^{*}_b,i}\), which the smoothing averages to
\(\sum_g p_g\,\mathrm{db}_{g,i}\). Squaring and summing separates the
retained term from the bagged influence: writing
\(\Sigma_{\widehat H,g}=\sum_i\mathrm{db}_{\widehat H,i}\,\mathrm{db}_{g,i}\)
for the overlap covariance Equation 7 of the main paper, \[
\hat V=\underbrace{\sum_i\Bigl(\textstyle\sum_{g}p_g\,\mathrm{db}_{g,i}\Bigr)^{2}}_{\text{bagged}}
\;+\;\underbrace{2\sum_{g}p_g\,\Sigma_{\widehat H,g}}_{\text{cross}}
\;+\;\underbrace{\sigma_{D,\widehat H}^{2}}_{\text{same-draws}} ,
\] the bagged term being the infinitesimal jackknife of the reduced base
\(t(\hat P^{*}_b)=\hat\beta(\widehat H)-D_{\widehat H^{*}_b}(b)\) of the
\emph{Bias removal} step --- consistent for
\(\operatorname{Var}\tilde\beta(\widehat H)\) and equal to
\(\sigma_{D,\widehat H}^{2}\) under separation. Retaining the same-draws
term adds the diagonal \(\sigma_{D,\widehat H}^{2}\ge0\) and the cross
term \(2\sum_{g}p_g\,\Sigma_{\widehat H,g}\). Membership overlap alone
does not sign
\(\Sigma_{\widehat H,g}=\sum_i\mathrm{db}_{\widehat H,i}\,\mathrm{db}_{g,i}\):
the dfbeta contributions are signed, and a subject shared by two regions
may influence the two fits in opposing directions. What signs the
\(p_g\)-weighted cross term is not equality of the influence vectors but
the local regime (A5-ii) together with the family guards. The bagged
term alone is consistent for
\(\operatorname{Var}\tilde\beta(\widehat H)\) --- it is the
infinitesimal jackknife of the reduced base
\(t(\hat P^{*}_b)=\hat\beta(\widehat H)-D_{\widehat H^{*}_b}(b)\) --- so
what remains to sign for conservatism is the retained mass
\(2\sum_g p_g\,\Sigma_{\widehat H,g}+\sigma_{D,\widehat H}^{2}\), whose
asymptotic non-negativity is assumption \textbf{(A6)}
(\(\sum_g p_g\,\Sigma_{\widehat H,g}\ge-\tfrac12\sigma_{D,\widehat H}^{2}\))
--- not the stronger \(\Sigma_{\widehat H,g}\ge0\) for every competitor.
In the local regime the re-selection mass concentrates on competitors
differing from \(\widehat H\) in a single cut, whose membership overlaps
\(\widehat H\)'s in all but a boundary layer; their influence vectors
are therefore positively, though not perfectly, correlated with
\(\widehat H\)'s, so each mass-carrying \(\Sigma_{\widehat H,g}\) is a
positive fraction of \(\sigma_{D,\widehat H}^{2}\)
(\(\Sigma_{\widehat H,\widehat H}=\sigma_{D,\widehat H}^{2}\) being the
limiting case as the boundary layer vanishes). Competitors far from
\(\widehat H\), whose covariance with \(\widehat H\) need not be
non-negative, carry vanishing weight \(p_g\to0\) under (A5-ii). On the
fully enumerated family the minimum-size and minimum-event guards and
the finite grid of distinct cuts keep \(\Sigma\succ0\) and the leading
candidates genuinely distinct (Section~\ref{sec-ap-delicate}), which
bounds the mass-carrying cross-covariances away from the
\(-\tfrac12\sigma_{D,\widehat H}^{2}\) floor; the weighted cross term is
then non-negative in the limit. This is assumption \textbf{(A6)}, a
sufficient condition on the \(p_g\)-weighted cross-covariance in the
local regime rather than an unconditional identity --- the paragraph
above shows the family guards enforce it --- and its finite-sample force
is read from Section 7 of the main paper rather than bounded here. It is
also checkable on the analysis at hand, at no additional fitting cost:
the re-selection frequencies \(\hat p_g\) are a by-product of the
multiplier draws and the cross-covariances \(\hat\Sigma_{\widehat H,g}\)
are inner products of the dfbeta influence vectors already assembled, so
the retained mass
\(2\sum_g\hat p_g\hat\Sigma_{\widehat H,g}+\hat\sigma_{D,\widehat H}^{2}\)
can be evaluated directly and reported alongside the interval as a
diagnostic. With it, \(\hat V\) over-states
\(\operatorname{Var}\tilde\beta(\widehat H)\): as one subgroup separates
(\(p_{\widehat H}\to1\)) the bagged term
\(\to\sigma_{D,\widehat H}^{2}\) and the cross term
\(\to2\sigma_{D,\widehat H}^{2}\), giving
\(\hat V\to4\sigma_{D,\widehat H}^{2}\) against a true
\(\sigma_{D,\widehat H}^{2}\), while in the local regime the inflation
is mild. Hence \(\hat V\) is \emph{conservative} rather than consistent
(the \emph{Variance} step). With a conservatively estimated scale the
interval does not under-cover in the separated regime, where the limit
is normal. In the local regime non-undercoverage is not automatic --- a
conservative variance and a non-Gaussian limit do not between them give
it --- and what is required is domination of the quantiles of
\(\lvert\Lambda\rvert\) by \(z_{1-\alpha/2}\sqrt V\), the residual bias
\(\mathbb E[\Lambda]\) of the \emph{Bias removal} step included.
Section~\ref{sec-ap-coverage} establishes this: the requirement is an
explicit functional of \((\mu,\Sigma,G)\), and it is met with a wide
margin, secured by the retained same-draws and cross terms rather than
by any tightness of \(\Lambda\). The conditional coverage of the
data-adaptive target \(\beta(\widehat H)\) --- the effect at the region
the search itself drew, the guarantee holding on average over the
selection (Hubbard et al. 2016) in the sense of Section 5.1 of the main
paper, distinct from the fixed-region coverage of selective inference
--- is therefore asymptotically \emph{at least} \(1-\alpha\) in the
local regime where selection is genuinely stochastic, and conservative
as one subgroup separates. The statement is asymptotic; a non-asymptotic
version, propagating the explicit Chernozhukov et al. (2015) rate
through the reductions of Section~\ref{sec-ap-delicate}, is not
attempted here. \(\qquad\square\)

\subsection{The local-regime residual: closed forms, the implemented
map, and effective competition}\label{sec-ap-residual}

This section quantifies the residual \(\mathbb E[\Lambda]\) of the
\emph{Bias removal} step. Every statement is analytic or seconds-scale
Monte Carlo on the limit objects \((\mu,\Sigma,G)\) --- no simulation
study is involved --- and is reproduced by the verification suite
accompanying the repository (27 automated checks; closed forms confirmed
by deterministic quadrature at relative tolerance \(10^{-9}\) and by two
independent Monte Carlo implementations).

\emph{Closed form, \(M=2\).} For \(\Sigma=I_2\) and the argmax map, with
\(\delta=w_1-w_2\),
\(m(w)=\sqrt2\,\phi(\delta/\sqrt2)=\pi^{-1/2}e^{-\delta^{2}/4}\),
maximized at the tie. Sweeping the separation \(\mu=(d,0)\),
\[\mathbb E[\Lambda](d)\;=\;\pi^{-1/2}e^{-d^{2}/4}\;-\;(2\pi)^{-1/2}e^{-d^{2}/8},\]
so at the tie \(m(0)=1/\sqrt\pi=0.5642\),
\(\mathbb E[m(\zeta)]=1/\sqrt{2\pi}=0.3989\), and
\(\mathbb E[\Lambda](0)=0.1652\): the correction removes exactly the
fraction \(2^{-1/2}=70.71\%\) of the optimism. The residual is positive
for \(|d|<d^{\star}=2\sqrt{\ln2}=1.6651\), zero there, and negative
beyond, with minimum \(-1/(8\sqrt\pi)=-0.0705\) at
\(|d|=\sqrt{12\ln2}=2.8841\) --- where \(\mathbb E[\Lambda]=-m(d)\)
exactly, the correction subtracting twice the true selection bias.
Numerically,
\(\mathbb E[\Lambda]=+0.1652,+0.1433,+0.0873,-0.0344,-0.0701\) at
\(d=0,0.5,1,2,3\). At the tie \(\operatorname{sd}(\Lambda)=0.916\), so
the retained bias is about \(0.18\) standard deviations of the limit ---
the margin the \emph{Coverage} step leans on.

\emph{Universal retention at ties.} The tie constant depends on neither
the covariance nor the family size. If \(G(cw)=G(w)\) for all \(c>0\)
--- the pure argmax, and the size-dominant limit of the implemented map
at a zero admission threshold --- then
\(z\sim N(0,2\Sigma)\stackrel{d}{=}\sqrt2\,z'\) with
\(z'\sim N(0,\Sigma)\) leaves the winner unchanged, so
\(m^{(2\Sigma)}(0)=\sqrt2\,m^{(\Sigma)}(0)\) and, by the projection
identity of the \emph{Bias removal} step,
\[\mathbb E[\Lambda]\;=\;\bigl(1-2^{-1/2}\bigr)\,m^{(\Sigma)}(0)\;>\;0\]
at every exact tie, for every positive semi-definite \(\Sigma\neq0\) and
every \(M\): the \(70.7\%/29.3\%\) split is universal, and \(\Sigma\)
and \(M\) enter only through the optimism scale \(m^{(\Sigma)}(0)\).
Under exchangeable correlation \(\rho\),
\(m^{(\Sigma)}(0)=\sqrt{1-\rho}\,c_M\) with
\(c_M=\mathbb E[\max_{g\le M}Z_g]\), \(Z_g\) i.i.d.~standard normal;
strong correlation shrinks optimism and residual proportionally, leaving
the retained fraction fixed. The lemma is confirmed numerically on the
limit objects for exchangeable, nested
(\(\operatorname{corr}=\sqrt{|A|/|B|}\)), overlapping
(\(\operatorname{corr}(A,B)=|A\cap B|/\sqrt{|A||B|}\)), and random
covariance structures, and under the size-dominant implemented map
(observed ratios \(0.2915\)--\(0.2943\) against \(1-2^{-1/2}=0.29289\)).
This is an exact statement about the limiting competition, not a
prediction of a finite-sample constant: the retained-bias fraction
measured in the simulations (Section 7 of the main paper) sits above
\(29.3\%\) and drifts with \(n\), as expected once the implemented map
departs from the argmax, the true subgroup occupies a modest share of
the population, and scoring conditions on detection. The universal
constant is the tie value of the limit; the simulations corroborate its
qualitative consequences --- positivity near ties and the sign reversal
above --- not the constant itself.

\emph{Effective competition.} Define \(M_{\mathrm{eff}}\) by
\(c_{M_{\mathrm{eff}}}=m^{(\Sigma)}(0)\), so that the tie residual is
\((1-2^{-1/2})\,c_{M_{\mathrm{eff}}}\) exactly: \(M_{\mathrm{eff}}\) is
the number of independent candidates producing the same optimism and the
same residual. Overlap compresses competition sharply --- exchangeable
\(M=10\) gives \(M_{\mathrm{eff}}=6.2,\,3.7,\,1.8\) at
\(\rho=0.3,0.6,0.9\), and a ten-candidate family of overlapping
covariate cells gives \(M_{\mathrm{eff}}=4.2\) --- so the nominal size
of the enumerated family is not the operative quantity. The empirical
\(\hat\Sigma\) is available at no additional cost from the dfbeta
influence vectors already computed,
\(\hat\Sigma_{jk}=n\sum_i\mathrm{db}_{g_j,i}\,\mathrm{db}_{g_k,i}\),
from which \(M_{\mathrm{eff}}\) and the implied tie residual follow
directly; the operative competition size is a property of
\(\hat\Sigma\), to be read from the data rather than equated with the
nominal count of enumerated candidates. One caution: a family closed
under complementation, or containing unions and the overall region, has
exactly linearly dependent estimators, so the full-family \(\Sigma\) is
singular; the residual formulas require only positive semi-definiteness
(degenerate Gaussian fields), while the \(\Sigma\succ0\) of (A5) is read
on the mass-carrying competitors, where the family guards keep the
leading candidates genuinely distinct (Section~\ref{sec-ap-delicate}).

\emph{The implemented map.} The implemented selection --- admission
thresholds, the splitting-consistency screen, and a size tie-break
within a \(10\%\) natural-scale effect neighbourhood of the maximum
(main paper) --- modifies \(m\) but not the structure of the argument.
The neighbourhood has no scale-free local limit: near
\(\mathrm{HR}\approx1\) a \(10\%\) natural-scale band is of order
\(\sqrt n\) wide on the local scale, so the limiting map is bracketed by
the pure argmax (band \(b=0\)) and the size-dominant rule (band
\(b=\infty\), the largest admissible region). The residual's sign and
order survive uniformly over \(b\): at the tie (\(M=2\), \(\Sigma=I\),
relative sizes \(1.0/1.6\), zero thresholds),
\(\mathbb E[\Lambda]=+0.164,+0.163,
+0.157,+0.142,+0.133,+0.141\) at \(b=0,0.25,0.5,1,2,\infty\). Under
separation the regimes part ways: the narrow band tracks the argmax
curve, including the sign reversal (\(-0.070\) at \(d=3\) for
\(b=0.25\)), while the size-dominant limit --- where competition is
threshold-crossing rather than argmax --- keeps a persistent positive
residual (\(+0.11\) to \(+0.14\) across \(d=0\)--\(3\)). Positive
admission thresholds truncate the noise field and increase both optimism
and residual; the splitting-consistency screen, modelled as a correlated
auxiliary field, leaves sign and order unchanged (the projection
identity holds conditionally on the auxiliary noise, hence
unconditionally). The \(o(n^{-1/2})\) conclusion therefore fails under
either limiting regime of the implemented map.

\emph{Relation to the simulations.} The qualitative prediction of the
sign law --- under-correction near ties reversing to over-correction as
competition separates --- is corroborated by the fixed-family
simulations, scored directly against the conditional target
\(\beta(\widehat H)\) (Section 7 of the main paper). Across three
scenarios of increasing harm separation, the mean residual bias of the
multiplier-resampling estimate against \(\beta(\widehat H)\) on the
forest-search family falls with separation and changes sign:

{\def\LTcaptype{none} % do not increment counter
\begin{longtable}[]{@{}
  >{\raggedright\arraybackslash}p{(\linewidth - 8\tabcolsep) * \real{0.2000}}
  >{\raggedright\arraybackslash}p{(\linewidth - 8\tabcolsep) * \real{0.2000}}
  >{\raggedright\arraybackslash}p{(\linewidth - 8\tabcolsep) * \real{0.2000}}
  >{\raggedright\arraybackslash}p{(\linewidth - 8\tabcolsep) * \real{0.2000}}
  >{\raggedright\arraybackslash}p{(\linewidth - 8\tabcolsep) * \real{0.2000}}@{}}
\toprule\noalign{}
\begin{minipage}[b]{\linewidth}\raggedright
\(\sqrt n\)-scaled residual
\end{minipage} & \begin{minipage}[b]{\linewidth}\raggedright
\(n=500\)
\end{minipage} & \begin{minipage}[b]{\linewidth}\raggedright
\(n=1000\)
\end{minipage} & \begin{minipage}[b]{\linewidth}\raggedright
\(n=1500\)
\end{minipage} & \begin{minipage}[b]{\linewidth}\raggedright
\(n=2000\)
\end{minipage} \\
\midrule\noalign{}
\endhead
\bottomrule\noalign{}
\endlastfoot
moderate (\(\theta^{\dagger}(H)=1.0\)) & \(+8.5\) & \(+8.4\) & \(+6.2\)
& \(+5.9\) \\
stronger (\(\theta^{\dagger}(H)=1.5\)) & \(+3.8\) & \(+1.0\) & \(-1.3\)
& \(-1.5\) \\
strongest (\(\theta^{\dagger}(H)=2.0\)) & \(+0.7\) & \(-0.9\) & \(-1.6\)
& \(-0.9\) \\
\end{longtable}
}

At the moderate separation the scaled residual does not decay: it holds
near \(8.5\) through \(n=1000\) and falls only to \(5.9\) by \(n=2000\),
whereas an \(o(n^{-1/2})\) bias reduction requires it to tend to zero.
Over this range the residual is therefore of order \(n^{-1/2}\), and the
gradual decline is itself what the two regimes predict --- a fixed
data-generating mechanism drifts toward the separated regime of (A5-i)
as \(n\) grows, where the residual does become \(o(n^{-1/2})\). At
stronger separation the scaled residual crosses from positive to
negative, the reversal the sign law predicts. The cells rest on \(438\)
to \(1{,}000\) detected replicates each, so the pattern is not Monte
Carlo noise; the ordering across scenarios holds at six of the seven
sample sizes, the exception being \(n=2000\), where the two separated
scenarios differ by about \(1.6\) Monte Carlo standard errors and are
not resolved. Two boundaries keep the reading honest. The magnitude is
not a clean match to the pure-argmax constant: the implemented map is
not the argmax (the size tie-break and admission thresholds modify
\(m\), as above), the true subgroup is a modest fraction of the
population so the within-subgroup coefficient carries its own
small-sample bias, and the scoring conditions on detection; the sign and
its reversal are the robust features, not any single constant. And the
corresponding DINA and GRF undershoot involves regenerating families,
outside the fixed-family scope of Theorem 2.

\subsection{Coverage in the local regime}\label{sec-ap-coverage}

The \emph{Coverage} step reduces non-undercoverage to a statement about
the limit objects. This section states it exactly, proves a sufficient
condition that is uniform in the local parameter, and records why the
obvious concentration route does not deliver it. As in
Section~\ref{sec-ap-residual}, everything is analytic or seconds-scale
Monte Carlo on \((\mu,\Sigma,G)\).

\emph{The criterion.} Since
\(\sqrt n\{\tilde\beta(\widehat H)-\beta(\widehat H)\}
\Rightarrow\Lambda\) and \(n\hat V\Rightarrow V\) with
\[V \;=\; \underbrace{p^{\top}\Sigma\,p}_{\text{bagged}}
\;+\;\underbrace{2\textstyle\sum_{g}p_g\,\Sigma_{Gg}}_{\text{cross}}
\;+\;\underbrace{\Sigma_{GG}}_{\text{same-draws}},\qquad
p_g=P_{\zeta^{*}}\!\bigl(G(\mu+\zeta+\zeta^{*})=g\bigr),\] the
asymptotic coverage of the infinitesimal-jackknife interval of Section 3
of the main paper for \(\beta(\widehat H)\) is
\[\lim_n P\bigl(\beta(\widehat H)\in\text{CI}\bigr)
\;=\;P\Bigl(\lvert\Lambda(\zeta)\rvert\le z_{1-\alpha/2}\sqrt{V(\zeta)}\Bigr),\]
both sides explicit functionals of \((\mu,\Sigma,G)\). Non-undercoverage
is therefore a \emph{checkable} property of the limiting competition,
not an assumption: note that \(V\) is itself random and co-moves with
\(\Lambda\) through the re-selection probabilities, so the requirement
is genuinely a joint one and not a comparison of two fixed scales.

\emph{A sufficient condition, uniform in \(\mu\).} Write
\[\mathcal K(\Sigma,\alpha)\;=\;q_{1-\alpha}\!\bigl(\max_g\lvert\zeta_g\rvert\bigr)
\;+\;m^{(\Sigma)}(0),\qquad
V_{\star}\;=\;\inf_{w}V(w).\]

\begin{quote}
\textbf{Proposition S4.1.} If \(\mathcal K(\Sigma,\alpha)\le
z_{1-\alpha/2}\sqrt{V_{\star}}\), then in the local regime the
interval's asymptotic coverage of \(\beta(\widehat H)\) is at least
\(1-\alpha\), uniformly over all local parameters
\(\mu\in\mathbb R^{M}\).
\end{quote}

\emph{Proof.} Three facts. First, \(m\ge0\): with
\(\psi(w)=\mathbb E\max_g(w_g+\zeta^{*}_g)\) one has
\(m(w)=\psi(w)-\mathbb E[w_{G}]\), and
\(\psi(w)\ge\max_g w_g\ge\mathbb E[w_{G}]\) by Jensen. Second,
\(\sup_w m(w)=m^{(\Sigma)}(0)=\mathbb E[\max_g\zeta_g]\): by the ray
identity of Section~\ref{sec-ap-residual},
\(\nabla m(w)=-\nabla^{2}\psi(w)(w-\bar w\mathbf 1)\) vanishes on the
tie line and \(\nabla^{2}\psi\succeq0\), so \(m\) is maximized there,
where the argmax of \(\zeta^{*}\) carries its own maximum. Third,
\(\lvert\zeta_{G}\rvert\le\max_g\lvert\zeta_g\rvert\), \(G\) being one
of the indices. Together these give the deterministic envelope
\[\lvert\Lambda(\zeta)\rvert\;=\;\lvert\zeta_{G}-m(\mu+\zeta)\rvert
\;\le\;\max_g\lvert\zeta_g\rvert+m^{(\Sigma)}(0)
\qquad\text{for every }\mu,\] whence
\(q_{1-\alpha}(\lvert\Lambda\rvert)\le\mathcal K\). Since
\(V\ge V_{\star}\) pointwise,
\(P(\lvert\Lambda\rvert\le z_{1-\alpha/2}\sqrt V)\ge
P(\lvert\Lambda\rvert\le z_{1-\alpha/2}\sqrt{V_{\star}})\ge1-\alpha\)
whenever \(z_{1-\alpha/2}\sqrt{V_{\star}}\ge\mathcal K\). Neither
\(\mathcal K\) nor \(V_{\star}\) depends on \(\mu\), so the bound is
uniform. \(\qquad\square\)

The envelope is deliberately crude --- it discards the cancellation
between \(\zeta_G\) and \(m\) that gives \(\Lambda\) its actual spread
--- and it closes when the effective competition of
Section~\ref{sec-ap-residual} is small: at
\(M_{\mathrm{eff}}\approx2\)--\(3\) the margin
\(z_{1-\alpha/2}\sqrt{V_{\star}}-\mathcal K\) is positive (\(+0.30\) at
\(M=2\), \(\Sigma=I\); \(+0.20\) at \(M=20\), \(\rho=0.8\)), and it is
inconclusive for large families of weakly correlated candidates. A
realized candidate family with many mildly correlated members sits in
the latter class, where the envelope of Proposition S4.1 is inconclusive
and the guarantee rests on the exact criterion; whether the
finite-sample interval covers is in any case the question the
simulations answer directly, and on the fixed forest-search family they
place conditional coverage of \(\beta(\widehat H)\) near nominal
(below).

\emph{What the criterion secures, and what it does not.} The criterion
says non-undercoverage is a property of the retained variance relative
to the residual bias, both explicit functionals of the limiting
competition. It is not a claim that coverage is far above nominal: the
retained same-draws and cross terms of the \emph{Variance} step widen
the interval enough to dominate the quantiles of \(\Lambda\) when the
residual is small against \(\sqrt V\), but the margin is finite and
shrinks as the residual grows. In the simulations of Section 7 of the
main paper, on the fixed forest-search family where the theorem applies,
this is what is observed: conditional coverage of \(\beta(\widehat H)\)
sits between \(0.93\) and \(0.99\) across sample sizes and across the
harm scenarios, near nominal rather than markedly conservative, with the
interval's estimated scale running above the empirical standard
deviation of the estimation error in every cell, by factors between
\(1.1\) and \(3.0\) that follow no stable pattern in either \(n\) or
separation. The direction the \emph{Variance} step predicts --- a
conservative variance --- is borne out; a specific inflation constant is
not, and none is claimed.

\emph{Why Gaussian concentration is not the route.} It is natural to
seek non-undercoverage from a sub-Gaussian bound on \(\Lambda\), and the
attempt fails twice. First, \(\Lambda\) is not globally Lipschitz:
crossing the \(j/k\) selection boundary changes the winner and
\(\Lambda\) jumps by \(\lvert\mu_j-\mu_k\rvert\), so concentration
applies to \(\Psi=\Lambda+\mu_{G}\), and transferring it back costs an
additive \(\max_g\mu_g-\min_g\mu_g\). Second, and more fundamentally, a
sub-Gaussian proxy equal to \(V\) does not give the normal quantile: it
yields only
\[P\bigl(\lvert\Lambda-\mathbb E\Lambda\rvert\le z_{1-\alpha/2}\sqrt V\bigr)
\;\ge\;1-2e^{-z_{1-\alpha/2}^{2}/2}\;=\;0.707\quad(\alpha=0.05),\] far
short of \(1-\alpha\). Closing at level \(\alpha\) requires proxy
variance at most \(z_{1-\alpha/2}^{2}/\{2\log(2/\alpha)\}=0.521\) times
\(V\) --- a variance-ratio condition, not a boundedness condition ---
and the Lipschitz constant of \(\Psi\) at \(\Sigma=I\), \(M=2\) gives a
proxy of \(1.80\) against an admissible \(1.30\). The concentration
route is thus genuinely insufficient here, which is why Proposition S4.1
proceeds by a deterministic envelope instead.

\emph{Uniformity in the local parameter.} Statements (ii) and (iii) of
Theorem 2 are pointwise in \(\mu\). Uniformity over compact
\(\{\lVert\mu\rVert\le C\}\) --- the arena in which local-asymptotic
claims about estimators with data-dependent corrections are ordinarily
framed (Ibragimov and Has'minskii 1981, sec. I.9, Thm II.12.1) ---
follows from the same bounds, \(m\) and its derivatives being bounded on
compacts, but is not claimed here. The coverage condition of Proposition
S4.1 is by contrast uniform in \(\mu\) outright, neither side of it
depending on the local parameter.

\emph{A regime where the interval under-covers.} The null scenario --- a
genuinely protective effect, so that no harm subgroup exists and every
detection is a false declaration --- behaves quite differently. The
detection rate falls with \(n\), as it should; the retained bias grows
toward the interval half-width; and conditional coverage on the
shrinking detected set declines well below nominal. Two cautions govern
how much this shows. The theorem conditions on the realized selection,
and here the conditioning event is an increasingly extreme tail of
noise-driven declarations, a configuration the local-sequence
idealization was not designed to describe; whether it falls inside or
outside the statement is a modelling judgment the simulation does not
settle, and the \emph{fixed}-alternative objection cannot be raised
against it alone, since the harm scenarios are fixed alternatives too.
Nor has the quantile-domination condition been evaluated on any of these
designs: it is a condition on the limiting objects, not a quantity the
simulations report. The contrast between this scenario and the harm
scenarios is therefore an observation about coverage, not a
demonstration of the condition holding in one place and failing in
another. What it does establish is that the conditional guarantee of
part (iv) is not automatic.

\subsection{What remains, and what carries it}\label{sec-ap-delicate}

Two steps in the local regime (A5-ii), where the limiting competition is
genuinely stochastic, sit above the routine estimating-equation
machinery. They are distinct, and separating them is what makes each
standard.

\emph{The re-selection laws agree (Theorem 1, part 1).} The argmax
\emph{location} converges by the argmax continuous-mapping theorem
(Vaart and Wellner 1996, Thm 3.2.2; Kosorok 2008, Thm 14.1), whose only
substantive hypothesis --- a unique limiting maximizer --- follows for
the finite family from \(\Sigma\succ 0\) alone, every pairwise Gaussian
difference being non-degenerate. Agreement of the multiplier and
bootstrap re-selection \emph{distributions} is the quantitative step,
supplied by the comparison and anti-concentration bounds for maxima of
Gaussian vectors (Chernozhukov et al. 2015, Thms 2--3), which are
dimension-free and degrade only as the minimum candidate-contrast
variance vanishes --- so \(\Sigma\succ 0\) is precisely their domain,
and near-duplicate candidates their limit.

\emph{The bagged functional is differentiable (Theorem 2).} The raw
value-at-argmax map is only directionally differentiable at a tie, where
the standard delta-method and bootstrap fail (Vaart and Wellner 1996,
Thms 3.9.4, 3.9.11; Dümbgen 1993; Fang and Santos 2019). Bagging removes
the obstruction: averaging the hard winner-indicator over the resampling
weights returns the smooth re-selection probabilities (Bühlmann and Yu
2002; Efron 2014, sec. 4), and the functional delta-method applies to
the smoothed map (Vaart and Wellner 1996, Thms 3.9.4, 3.9.11). In the
local regime the re-selection probabilities are interior, which is what
makes them an admissible gradient; the same regime is where the
correction is first-order material, so the smoothing is most effective
exactly where it is most needed. The smoothing controls the rate, not
the mean: the first-order gradient is blind to the curvature of the
optimism functional, whose Jensen gap is the residual of
Section~\ref{sec-ap-residual}.

What the two steps take as given is the proof of each cited master
result --- the argmax and delta-method theorems, the Gaussian-maximum
comparison and anti-concentration bounds, the bagging-smoothing identity
--- not their conclusions; the reductions above are carried out in full.
Theorem 2 asserts \emph{conservative} coverage of the interval ---
outright under separation, under the quantile-domination proviso of its
part (iv) in the local regime --- --- not consistency of the bagged
variance, and not an ordering of that variance against the naive robust
and full-bootstrap variances: near a selection boundary the variance
functional is volatile (Efron 2014, Remark K) and its finite-sample
behaviour is identifier-dependent --- above the naive for a fixed
candidate family, below it for regenerating identifiers --- so the paper
reports it empirically in Section 7 rather than ordering it here.

\section{Interpreting the regimes for subgroup selection: separation,
local alternatives, and near-ties}\label{sec-ap-regimes}

The assumptions of Section~\ref{sec-ap-assumptions} are stated in the
language of empirical-process limits; this subsection reads them back
into the subgroup-identification problem they describe, so that a reader
who is not a specialist in that theory can see what each condition asks
of the data and what it implies for the correction. The four objects
that organize the proofs --- \emph{separation}, the \emph{local} regime,
the margin sequence \(\eta_n\), and the \emph{no-tie} condition --- are
all statements about one underlying thing: how clearly the most harmful
subgroup stands apart from its competitors in the candidate family.

\textbf{Separation: a subgroup that clearly stands apart} Separation is
the assumption that one candidate region \(g^{\star}\) has a population
effect strictly larger than every other admitted candidate's, by a
margin \(\eta > 0\) that does not shrink as the trial grows. In subgroup
terms there is a genuinely most-harmful region, and it sits far enough
above the runners-up that the data can resolve the gap: re-run the
search on an independent sample and it returns the same region with
probability approaching one. Because the winner is, in the limit,
predetermined, selecting the maximum injects no optimism --- the
winner's-curse bias is \(o(n^{-1/2})\) --- and the de-biased estimate
\(\tilde\beta(\widehat H)\) barely moves the naive
\(\hat\beta(\widehat H)\), while the selection-aware interval contracts
to the ordinary within-subgroup robust interval. Here the correction is
a finite-sample refinement, not a structural repair. Empirically,
separation is the regime of a high and stable detection rate with
different identifiers agreeing on the same region.

\textbf{The local regime: several near-equal candidates and a genuine
contest.} The local regime is the opposite posture: the competing
candidates' population effects differ only at the \(n^{-1/2}\) scale, so
the gaps between the top regions are the same order as the sampling
error in each region's own estimate. No single subgroup dominates;
several are about equally --- and modestly --- harmful in truth. Which
region posts the \emph{largest estimated} effect on a given trial is
then settled largely by sampling noise, and a fresh sample could easily
crown a different region. This is precisely the configuration in which
selecting the maximum inflates the reported effect (the winner's curse),
and the inflation is of order \(n^{-1/2}\), the same order as the
estimate's own error --- hence a \emph{leading-order} bias that the
interval must account for to cover. The ``local'' terminology is the
local-alternative (Pitman-drift) scaling standard in
inference-after-selection: the candidate effects are taken to approach a
common value at rate \(n^{-1/2}\) precisely so that selection remains a
live, non-degenerate contest in the limit --- the maximum of a
correlated Gaussian vector \(\max_g Z_g\), \(Z \sim N(0,\Sigma)\) ---
rather than collapsing to a foregone winner. This is the regime the
correction is built for, and the one Section 5.1 of the main paper takes
as its headline.

\textbf{The margin sequence \(\eta_n\) and its boundary} Writing the
margin as a sequence \(\eta_n\) makes the two regimes endpoints of a
single axis: \(\eta_n\) bounded away from zero is separation, and
\(\eta_n \asymp n^{-1/2}\) is the local regime. The remaining case ---
\(\eta_n = o(n^{-1/2})\), the margin shrinking \emph{faster} than
\(n^{-1/2}\) --- is a \emph{third} regime, and (A5) deliberately
excludes it. In subgroup terms it describes competing regions collapsing
onto one another faster than the data can tell them apart: an identifier
so noisy that the regions it pushes to the top of the ranking are, even
at the \(n^{-1/2}\) resolution, statistically indistinguishable from
each other. A ``very noisy identifier producing near-ties'' is exactly
this case. Outside (A5) the clean argmax theory no longer applies ---
the limiting Gaussian maximum is attained by more than one candidate,
the selection law loses the continuity the pushforward of Theorem 1
relies on, and the infinitesimal-jackknife variance can understate how
much the selected region moves from sample to sample, so the interval
can fall below nominal. The point to register is that this undercoverage
is what the theory predicts \emph{at the boundary it excludes}, not a
failure within the regime it certifies. One caveat: \(\eta_n\) is a
margin among the candidates of a \emph{fixed} family; for a model-based
front-end whose family is itself resampled, near-ties are only one of
two distinct fragilities.

\textbf{No-tie regularity} The no-tie condition in (A5) requires that
the limiting effect-vector \(Z \sim N(0,\Sigma)\) have a unique
maximizer almost surely --- no two candidates tying for the largest
limiting effect --- and that no candidate sit exactly on its admission
threshold. When the overlap covariance \(\Sigma\) (Equation 7 of the
main paper) is non-degenerate this holds automatically, by the
anti-concentration of the Gaussian maximum (Chernozhukov et al. 2015): a
non-degenerate Gaussian maximum places no probability on ties. In
subgroup terms the condition asks only that no two candidate regions be
statistical duplicates --- distinct definitions whose effect estimates
move in lockstep. Provided that holds, there is in the limit an
unambiguous most-harmful-estimated region on each perturbed draw, and
the bootstrap re-selection law converges; this is the hinge that lets
``re-run the entire search'' collapse to ``re-rank the perturbed
effects'' (Theorem 1) and lets the selection-aware interval be read off
the same draws. The condition can be strained by near-duplicate cuts ---
for example \(\{nodes \le 3\}\) and \(\{nodes \le 4\}\) carve out the
same subjects when no patient has exactly four nodes --- so that a
cluster of near-identical top candidates leaves ``which won'' a
coin-flip the limiting theory cannot resolve. This is precisely the
\(\eta_n = o(n^{-1/2})\) failure of the previous paragraph, and the
delicate step of Section~\ref{sec-ap-delicate} is exactly where it would
manifest. On the fully enumerated family the minimum-size and
minimum-event guards and the finite grid of distinct cuts keep
\(\Sigma\) non-degenerate and the leading candidates genuinely distinct,
so the no-tie condition holds in the regime the paper recommends; it is
model-based front-ends that can manufacture near-duplicate top
candidates.

\section{GBSG application: de-biased subgroup hazard-ratio
table}\label{sec-supp-gbsg-table}

Table~\ref{tbl-gbsg-multimethod} accompanies the German Breast Cancer
Study Group analysis of Section 6 of the main paper.

\begin{table}[H]

\caption{\label{tbl-gbsg-multimethod}De-biased subgroup hazard ratios
for the GBSG trial; parenthetical values are 95\% confidence intervals.
Here \(\widehat{H}\) is \({er \le 0} \cap {size \le 35}\) for FS,
\({er \le 0}\) for GRF, and \({grade3 \ge 1} \cap {pgr \le 10}\) for
DINA.}

\centering{

\centering
\begin{tabular}[t]{lccccc}
\toprule
Method & SG & $n$ (\%) & Naive & Full bootstrap & Multiplier\\
\midrule
FS & $\widehat{H}$ & 61 (9) & 2.54 (1.25, 5.17) & 1.94 (0.89, 4.21) & 1.44 (0.62, 3.36)\\
 & $\widehat{H}^{c}$ & 625 (91) & 0.61 (0.47, 0.79) & 0.63 (0.43, 0.93) & 0.64 (0.40, 1.03)\\
\midrule
DINA & $\widehat{H}$ & 89 (13) & 1.60 (0.88, 2.91) & 0.89 (0.38, 2.07) & 1.11 (0.45, 2.75)\\
 & $\widehat{H}^{c}$ & 597 (87) & 0.63 (0.48, 0.82) & 0.66 (0.40, 1.11) & 0.66 (0.40, 1.09)\\
\midrule
GRF & $\widehat{H}$ & 82 (12) & 1.95 (1.04, 3.67) & 1.04 (0.49, 2.24) & 1.05 (0.47, 2.34)\\
 & $\widehat{H}^{c}$ & 604 (88) & 0.61 (0.47, 0.80) & 0.66 (0.39, 1.13) & 0.66 (0.41, 1.06)\\
\bottomrule
\end{tabular}

}

\end{table}%

\section{ACTG175 application: the odds-ratio
path}\label{sec-supp-actg175}

The odds-ratio companion to the GBSG analysis exercises the GLM path of
the same machinery (Section 1 of the main paper). AIDS Clinical Trials
Group 175 (Hammer et al. 1996) randomized \(n =\) 1,083 patients to
zidovudine plus didanosine, the experimental arm, against didanosine
monotherapy, the control; the dataset is accessed here through the
\texttt{speff2trial} package. The outcome here is an indicator of no CD4
improvement, coded so that an odds ratio above the threshold signals
harm. The candidate covariates were the trial's baseline
characteristics: age, weight (wtkg, \(J=10\) cuts), Karnofsky
performance score (karnof), days of prior antiretroviral therapy
(preanti), and baseline CD4 (cd40, \(J=10\) cuts) and CD8 (cd80) counts,
together with binary indicators for hemophilia, homosexual activity, a
history of intravenous drug use, race, sex, and symptomatic status.

\begin{table}

\caption{\label{tbl-actg175-multimethod}De-biased subgroup odds ratios
for the ACTG175 trial (binary outcome, OR scale); parenthetical values
are 95\% confidence intervals. Here \(\widehat{H}\) is
\({wtkg > 86} \cap {cd40 > 380}\) for FS,
\({preanti \ge 849.4} \cap {cd40 \ge 338}\) for DINA, and
\({wtkg > 84.37} \cap {cd40 > 368.2}\) for GRF.}

\centering{

\centering
\begin{tabular}[t]{lccccc}
\toprule
Method & SG & $n$ (\%) & Naive & Full bootstrap & Multiplier\\
\midrule
FS & $\widehat{H}$ & 72 (7) & 3.58 (1.31, 9.77) & 2.43 (0.80, 7.33) & 1.79 (0.63, 5.08)\\
 & $\widehat{H}^{c}$ & 1011 (93) & 0.59 (0.45, 0.76) & 0.61 (0.41, 0.89) & 0.62 (0.37, 1.03)\\
\midrule
DINA & $\widehat{H}$ & 98 (9) & 1.94 (0.84, 4.47) & 0.85 (0.23, 3.11) & 0.95 (0.29, 3.10)\\
 & $\widehat{H}^{c}$ & 985 (91) & 0.60 (0.46, 0.78) & 0.63 (0.40, 1.02) & 0.63 (0.38, 1.06)\\
\midrule
GRF & $\widehat{H}$ & 95 (9) & 2.25 (0.97, 5.20) & 1.09 (0.40, 3.01) & 1.10 (0.43, 2.80)\\
 & $\widehat{H}^{c}$ & 988 (91) & 0.58 (0.45, 0.76) & 0.63 (0.38, 1.04) & 0.62 (0.37, 1.04)\\
\bottomrule
\end{tabular}

}

\end{table}%

The results displayed in Table~\ref{tbl-actg175-multimethod} are as
follows. FS and GRF identify similar high-weight and high-CD4 regions
\(\{wtkg > 86\} \cap \{cd40 > 380\}\) and
\(\{wtkg > 84.37\} \cap \{cd40 > 368.2\}\), respectively --- close but
not identical cuts, agreeing on 1,060 of 1,083 patients (97.9\%, Cohen's
\(\kappa = 0.85\)) --- where 67.6\% of treated patients show no CD4
improvement against 36.8\% on control, and the naive odds ratio of 3.58
corrects to 2.43 (FB) and 1.79 (MR) but stays well above the threshold.
DINA selects \(\{preanti \ge 849.4\} \cap \{cd40 \ge 338\}\), whose
apparent harm substantially diminishes under correction --- naive 1.94
to 0.85 (FB) and 0.95 (MR).

As noted above there is considerable overlap between the identified
subgroups of FS and GRF, and both corrections remain elevated (above
\(1.0\)) but there are differences in magnitude between the FB (2.43
{[}FS{]} vs 1.09 {[}GRF{]}) and MR corrections (1.79 {[}FS{]} vs 1.10
{[}GRF{]}).

The cost advantage holds for this larger study --- the full bootstrap in
minutes against the refit-free procedure in seconds (roughly 415 to
1,660 times faster across the three identifiers).

\section{Additional simulation results}\label{sec-additional-results}

\textbf{Overview of the additional simulations}

We expand on the fixed-family study of Section 7 of the main paper along
two axes --- across identifiers and across harm strength --- on the same
GBSG-based accelerated-failure-time super-population (administrative
censoring at 84 months; true harm subgroup
\(H = \{\text{er} \le 8\} \cap \{\text{meno} = 0\}\), prevalence
\(\approx 12.5\%\)). The complement marginal hazard ratio is held fixed
at \(\theta^{\dagger}(H^{c}) \approx 0.66\) throughout; only the
harm-subgroup marginal hazard ratio \(\theta^{\dagger}(H)\) is varied,
which at fixed prevalence also moves the overall ITT effect. Three
identifiers --- forest search (the enumerated consistency search), DINA,
and GRF --- each rediscover \(H\) from the raw covariates, and every
summary is conditional on a subgroup being detected.

\textbf{Identifier settings and thresholds.} To stress false
identification, all three identifiers run with deliberately sub-null
effect thresholds: \(c_{\mathrm{screen}} = \log(0.9)\) and
\(c_{\mathrm{cons}} = \log(0.8)\) for forest search and DINA, and a
floor of \(0\) for GRF. These admit more false harm declarations than
the defaults would; raising them mitigates false discovery, but the aim
of the design is to probe how far de-biasing alone restores robust
inference. The default thresholds declare far fewer subgroups, and the
correction removes the same selection optimism whenever a declaration is
made. Forest search is additionally offered pre-specified candidates for
positive oestrogen and positive progesterone status (default cuts
elsewhere; \(J = 10\) cuts for oestrogen, per Section 2.1 of the main
paper). The full nonparametric bootstrap re-runs the entire pipeline ---
enumerate, screen, select --- on each resample, whereas multiplier
resampling perturbs and re-selects on the fixed observed-data family.

\begin{figure}

\centering{

\pandocbounded{\includegraphics[keepaspectratio]{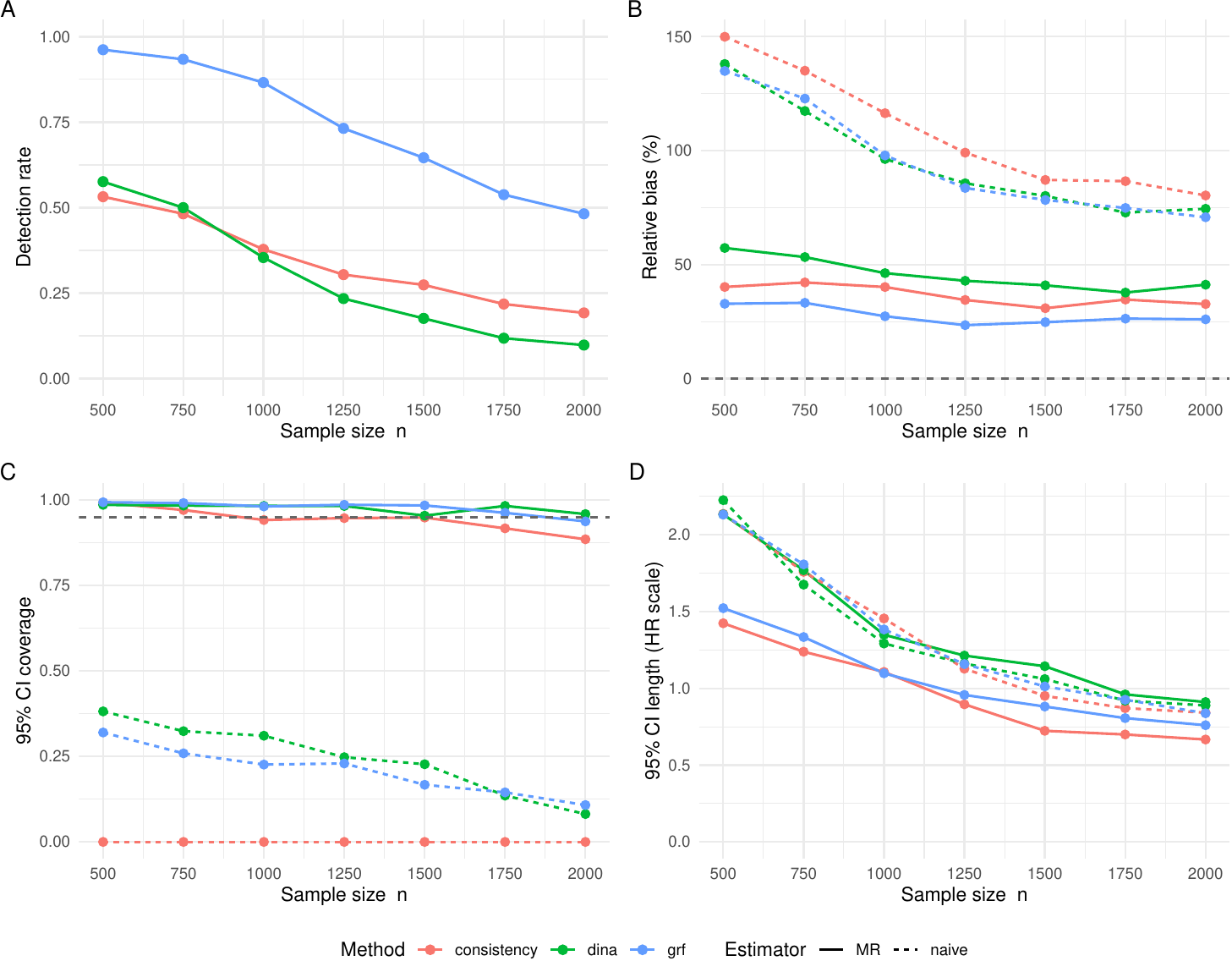}}

}

\caption{\label{fig-mr-coverage-h075c}(A) Subgroup-detection rate vs n
by detector, 500 simulations per cell. (B) Bias of the MR \& naive
estimator (line type) relative to the conditional subgroup hazard ratio
θ(Ĥ), as a percentage of the target, on the harm subgroup Ĥ, vs n; the
dashed line at 0 marks no bias. (C) Empirical 95\% CI coverage of the
conditional subgroup hazard ratio θ(Ĥ) for the MR \& naive estimator
(line type), on the harm subgroup Ĥ, vs per-trial sample size n (true
\(\theta^{\dagger}(H) = 0.75\), \(\theta^{\dagger}(H^{c}) = 0.66\)
(marginal HR); \(\theta^{\ddagger}(H) = 0.70\),
\(\theta^{\ddagger}(H^{c}) = 0.58\) (CDE); 500 simulations per cell,
NULL MR draws; detectors: consistency, dina, grf). Dashed line = 0.95
nominal. Uncertainty bands omitted for legibility (select a single
estimator to show them). All rates conditional on identifying a
subgroup. (D) Median 95\% CI width for the MR \& naive estimator (line
type) on the harm subgroup Ĥ, vs n (HR scale).}

\end{figure}

\begin{figure}

\centering{

\pandocbounded{\includegraphics[keepaspectratio]{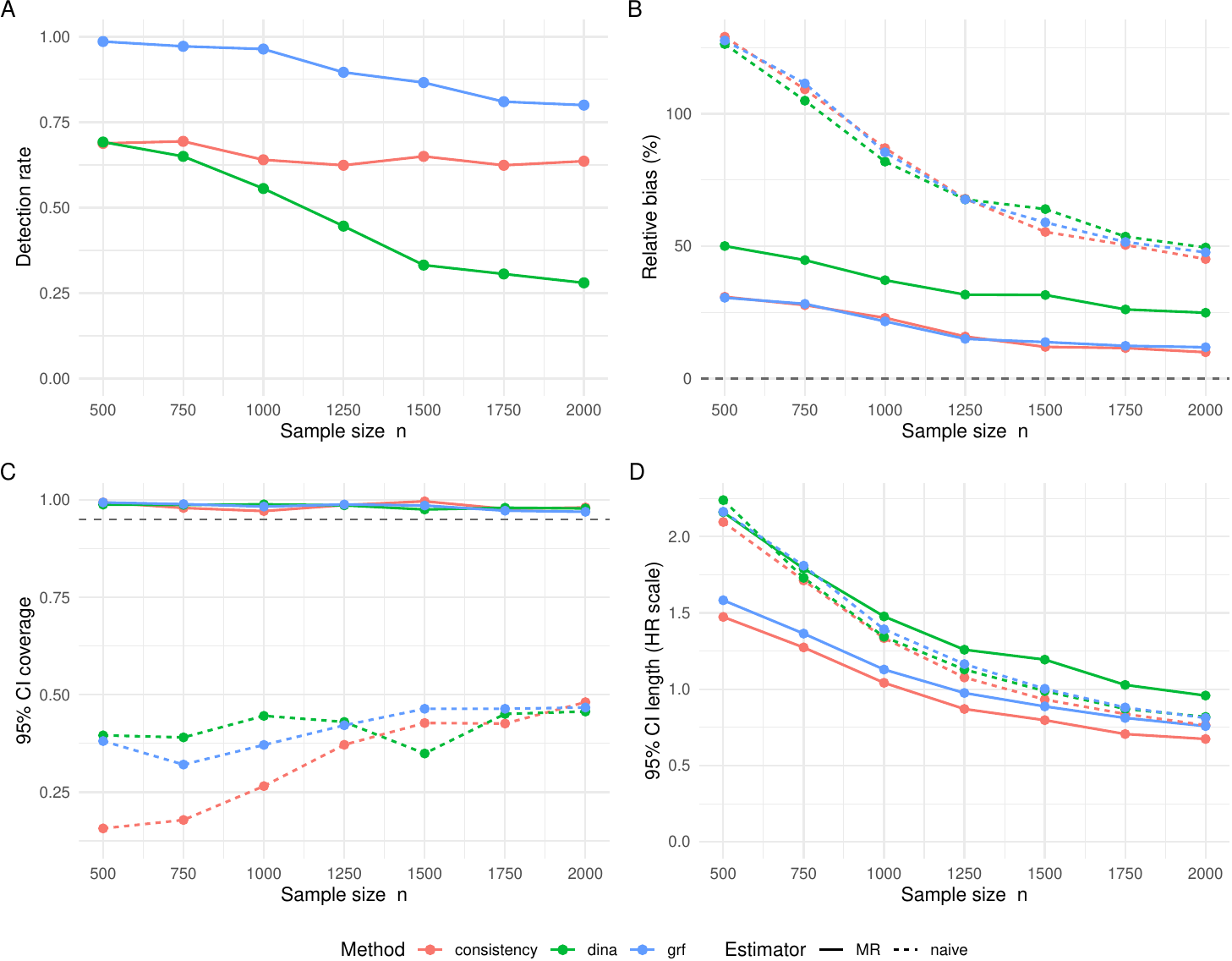}}

}

\caption{\label{fig-mr-coverage-h10c}(A) Subgroup-detection rate vs n by
detector, 500 simulations per cell. (B) Bias of the MR \& naive
estimator (line type) relative to the conditional subgroup hazard ratio
θ(Ĥ), as a percentage of the target, on the harm subgroup Ĥ, vs n; the
dashed line at 0 marks no bias. (C) Empirical 95\% CI coverage of the
conditional subgroup hazard ratio θ(Ĥ) for the MR \& naive estimator
(line type), on the harm subgroup Ĥ, vs per-trial sample size n (true
\(\theta^{\dagger}(H) = 1.00\), \(\theta^{\dagger}(H^{c}) = 0.66\)
(marginal HR); \(\theta^{\ddagger}(H) = 1.00\),
\(\theta^{\ddagger}(H^{c}) = 0.58\) (CDE); 500 simulations per cell,
NULL MR draws; detectors: consistency, dina, grf). Dashed line = 0.95
nominal. Uncertainty bands omitted for legibility (select a single
estimator to show them). All rates conditional on identifying a
subgroup. (D) Median 95\% CI width for the MR \& naive estimator (line
type) on the harm subgroup Ĥ, vs n (HR scale).}

\end{figure}

\begin{figure}

\centering{

\pandocbounded{\includegraphics[keepaspectratio]{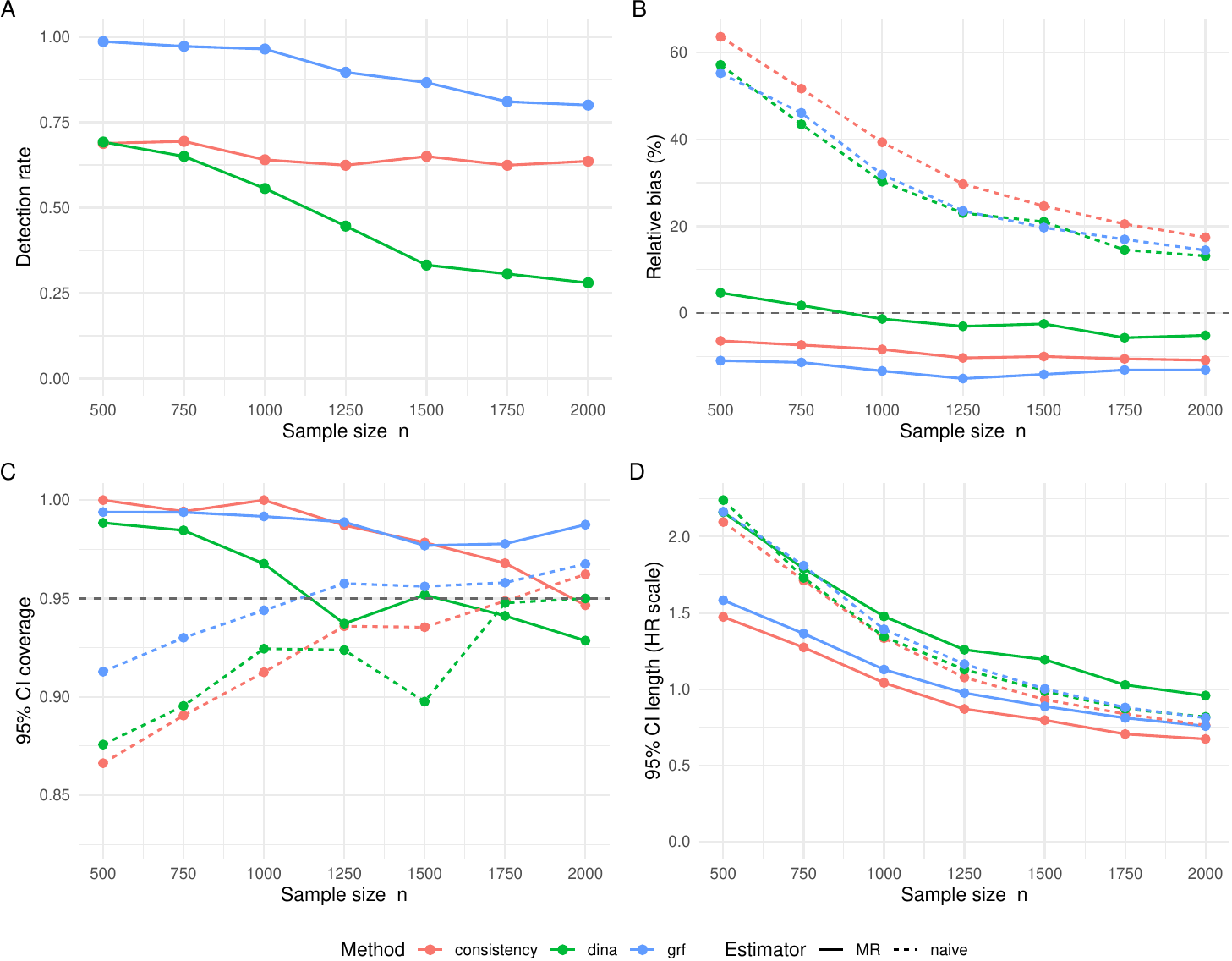}}

}

\caption{\label{fig-mr-coverage-h10m}(A) Subgroup-detection rate vs n by
detector, 500 simulations per cell. (B) Bias of the MR \& naive
estimator (line type) relative to the marginal subgroup hazard ratio
(C†), as a percentage of the target, on the harm subgroup Ĥ, vs n; the
dashed line at 0 marks no bias. (C) Empirical 95\% CI coverage of the
marginal subgroup hazard ratio (C†) for the MR \& naive estimator (line
type), on the harm subgroup Ĥ, vs per-trial sample size n (true
\(\theta^{\dagger}(H) = 1.00\), \(\theta^{\dagger}(H^{c}) = 0.66\)
(marginal HR); \(\theta^{\ddagger}(H) = 1.00\),
\(\theta^{\ddagger}(H^{c}) = 0.58\) (CDE); 500 simulations per cell,
NULL MR draws; detectors: consistency, dina, grf). Dashed line = 0.95
nominal. Uncertainty bands omitted for legibility (select a single
estimator to show them). All rates conditional on identifying a
subgroup. (D) Median 95\% CI width for the MR \& naive estimator (line
type) on the harm subgroup Ĥ, vs n (HR scale).}

\end{figure}

\begin{figure}

\centering{

\pandocbounded{\includegraphics[keepaspectratio]{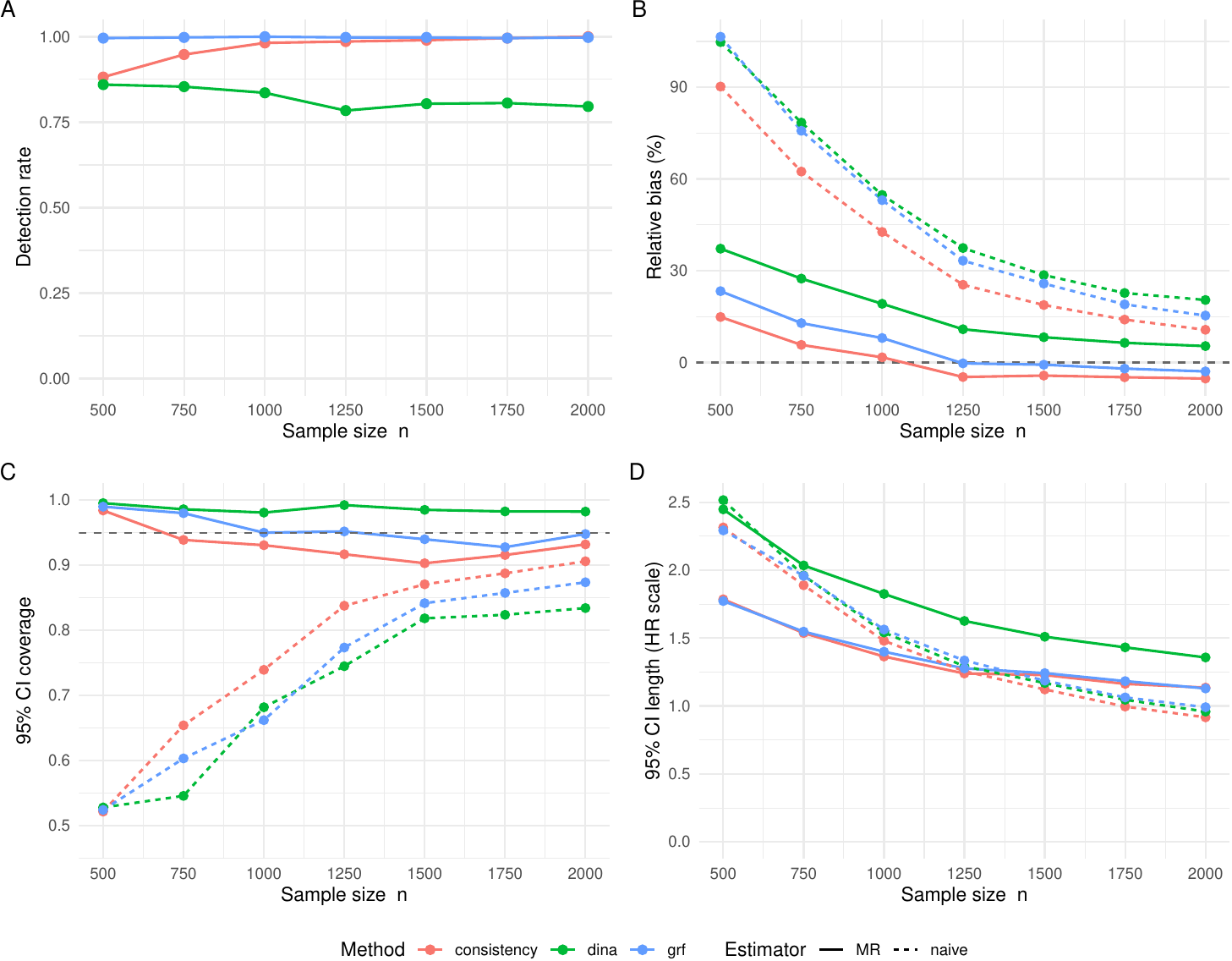}}

}

\caption{\label{fig-mr-coverage-h15c}(A) Subgroup-detection rate vs n by
detector, 500 simulations per cell. (B) Bias of the MR \& naive
estimator (line type) relative to the conditional subgroup hazard ratio
θ(Ĥ), as a percentage of the target, on the harm subgroup Ĥ, vs n; the
dashed line at 0 marks no bias. (C) Empirical 95\% CI coverage of the
conditional subgroup hazard ratio θ(Ĥ) for the MR \& naive estimator
(line type), on the harm subgroup Ĥ, vs per-trial sample size n (true
\(\theta^{\dagger}(H) = 1.51\), \(\theta^{\dagger}(H^{c}) = 0.66\)
(marginal HR); \(\theta^{\ddagger}(H) = 1.67\),
\(\theta^{\ddagger}(H^{c}) = 0.58\) (CDE); 500 simulations per cell,
NULL MR draws; detectors: consistency, dina, grf). Dashed line = 0.95
nominal. Uncertainty bands omitted for legibility (select a single
estimator to show them). All rates conditional on identifying a
subgroup. (D) Median 95\% CI width for the MR \& naive estimator (line
type) on the harm subgroup Ĥ, vs n (HR scale).}

\end{figure}

\begin{figure}

\centering{

\pandocbounded{\includegraphics[keepaspectratio]{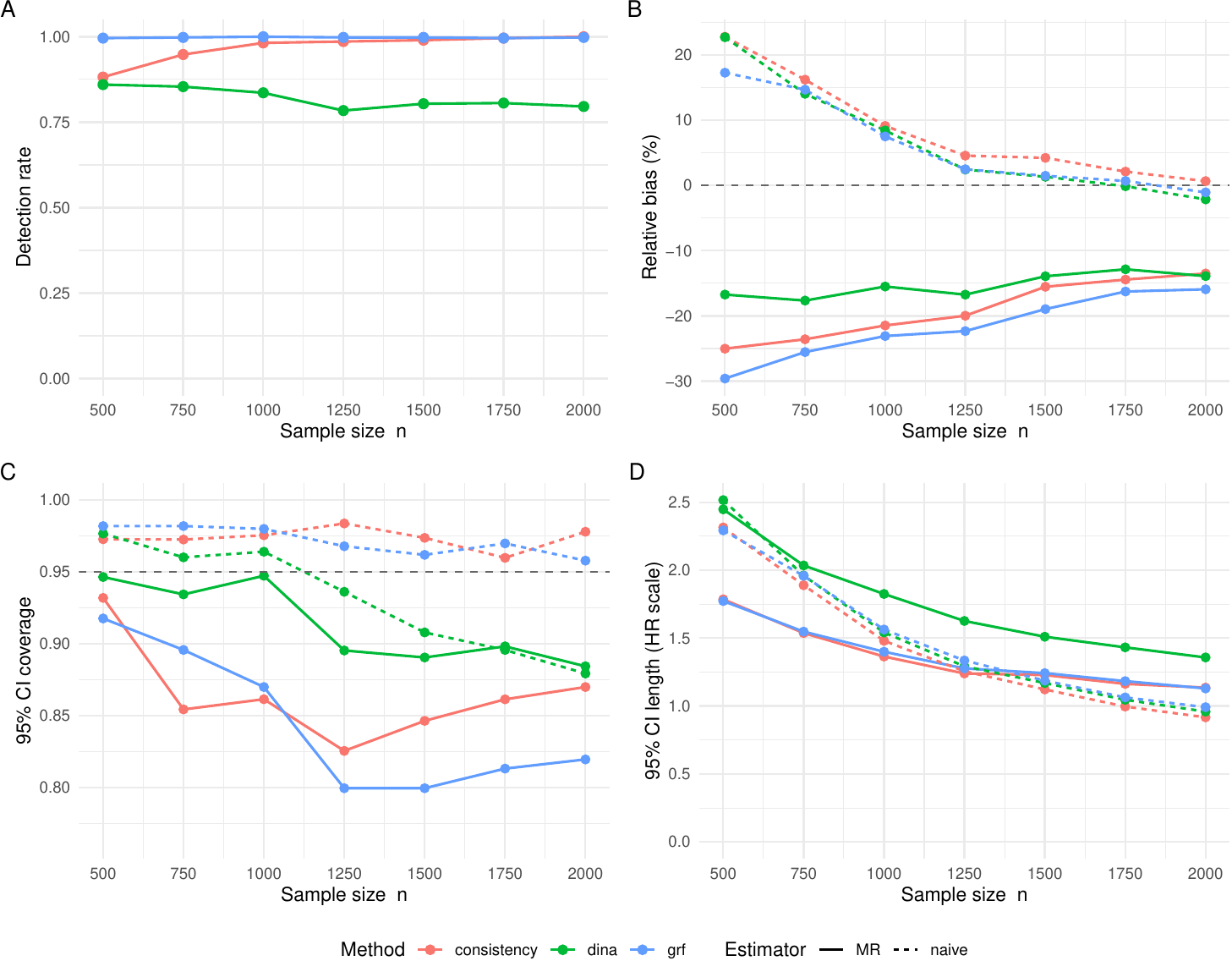}}

}

\caption{\label{fig-mr-coverage-h15m}(A) Subgroup-detection rate vs n by
detector, 500 simulations per cell. (B) Bias of the MR \& naive
estimator (line type) relative to the marginal subgroup hazard ratio
(C†), as a percentage of the target, on the harm subgroup Ĥ, vs n; the
dashed line at 0 marks no bias. (C) Empirical 95\% CI coverage of the
marginal subgroup hazard ratio (C†) for the MR \& naive estimator (line
type), on the harm subgroup Ĥ, vs per-trial sample size n (true
\(\theta^{\dagger}(H) = 1.51\), \(\theta^{\dagger}(H^{c}) = 0.66\)
(marginal HR); \(\theta^{\ddagger}(H) = 1.67\),
\(\theta^{\ddagger}(H^{c}) = 0.58\) (CDE); 500 simulations per cell,
NULL MR draws; detectors: consistency, dina, grf). Dashed line = 0.95
nominal. Uncertainty bands omitted for legibility (select a single
estimator to show them). All rates conditional on identifying a
subgroup. (D) Median 95\% CI width for the MR \& naive estimator (line
type) on the harm subgroup Ĥ, vs n (HR scale).}

\end{figure}

Table~\ref{tbl-sim-estimation-dina-h10},
Table~\ref{tbl-sim-estimation-grf-h10}, and
Table~\ref{tbl-sim-estimation-fs-h15} give estimation and coverage at a
single sample size (\(B = 300\) full-bootstrap resamples, 5,000
multiplier draws, 500 simulations):
Table~\ref{tbl-sim-estimation-dina-h10} for DINA and
Table~\ref{tbl-sim-estimation-grf-h10} for GRF, both at the
borderline-null \(\theta^{\dagger}(H) = 1.00\) and \(n = 500\) (the
false-positive stress test of Section 7 of the main paper), and
Table~\ref{tbl-sim-estimation-fs-h15} for forest search at a
genuine-harm \(\theta^{\dagger}(H) = 1.51\)
(\(\theta^{\ddagger}(H) = 1.67\)) and \(n = 1{,}000\). The sample-size
sweeps then trace operating characteristics against per-trial sample
size (\(n = 500\)--\(2000\); 1,000 simulations per cell) at three harm
strengths --- \(\theta^{\dagger}(H) = 0.75\) (protective, below the
null), 1.00, and 1.51 --- each rendered twice: once scored against the
marginal hazard ratio \(\theta^{\dagger}\)
(Figure~\ref{fig-mr-coverage-h075m}, Figure~\ref{fig-mr-coverage-h10m},
Figure~\ref{fig-mr-coverage-h15m}) and once against the conditional
estimand \(\theta(\widehat H)\) that Theorem 2 of the main paper
characterizes (Figure~\ref{fig-mr-coverage-h075c},
Figure~\ref{fig-mr-coverage-h10c}, Figure~\ref{fig-mr-coverage-h15c}).
Each panel reports detection rate, relative bias, 95\% coverage, and
median CI width for the multiplier-resampling and naive estimators (the
full bootstrap is not swept, owing to its computational cost).

\textbf{High-level findings.} The winner's curse dominates the naive
analysis, and within a single identifier it weakens as the signal
strengthens: for forest search the naive plug-in overstates the marginal
hazard ratio by +63\% at the borderline null (main-text estimation
table) and by only +10\% once a genuine effect is present
(Table~\ref{tbl-sim-estimation-fs-h15}). At the null the model-based
identifiers overstate the marginal by a comparable margin --- +58\%
(DINA) and +57\% (GRF). Multiplier resampling removes the bulk of this
optimism, returning DINA and GRF to within a few percent of the oracle
at the null (Table~\ref{tbl-sim-estimation-dina-h10},
Table~\ref{tbl-sim-estimation-grf-h10}), though as the harm signal
strengthens its correction can tip mildly negative against the marginal
target (-20.7\% for forest search at \(\theta^{\dagger}(H) = 1.51\); the
same intervals remain near-nominal against the conditional estimand,
Section~\ref{sec-sim-estimation-fs-h15}). The full bootstrap behaves
differently according to how the candidate family is formed. On the
fixed, enumerated forest-search family the two corrections target the
same estimand and remove the bulk of the selection bias, with the
multiplier intervals attaining nominal coverage; the residual difference
between them at these sample sizes is a finite-sample effect that
conflates two sources --- the genuine higher-order residual between the
two corrections and the Monte-Carlo difference in their resample counts
(\(B = 300\) full-bootstrap against 5,000 multiplier draws) --- which
these settings do not separate; isolating the first-order residual would
require matching the counts, and the first-order agreement of Theorem 1
of the main paper is in any case the large-\(n\) statement. On the
data-driven DINA and GRF families, by contrast, the full bootstrap
regenerates the family on each resample, credits a freshly-found extreme
region as selection optimism, and over-corrects --- undershooting the
marginal target by roughly a quarter (-25.9\% DINA, -24.1\% GRF). This
is the conditional-versus-unconditional distinction of Section 5 of the
main paper made empirical.

Across the sample-size sweeps, detection declines gently in \(n\) at and
below the null and flattens near one under strong harm, and
multiplier-resampling coverage of the marginal target improves toward
nominal as \(n\) grows, with several of the cell-to-cell movements lying
within the Monte-Carlo error shown by the Wilson bands. Coverage of the
conditional estimand \(\theta(\widehat H)\) shows the sharper contrast
the theory predicts: near-nominal and stable across \(n\) for the
fixed-family forest search, but strained for the data-driven
identifiers, whose coverage of \(\theta(\widehat H)\) recovers toward
nominal only as the proposed family stabilizes with \(n\) --- the
variance-deficit signature of Section 5 of the main paper. The bias
panels of the conditional-estimand sweeps
(Figure~\ref{fig-mr-coverage-h10c}, Figure~\ref{fig-mr-coverage-h15c})
show the point side of the same statement: the multiplier bias against
\(\theta(\widehat H)\) contracts with \(n\) for the fixed-family forest
search --- slowly at the borderline null, where the persistent near-tie
holds \(\theta(\widehat H)\) away from the marginal, and faster as the
signal separates --- the large-sample face of the point-consistency of
Theorem 2 of the main paper, distinct from the conservative coverage the
interval is built to attain. These coverage figures are not directly
comparable across identifiers, however: because every summary conditions
on detection and detection differs sharply between identifiers, each
identifier's coverage is averaged over a different subset of replicates.

\textbf{On the full-bootstrap interval lengths
(Table~\ref{tbl-sim-estimation-dina-h10} and
Table~\ref{tbl-sim-estimation-grf-h10}).} On a regenerating family the
full bootstrap occasionally re-selects a near-degenerate region --- a
small subgroup with few events in an arm --- whose within-subgroup Cox
interval is very wide, so the arithmetic mean CI length over resamples
is dominated by those rare draws while the median is unaffected. The
reported mean full-bootstrap CI length is accordingly far larger than
the well-behaved median (mean 82.0 against median 1.70 for DINA; mean
408.3 against median 1.75 for GRF), and these means are retained
deliberately rather than being transcription errors. The fixed-family
forest search shows no such behaviour, its full-bootstrap mean length
being comparable to the others' medians. The contrast is itself a
plausible finding: the instability of the unconditional full bootstrap
on model-selected families --- the very fragility the conditional
multiplier resampling avoids by holding the family fixed.

\textbf{Binary outcomes on the odds-ratio scale.}

\begin{figure}

\centering{

\pandocbounded{\includegraphics[keepaspectratio]{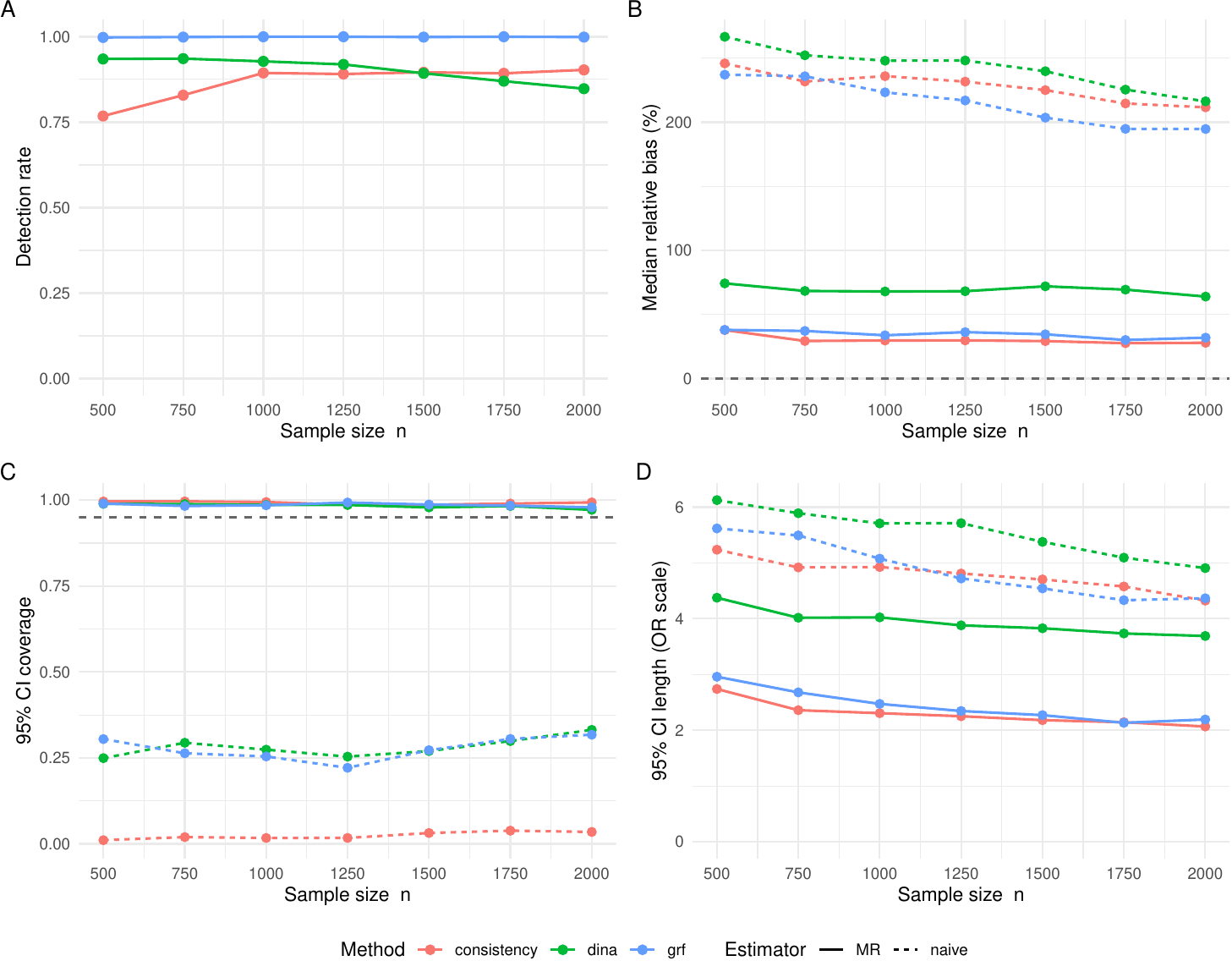}}

}

\caption{\label{fig-mr-coverage-or075c}(A) Subgroup-detection rate vs n
by detector; ACTG175 binary (1,000 simulations per cell). (B) Median
bias of the MR \& naive estimator (line type) relative to the
conditional subgroup odds ratio θ(Ĥ), as a percentage of the target, on
the harm subgroup Ĥ, vs n; the dashed line at 0 marks no bias. (C)
Empirical 95\% CI coverage of the conditional subgroup odds ratio θ(Ĥ)
for the MR \& naive estimator (line type), on the harm subgroup Ĥ, vs
per-trial sample size n (ACTG175 binary design, true
\(\theta^{\dagger}(H) = 0.75\), \(\theta^{\dagger}(H^{c}) = 0.66\)
(marginal OR); \(\theta^{\ddagger}(H) = 0.73\),
\(\theta^{\ddagger}(H^{c}) = 0.63\) (CDE); 1,000 simulations per cell,
NULL MR draws; detectors: consistency, dina, grf). Dashed line = 0.95
nominal. Uncertainty bands omitted for legibility (select a single
estimator to show them). All rates conditional on identifying a
subgroup. (D) Median 95\% CI width for the MR \& naive estimator (line
type) on the harm subgroup Ĥ, vs n (OR scale).}

\end{figure}

\begin{figure}

\centering{

\pandocbounded{\includegraphics[keepaspectratio]{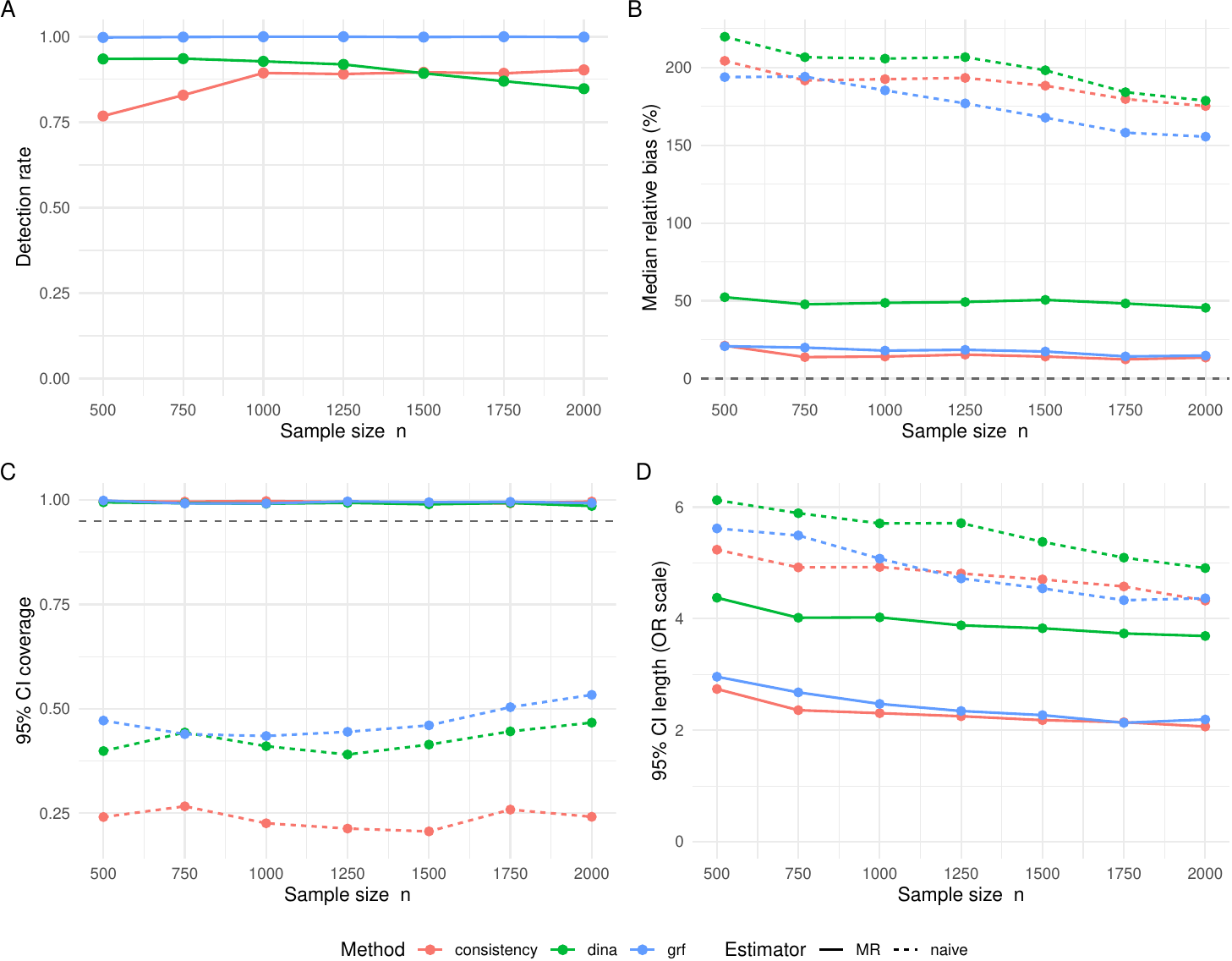}}

}

\caption{\label{fig-mr-coverage-or075m}(A) Subgroup-detection rate vs n
by detector; ACTG175 binary (1,000 simulations per cell). (B) Median
bias of the MR \& naive estimator (line type) relative to the marginal
subgroup odds ratio (C†), as a percentage of the target, on the harm
subgroup Ĥ, vs n; the dashed line at 0 marks no bias. (C) Empirical 95\%
CI coverage of the marginal subgroup odds ratio (C†) for the MR \& naive
estimator (line type), on the harm subgroup Ĥ, vs per-trial sample size
n (ACTG175 binary design, true \(\theta^{\dagger}(H) = 0.75\),
\(\theta^{\dagger}(H^{c}) = 0.66\) (marginal OR);
\(\theta^{\ddagger}(H) = 0.73\), \(\theta^{\ddagger}(H^{c}) = 0.63\)
(CDE); 1,000 simulations per cell, NULL MR draws; detectors:
consistency, dina, grf). Dashed line = 0.95 nominal. Uncertainty bands
omitted for legibility (select a single estimator to show them). All
rates conditional on identifying a subgroup. (D) Median 95\% CI width
for the MR \& naive estimator (line type) on the harm subgroup Ĥ, vs n
(OR scale).}

\end{figure}

A companion study carries the correction onto the binary, odds-ratio
path. Its data-generating mechanism is built on the ACTG175 covariate
distribution: baseline characteristics from the
zidovudine-plus-didanosine versus didanosine trial define the
super-population, treatment is randomized, and the binary outcome --- no
CD4 improvement --- is generated with a known subgroup
\(H = \{\textit{wtkg} > 70\text{th percentile}\} \cap \{\textit{cd40} > 70\text{th percentile}\}\),
the high-weight/high-CD4 region echoing the ACTG175 application,
carrying a marginal odds ratio \(\theta^{\dagger}(H) = 0.75\) against a
fixed complement \(\theta^{\dagger}(H^{c}) \approx 0.66\). This is a
\emph{protective} region --- the binary analogue of the survival
\(\theta^{\dagger}(H) = 0.75\) design point --- so it stresses the most
consequential error in this setting: declaring a protective subgroup
harmful. Forest search, DINA, and GRF each rediscover \(H\) from the raw
covariates over per-trial sample sizes \(n = 500\)--\(2000\) (500
simulations per cell, 5,000 multiplier draws; the full bootstrap is
again omitted for cost), every summary conditional on detection;
Figure~\ref{fig-mr-coverage-or075m} (marginal odds ratio) and
Figure~\ref{fig-mr-coverage-or075c} (conditional estimand) report
detection, bias, coverage, and median CI width as \(2 \times 2\) panels.

The odds-ratio scale reproduces the hazard-ratio reversal story.
Detection rises with \(n\) for forest search (0.77 to 0.90) and stays
high otherwise --- about 0.90 for DINA and near 1.00 for GRF. Because
the region is protective but is selected on apparent excess risk, the
naive within-subgroup odds ratio is driven above the protective truth,
and its interval covers the marginal target poorly (naive coverage of
\(\theta^{\dagger}(H)\) is 0.21-0.27 across the sweep). Multiplier
resampling removes the bulk of this reversal and restores near-nominal
coverage of the marginal target (0.99-1.00). The oracle, refit on the
true \(H\), is near-unbiased with nominal coverage throughout,
confirming that the residual movement is selection, not the
within-subgroup fit.

\textbf{Classification accuracy.}

\begin{figure}

\centering{

\pandocbounded{\includegraphics[keepaspectratio]{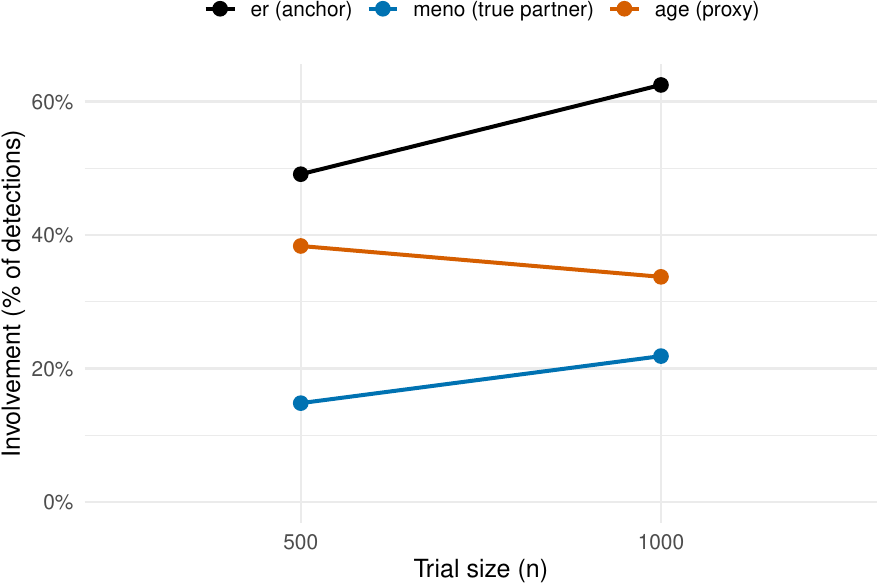}}

}

\caption{\label{fig-fs-id-involvement}Covariate involvement in the
FS-identified harm subgroup as a function of trial size, 500 replicates
per size, at the borderline null (\(\theta^{\dagger}(H) = 1.00\); M1
generator; true subgroup
\(\{\mathrm{er} \le 8\} \cap \{\mathrm{meno} = 0\}\)). The
oestrogen-receptor anchor (\(\mathrm{er}\)) and the true binary partner
(\(\mathrm{meno}\)) are recovered more often as \(n\) grows, whereas the
correlated continuous covariate \(\mathrm{age}\), a proxy for
\(\mathrm{meno}\), stays flat---the proxy does not give way with more
data.}

\end{figure}

\begin{figure}

\centering{

\pandocbounded{\includegraphics[keepaspectratio]{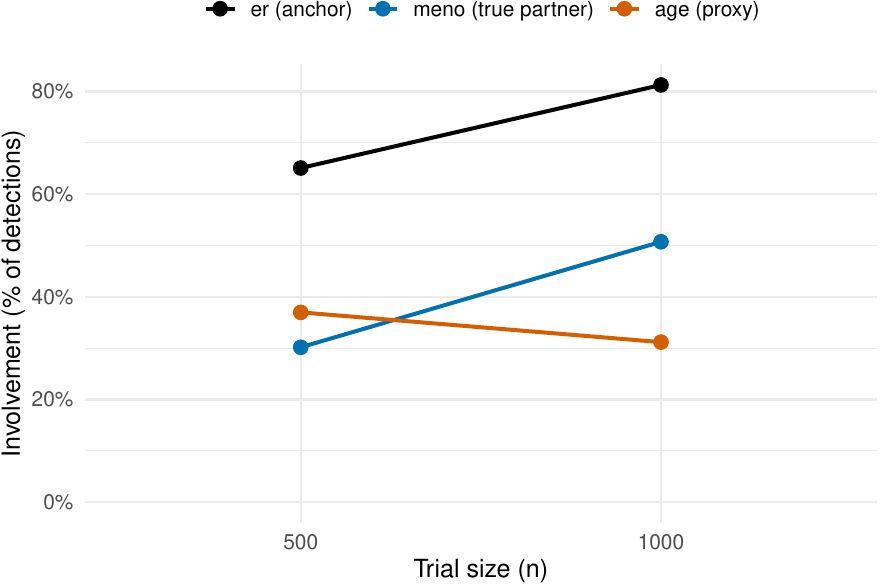}}

}

\caption{\label{fig-fs-id-involvement-h15}Covariate involvement in the
FS-identified harm subgroup as a function of trial size, 500 replicates
per size, under a genuine harm signal (\(\theta^{\dagger}(H) = 1.51\);
contrast the borderline-null \(\theta^{\dagger}(H) = 1.00\) case). M1
generator; true subgroup
\(\{\mathrm{er} \le 8\} \cap \{\mathrm{meno} = 0\}\). The
oestrogen-receptor anchor (\(\mathrm{er}\)) and the true binary partner
(\(\mathrm{meno}\)) are both recovered more often as \(n\) grows; unlike
the null case, \(\mathrm{meno}\) involvement rises steeply while the
\(\mathrm{age}\) proxy stays flat, so the true partner increasingly
fills the second slot once a real effect is present.}

\end{figure}

\begin{figure}

\centering{

\pandocbounded{\includegraphics[keepaspectratio]{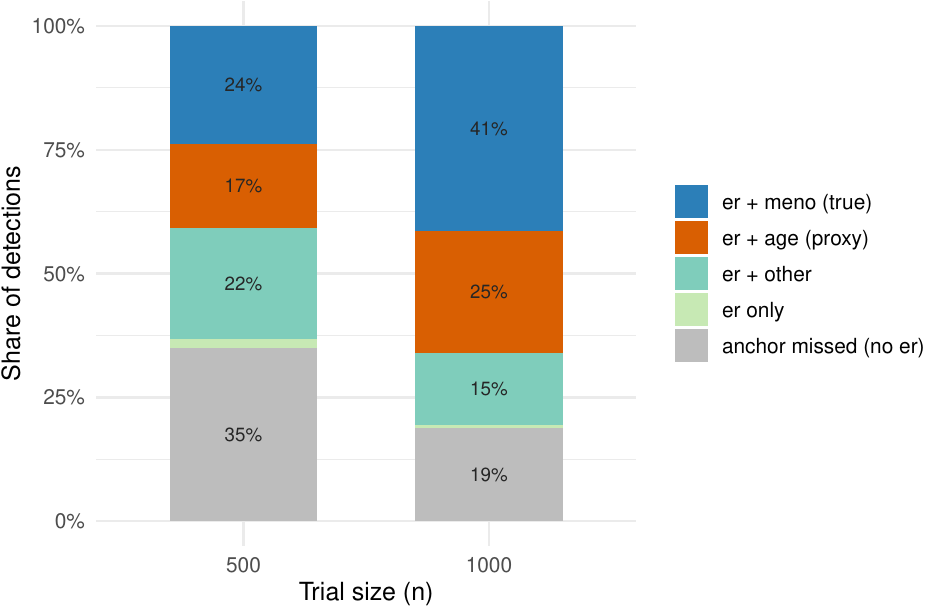}}

}

\caption{\label{fig-fs-id-structure-h15}Composition of the FS-identified
subgroup by structure, at \(n=500\) and \(n=1000\) (500 replicates
each), under a genuine harm signal (\(\theta^{\dagger}(H) = 1.51\)). As
\(n\) grows the anchor is missed less often and spurious second factors
recede, and---in contrast to the borderline-null case---the share
pairing the anchor with the true \(\mathrm{meno}\) partner pulls further
ahead of the \(\mathrm{age}\)-proxy share: the near-tie that traps
identification at the null resolves toward the truth once the effect
separates.}

\end{figure}

The bias and coverage performance depends on the classification accuracy
of the identifier. Recall that the true subgroup in the simulations
involves two factors,
\(H = \{\text{er} \le 8\} \cap \{\text{meno} = 0\}\). Forest search
locates the oestrogen-receptor anchor with increasing reliability as the
trial grows (Figure~\ref{fig-fs-id-involvement}), but the partner factor
of the identified pair divides between the true binary modifier
\(\text{meno}\) and the correlated continuous covariate \(\text{age}\)
--- premenopausal patients are younger, so a data-driven cut
\(\{\text{age} \le c\}\) approximates \(\{\text{meno} = 0\}\).
Conditional on the search reaching this contest, the division barely
moves as the sample size doubles (Table~\ref{tbl-fs-id-structure}); the
proxy does not give way with more data. In the language of Section 5.1
of the main paper this is not the separated regime but the near-tie
boundary that (A5) excludes (\(\eta_n = o(n^{-1/2})\)) --- two candidate
regions whose population effects remain indistinguishable at the
\(n^{-1/2}\) resolution, here because the proxy cut tracks the causal
partition closely enough that no sample size resolves the margin. The
anchor approaches separation while the partner sits at the excluded
boundary. The consequence is interpretive rather than inferential:
because the multiplier correction conditions on whichever region is
realized, \(\beta(\widehat{H})\) remains the estimand whether
\(\widehat{H}\) names \(\text{meno}\) or its \(\text{age}\) proxy, but
the covariate naming a discovered subgroup may be a correlate of the
effect modifier rather than the modifier itself --- an ambiguity more
data does not dispel at this borderline-null strength. Under a genuine
harm signal (\(\theta^{\dagger}(H) = 1.51\)), by contrast, the same
identification displays (Figure~\ref{fig-fs-id-involvement-h15},
Figure~\ref{fig-fs-id-structure-h15},
Table~\ref{tbl-fs-id-structure-h15}) show the true \(\text{meno}\)
partner increasingly recovered and the \(\text{age}\) proxy receding as
\(n\) grows---the conditional partner split shifting toward the
truth---so the ambiguity resolves once the effect separates.

\subsection{Estimation and coverage: full bootstrap (FB) versus
multiplier resampling (MR) de-biased
procedures}\label{sec-sim-estimation}

Relative bias and \(95\%\) coverage are reported against four targets
--- the oracle (\(b^{\text{oracle}}\), \(C^{\text{oracle}}\)), the
conditional estimand \(\theta(\widehat H) = \exp\{\beta(\widehat H)\}\)
whose coverage Theorem 2 characterizes (\(b^{\theta}\), \(C^{\theta}\)),
the controlled direct effect (\(b^{\ddagger}\), \(C^{\ddagger}\)), and
the marginal subgroup hazard ratio (\(b^{\dagger}\), \(C^{\dagger}\)).
The empirical standard deviation \(\text{SD}_{\beta}\) and the mean
model standard error \(\widehat{\text{SD}}_{\beta}\) are reported on the
log-hazard-ratio (\(\beta\)) scale. The reported length is the mean
confidence-interval length. The same conventions apply to the analogous
tables in the supplementary material.

Table~\ref{tbl-sim-estimation-fs-h10} reports estimation and coverage
where forest search identified a subgroup in 66\% of the 500
simulations. The naive plug-in at \(\widehat H\) inherits the selection
bias of the identified subgroup, over-estimating the standard
within-subgroup analysis of the true harm subgroup (the oracle) by 56\%,
and the population effect at the realized region \(\theta(\widehat H)\)
by 129\%. Relative to the oracle, both FB and MR remove the bulk of it,
and on this fixed family they behave similarly, with MR providing a more
pronounced bias reduction. The conditional estimand is itself a
per-replicate random target, varying with the realized region
\(\widehat H\). Because the identifier's sensitivity and specificity are
modest at this sample size, \(\widehat H\) typically absorbs subjects
from the complement, for whom the effect is a substantial benefit
relative to the null marginal hazard ratio; the population hazard ratio
\(\theta(\widehat H)\) is thereby pulled below one, and in this
borderline-null configuration the region forest search labels harmful
is, in the population, mildly protective. While a detection rate of 66\%
would generally be worrisome in this scenario, the FB and MR de-biasing
procedures provide a reasonable degree of mitigation.

The oracle row carries no entry in the \(b^{\theta}\) and \(C^{\theta}\)
columns. It refits on the true region \(H\), so it estimates the effect
there rather than the conditional estimand at the realized region; those
entries would measure the mismatch between the two regions, not a bias
or a coverage rate. The mismatch is reported instead as the
identification factor \(\kappa = \beta(H)/\beta(\widehat H) = 1.42\) ---
the true harm region carries a population effect 42 per cent above the
effect at the region the search returned, the effect-scale counterpart
to the classification rates above. The complement confirms the reading:
\(\widehat H^{c}\) is 83.9\% of the trial and nearly coincides with
\(H^{c}\), and there the suppressed entry was -0.6\%.

\begin{table}

\caption{\label{tbl-sim-estimation-fs-h10}Estimation and coverage for
forest search over the 329 detected replicates (66\% of 500 simulations;
n = 500). Scenario hazard-ratio effects \(\theta^{\dagger}(H) = 1.00\),
\(\theta^{\dagger}(H^{c}) = 0.66\), \(\theta^{\ddagger}(H) = 1.00\),
\(\theta^{\ddagger}(H^{c}) = 0.58\); the conditional estimand
\(\theta(\widehat H)\) has mean 0.72, median 0.70 (hazard-ratio scale)
and SD 0.14 (\(\beta\) scale). Classification of \(\widehat H\) vs the
true harm subgroup: sens 0.36, spec 0.87, PPV 0.30, NPV 0.91. The oracle
refits on the true region \(H\), so it does not estimate the conditional
estimand \(\theta(\widehat H)\); its \(b^{\theta}\) and \(C^{\theta}\)
entries are a region mismatch rather than a bias or coverage rate and
are suppressed. That mismatch is the identification factor
\(\kappa = \beta(H)/\beta(\widehat H)\) = 1.42 (median 1.43), with
\(\beta(H)\) = 1.003.}

\centering{

\centering\begingroup\fontsize{10}{12}\selectfont

\begin{tabular}[t]{lcccccccccccc}
\toprule
\multicolumn{5}{c}{ } & \multicolumn{4}{c}{Relative bias (\%)} & \multicolumn{4}{c}{95\% coverage} \\
\cmidrule(l{3pt}r{3pt}){6-9} \cmidrule(l{3pt}r{3pt}){10-13}
Estimator & Avg & \(\text{SD}_{\beta}\) & \(\widehat{\text{SD}}_{\beta}\) & Length & \(b^{\text{oracle}}\) & \(b^{\theta}\) & \(b^{\ddagger}\) & \(b^{\dagger}\) & \(C^{\text{oracle}}\) & \(C^{\theta}\) & \(C^{\ddagger}\) & \(C^{\dagger}\)\\
\midrule
\addlinespace[0.3em]
\multicolumn{13}{l}{\textbf{Harm Ĥ (mean |Ĥ| = 80, 16.1\% ITT)}}\\
\hspace{1em}oracle & 1.13 & 0.31 & 0.33 & 1.55 & +0.0 & — & +13.3 & +13.2 & 1.00 & — & 0.96 & 0.96\\
\hspace{1em}naive & 1.63 & 0.17 & 0.32 & 2.22 & +55.7 & +129.3 & +62.9 & +62.8 & 0.78 & 0.13 & 0.86 & 0.86\\
\hspace{1em}FB & 1.22 & 0.19 & 0.36 & 1.89 & +16.6 & +71.9 & +22.2 & +22.2 & 0.97 & 0.81 & 0.99 & 0.99\\
\hspace{1em}MR & 0.93 & 0.24 & 0.39 & 1.63 & -11.6 & +31.1 & -6.5 & -6.6 & 0.97 & 0.98 & 0.99 & 0.99\\
\addlinespace[0.3em]
\multicolumn{13}{l}{\textbf{Complement Ĥᶜ (mean |Ĥᶜ| = 420, 83.9\% ITT)}}\\
\hspace{1em}oracle & 0.66 & 0.13 & 0.14 & 0.36 & +0.0 & — & +12.4 & +0.0 & 1.00 & — & 0.89 & 0.96\\
\hspace{1em}naive & 0.60 & 0.13 & 0.14 & 0.33 & -7.9 & -8.5 & +3.3 & -8.0 & 1.00 & 0.91 & 0.95 & 0.92\\
\hspace{1em}FB & 0.63 & 0.13 & 0.19 & 0.49 & -3.7 & -4.4 & +8.0 & -3.8 & 1.00 & 0.98 & 0.98 & 0.98\\
\hspace{1em}MR & 0.65 & 0.14 & 0.25 & 0.67 & -1.3 & -1.9 & +10.8 & -1.4 & 1.00 & 1.00 & 0.99 & 1.00\\
\bottomrule
\end{tabular}
\endgroup{}

}

\end{table}%

\subsection{DINA at the borderline null: estimation and
coverage}\label{sec-sim-estimation-dina-h10}

DINA enters as a candidate \emph{generator}: it proposes regions from
its fitted effect surface, and those regions are re-ranked on the
inferential coefficient \(\hat\beta(g)\)
(Section~\ref{sec-appendix-dina}). The candidate family it proposes is
therefore not the fixed enumerated family of forest search, so the two
de-biased procedures need not target the same estimand here.
Table~\ref{tbl-sim-estimation-dina-h10} reports estimation and coverage
where DINA declared a (false) harm subgroup in 70\% of the 500
simulations. The naive plug-in at \(\widehat H\) inherits the selection
bias of the identified subgroup, over-estimating the standard
within-subgroup analysis of the true harm subgroup (the oracle) by 56\%
and the population effect at the realized region \(\theta(\widehat H)\)
by 129\%. Multiplier resampling removes the bulk of this optimism
against the oracle and marginal targets (MR bias against the oracle
+5.2\%), while the full bootstrap -- regenerating the DINA family on
each resample and crediting a freshly-found extreme region as selection
optimism -- over-corrects, undershooting the marginal target (FB bias
-25.9\%). This is the conditional-versus-unconditional distinction of
Section 5 of the main paper made empirical. Because the family
regenerates under resampling, DINA does not attain the fixed-family
coverage of \(\theta(\widehat H)\) that forest search does; its coverage
against \(\theta(\widehat H)\) improves toward nominal with \(n\) as the
proposed family stabilizes, the variance-deficit signature of Section 5
of the main paper traced across sample size in
Figure~\ref{fig-mr-coverage-h10c}.

\begin{table}

\caption{\label{tbl-sim-estimation-dina-h10}Estimation and coverage for
DINA over the 350 detected replicates (70\% of 500 simulations; n =
500). Scenario hazard-ratio effects \(\theta^{\dagger}(H) = 1.00\),
\(\theta^{\dagger}(H^{c}) = 0.66\), \(\theta^{\ddagger}(H) = 1.00\),
\(\theta^{\ddagger}(H^{c}) = 0.58\); the conditional estimand
\(\theta(\widehat H)\) has mean 0.70, median 0.69 (hazard-ratio scale)
and SD 0.12 (\(\beta\) scale). Classification of \(\widehat H\) vs the
true harm subgroup: sens 0.28, spec 0.88, PPV 0.24, NPV 0.90.
Estimators, targets, and SD scales are defined with the main-text
estimation tables (Section 7). The oracle refits on the true region
\(H\), so it does not estimate the conditional estimand
\(\theta(\widehat H)\); its \(b^{\theta}\) and \(C^{\theta}\) entries
are a region mismatch rather than a bias or coverage rate and are
suppressed. That mismatch is the identification factor
\(\kappa = \beta(H)/\beta(\widehat H)\) = 1.45 (median 1.46), with
\(\beta(H)\) = 1.003.}

\centering{

\centering\begingroup\fontsize{10}{12}\selectfont

\begin{tabular}[t]{lcccccccccccc}
\toprule
\multicolumn{5}{c}{ } & \multicolumn{4}{c}{Relative bias (\%)} & \multicolumn{4}{c}{95\% coverage} \\
\cmidrule(l{3pt}r{3pt}){6-9} \cmidrule(l{3pt}r{3pt}){10-13}
Estimator & Avg & \(\text{SD}_{\beta}\) & \(\widehat{\text{SD}}_{\beta}\) & Length & \(b^{\text{oracle}}\) & \(b^{\theta}\) & \(b^{\ddagger}\) & \(b^{\dagger}\) & \(C^{\text{oracle}}\) & \(C^{\theta}\) & \(C^{\ddagger}\) & \(C^{\dagger}\)\\
\midrule
\addlinespace[0.3em]
\multicolumn{13}{l}{\textbf{Harm Ĥ (mean |Ĥ| = 71, 14.3\% ITT)}}\\
\hspace{1em}oracle & 1.09 & 0.32 & 0.33 & 1.50 & +0.0 & — & +9.4 & +9.4 & 1.00 & — & 0.95 & 0.95\\
\hspace{1em}naive & 1.58 & 0.28 & 0.36 & 2.58 & +56.5 & +129.0 & +58.3 & +58.2 & 0.86 & 0.37 & 0.88 & 0.88\\
\hspace{1em}FB & 0.74 & 0.31 & 0.68 & 81.99 & -27.2 & +6.0 & -25.9 & -25.9 & 0.93 & 1.00 & 0.96 & 0.96\\
\hspace{1em}MR & 1.07 & 0.30 & 0.49 & 2.63 & +5.2 & +53.8 & +6.6 & +6.5 & 0.99 & 0.99 & 0.99 & 0.99\\
\addlinespace[0.3em]
\multicolumn{13}{l}{\textbf{Complement Ĥᶜ (mean |Ĥᶜ| = 429, 85.7\% ITT)}}\\
\hspace{1em}oracle & 0.66 & 0.13 & 0.14 & 0.36 & +0.0 & — & +12.1 & -0.2 & 1.00 & — & 0.90 & 0.97\\
\hspace{1em}naive & 0.62 & 0.13 & 0.14 & 0.34 & -4.6 & -6.0 & +6.7 & -5.0 & 1.00 & 0.93 & 0.95 & 0.95\\
\hspace{1em}FB & 0.66 & 0.14 & 0.24 & 0.65 & +0.7 & -0.6 & +12.8 & +0.4 & 1.00 & 1.00 & 0.99 & 1.00\\
\hspace{1em}MR & 0.65 & 0.13 & 0.26 & 0.70 & -0.0 & -1.3 & +12.0 & -0.3 & 1.00 & 1.00 & 1.00 & 1.00\\
\bottomrule
\end{tabular}
\endgroup{}

}

\end{table}%

\subsection{GRF at the borderline null: estimation and
coverage}\label{sec-sim-estimation-grf-h10}

Like DINA, the causal forest (GRF) enters as a candidate
\emph{generator}, its qualifying regions re-ranked on the inferential
coefficient \(\hat\beta(g)\) (Section~\ref{sec-appendix-grf}); the
family it proposes regenerates under resampling rather than being held
fixed. Table~\ref{tbl-sim-estimation-grf-h10} reports estimation and
coverage where GRF declared a (false) harm subgroup in 98\% of the 500
simulations -- the highest detection rate of the three identifiers at
this borderline null. The naive plug-in at \(\widehat H\) over-estimates
the standard within-subgroup analysis of the true harm subgroup (the
oracle) by 62\% and the population effect at the realized region
\(\theta(\widehat H)\) by 131\%. Multiplier resampling returns the
estimate to within a few percent of the oracle (MR bias against the
oracle -7.2\%), whereas the full bootstrap over-corrects on the
regenerating family, undershooting the marginal target (FB bias -24.1\%)
-- the conditional-versus-unconditional distinction of Section 5 of the
main paper. As with DINA, coverage of the conditional estimand
\(\theta(\widehat H)\) falls below the fixed-family level forest search
attains and improves toward nominal with \(n\); the sample-size
behaviour is shown in Figure~\ref{fig-mr-coverage-h10c}.

\begin{table}

\caption{\label{tbl-sim-estimation-grf-h10}Estimation and coverage for
GRF over the 490 detected replicates (98\% of 500 simulations; n = 500).
Scenario hazard-ratio effects \(\theta^{\dagger}(H) = 1.00\),
\(\theta^{\dagger}(H^{c}) = 0.66\), \(\theta^{\ddagger}(H) = 1.00\),
\(\theta^{\ddagger}(H^{c}) = 0.58\); the conditional estimand
\(\theta(\widehat H)\) has mean 0.69, median 0.67 (hazard-ratio scale)
and SD 0.11 (\(\beta\) scale). Classification of \(\widehat H\) vs the
true harm subgroup: sens 0.26, spec 0.87, PPV 0.22, NPV 0.89.
Estimators, targets, and SD scales are defined with the main-text
estimation tables (Section 7). The oracle refits on the true region
\(H\), so it does not estimate the conditional estimand
\(\theta(\widehat H)\); its \(b^{\theta}\) and \(C^{\theta}\) entries
are a region mismatch rather than a bias or coverage rate and are
suppressed. That mismatch is the identification factor
\(\kappa = \beta(H)/\beta(\widehat H)\) = 1.48 (median 1.50), with
\(\beta(H)\) = 1.003.}

\centering{

\centering\begingroup\fontsize{10}{12}\selectfont

\begin{tabular}[t]{lcccccccccccc}
\toprule
\multicolumn{5}{c}{ } & \multicolumn{4}{c}{Relative bias (\%)} & \multicolumn{4}{c}{95\% coverage} \\
\cmidrule(l{3pt}r{3pt}){6-9} \cmidrule(l{3pt}r{3pt}){10-13}
Estimator & Avg & \(\text{SD}_{\beta}\) & \(\widehat{\text{SD}}_{\beta}\) & Length & \(b^{\text{oracle}}\) & \(b^{\theta}\) & \(b^{\ddagger}\) & \(b^{\dagger}\) & \(C^{\text{oracle}}\) & \(C^{\theta}\) & \(C^{\ddagger}\) & \(C^{\dagger}\)\\
\midrule
\addlinespace[0.3em]
\multicolumn{13}{l}{\textbf{Harm Ĥ (mean |Ĥ| = 72, 14.4\% ITT)}}\\
\hspace{1em}oracle & 1.05 & 0.32 & 0.33 & 1.43 & +0.0 & — & +4.8 & +4.7 & 1.00 & — & 0.96 & 0.96\\
\hspace{1em}naive & 1.57 & 0.23 & 0.37 & 2.59 & +62.1 & +131.2 & +56.9 & +56.8 & 0.83 & 0.37 & 0.93 & 0.93\\
\hspace{1em}FB & 0.76 & 0.28 & 0.68 & 408.32 & -21.8 & +11.2 & -24.1 & -24.1 & 0.94 & 0.99 & 0.99 & 0.99\\
\hspace{1em}MR & 0.90 & 0.28 & 0.46 & 1.98 & -7.2 & +32.8 & -9.7 & -9.8 & 0.98 & 0.98 & 1.00 & 1.00\\
\addlinespace[0.3em]
\multicolumn{13}{l}{\textbf{Complement Ĥᶜ (mean |Ĥᶜ| = 428, 85.6\% ITT)}}\\
\hspace{1em}oracle & 0.63 & 0.13 & 0.14 & 0.35 & +0.0 & — & +8.4 & -3.5 & 1.00 & — & 0.92 & 0.94\\
\hspace{1em}naive & 0.60 & 0.13 & 0.14 & 0.33 & -5.1 & -9.7 & +2.7 & -8.6 & 1.00 & 0.88 & 0.96 & 0.90\\
\hspace{1em}FB & 0.65 & 0.13 & 0.26 & 0.68 & +2.9 & -2.1 & +11.3 & -0.9 & 1.00 & 1.00 & 1.00 & 1.00\\
\hspace{1em}MR & 0.65 & 0.13 & 0.26 & 0.68 & +2.6 & -2.4 & +11.0 & -1.2 & 1.00 & 1.00 & 1.00 & 1.00\\
\bottomrule
\end{tabular}
\endgroup{}

}

\end{table}%

\subsection{Simulation synthesis: per-analysis bias and coverage (h =
1.0)}\label{sec-supp-sim-synth}

These figures and the detailed discussion accompany the simulation
highlights of Section 7 of the main paper.

\begin{figure}

\centering{

\includegraphics[width=1\linewidth,height=\textheight,keepaspectratio]{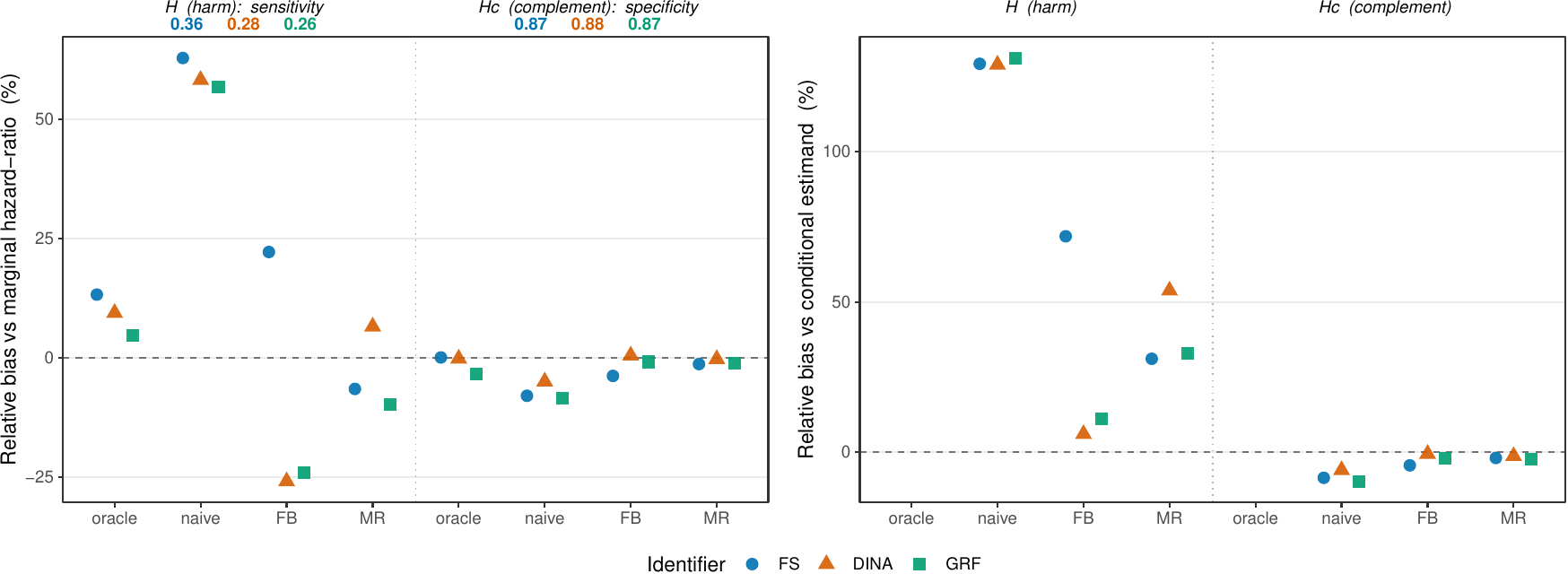}

}

\caption{\label{fig-sim-synthesis-h10-sidebyside}Relative bias of the
within-subgroup hazard ratio against two targets, for forest search
(FS), DINA, and GRF under the simulation setting corresponding to
Table~\ref{tbl-sim-estimation-fs-h10}: the marginal hazard-ratio
\(\theta^{\dagger}\) in the \emph{true} subgroup (left) and the
conditional estimand \(\theta(\widehat H)\) at the \emph{realized}
region (right). At each of the eight analyses (oracle, naive, full
bootstrap (FB), and multiplier resampling (MR) within the identified
harm subgroup \(\widehat H\) and its complement \(\widehat H^{c}\)) the
three identifiers are shown side by side; the dashed line marks zero
bias. The colour-coded readout above the left panel reports each
identifier's recovery of the true membership (sensitivity on
\(\widehat H\), specificity on \(\widehat H^{c}\)) and applies to both
panels. The oracle appears in the left (marginal-target) panel only: it
refits on the true subgroup \(H\), so it does not estimate the
conditional estimand \(\theta(\widehat H)\) and an entry against that
target would report the mismatch between the true and realized regions
rather than a bias. Summaries are over the detected replicates (66\%,
70\%, and 98\% for FS, DINA, and GRF).}

\end{figure}%

\begin{figure}

\centering{

\includegraphics[width=1\linewidth,height=\textheight,keepaspectratio]{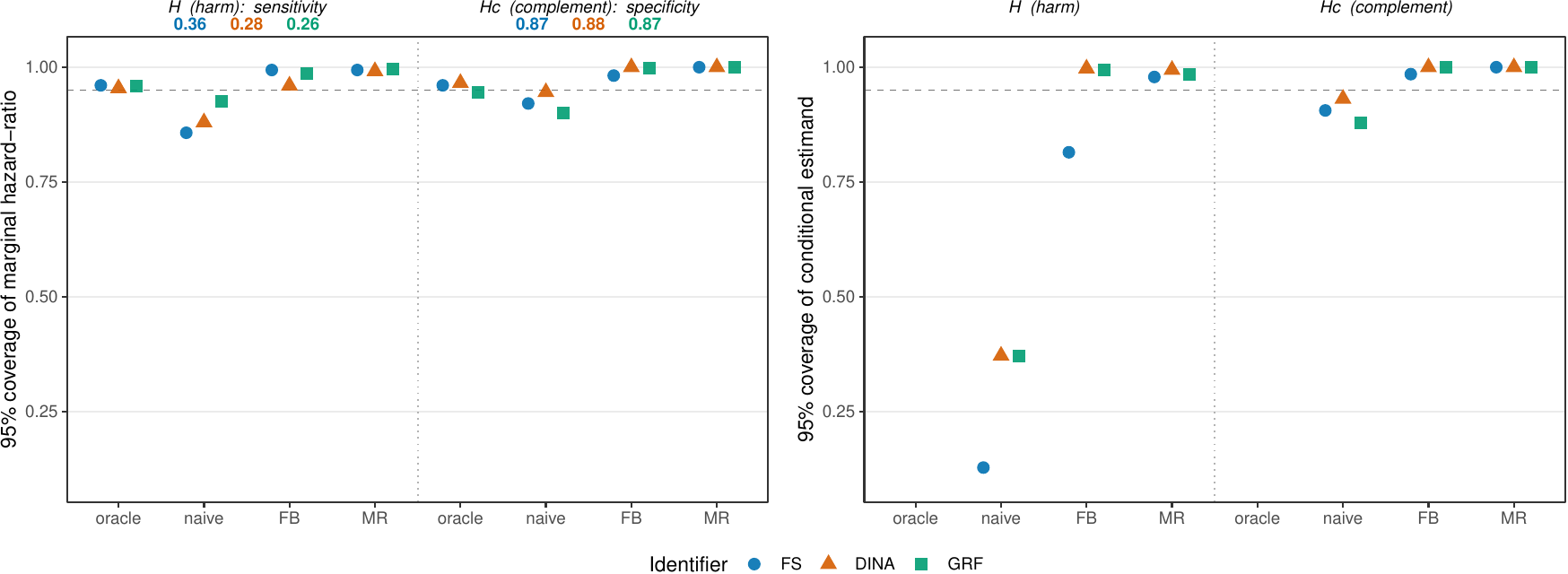}

}

\caption{\label{fig-sim-synthesis-h10-betaHhat}Coverage of the nominal
95\% interval against two estimands, for forest search (FS), DINA, and
GRF at the eight analyses of
Figure~\ref{fig-sim-synthesis-h10-sidebyside}. The oracle appears in the
left (marginal-target) panel only, since it refits on the true subgroup
\(H\) and so does not estimate the conditional estimand. The dashed line
marks 0.95 and summaries are over the detected replicates.}

\end{figure}%

The winner's curse is the dominant feature of the naive analysis and the
correction removes most of it: on the harm subgroup the naive plugin
overstates the marginal hazard ratio by 63\% (forest search), 58\%
(DINA), and 57\% (GRF), while the oracle --- refit on the true subgroup
--- carries only a small upward residual (+13\%, +9\%, and +5\%). This
residual combines the ordinary finite-sample inflation of a Cox fit on a
small subgroup, present even without selection, with a muted spillover
of the selection optimism onto the true-subgroup fit through the overlap
between candidate and true regions (Equation 7 of the main paper); it is
largest for forest search, whose lower detection rate makes the
conditioning event most restrictive, and shrinks as detection broadens
(DINA 70\%, GRF 98\%). Relative to the oracle's own residuals,
multiplier resampling brings each identifier close to the marginal
hazard-ratio --- forest search to -6.6\% (oracle +13.2\%), DINA to
+6.5\% (+9.4\%), and GRF to -9.8\% (+4.7\%).

The full bootstrap is the \emph{unconditional} comparator and behaves
differently by identifier. For forest search, whose candidate family is
fixed, it moves with the multiplier --- both pull the naive plugin back
toward the target (+22.2\% against -6.6\%). For DINA and GRF, whose
families regenerate under resampling, it instead \emph{undershoots} ---
-25.9\% and -24.1\%. In the complement all three are essentially
unbiased under every estimator (\(|b^{\dagger}| \le 9\%\)). The capture
rates above the panel locate the difficulty directly: the complement is
recovered fairly well (specificity 0.87--0.87, negative predictive value
0.89--0.91), the harm subgroup only partially (sensitivity 0.26--0.36,
positive predictive value 0.22--0.30), with forest search the most
sensitive (DINA and GRF are similar).

In terms of coverage the clinically relevant target is the marginal
hazard ratio \(\theta^{\dagger}(H)\) in the true subgroup --- the effect
the oracle estimates and a pre-specified analysis would report --- and
the multiplier-resampling intervals cover it at or above nominal in the
complement for all three identifiers, and at nominal in the harm
subgroup for forest search and GRF; the one shortfall is DINA in the
harm subgroup (0.99). The full bootstrap, with its wider unconditional
intervals, covers at or above nominal throughout --- including DINA's
harm subgroup (0.96).

The conditional estimand \(\theta(\widehat H)\) is the parameter Theorem
2 characterizes. The low positive predictive value fixes where it falls:
the realized region mixes the null within-harm effect on its true
members with the protective complement effect on the false positives it
admits, so \(\theta(\widehat H)\) sits below the marginal hazard ratio
--- near a positive-predictive-value-weighted average of
\(\theta^{\dagger}(H)\) and \(\theta^{\dagger}(H^{c})\) --- and every
analysis that estimates it (naive, full bootstrap, and multiplier
resampling) shows positive bias against it in the right panel of
Figure~\ref{fig-sim-synthesis-h10-sidebyside}; the oracle is absent from
that panel, since it refits on the true region \(H\) and so does not
estimate \(\theta(\widehat H)\). Theorem 2's guarantee for that estimand
is asymptotic and conditional: in the limit the interval covers
\(\theta(\widehat H)\) at rate at least \(1 - \alpha\) where the
quantile-domination condition of its part (iv) holds. What the
simulations report is the coverage itself rather than that condition,
and they find it near or above nominal across the sample sizes studied;
point agreement with \(\theta(\widehat H)\), by contrast, is a sharp
large-sample property, approached only as the competition separates.
That coverage (Figure~\ref{fig-sim-synthesis-h10-betaHhat}) tracks the
scope of the theorem: near-nominal for forest search (0.98), whose
candidate family is fixed, as the theorem requires, and lower for DINA
and GRF (0.99, 0.98), whose families regenerate under resampling and
fall outside the fixed-family guarantee.

\subsection{Estimation and coverage: full bootstrap (FB) versus
multiplier resampling (MR) de-biased
procedures}\label{sec-sim-estimation}

Relative bias and \(95\%\) coverage are reported against four targets
--- the oracle (\(b^{\text{oracle}}\), \(C^{\text{oracle}}\)), the
conditional estimand \(\theta(\widehat H) = \exp\{\beta(\widehat H)\}\)
whose coverage Theorem 2 characterizes (\(b^{\theta}\), \(C^{\theta}\)),
the controlled direct effect (\(b^{\ddagger}\), \(C^{\ddagger}\)), and
the marginal subgroup hazard ratio (\(b^{\dagger}\), \(C^{\dagger}\)).
The empirical standard deviation \(\text{SD}_{\beta}\) and the mean
model standard error \(\widehat{\text{SD}}_{\beta}\) are reported on the
log-hazard-ratio (\(\beta\)) scale. The reported length is the mean
confidence-interval length. The same conventions apply to the analogous
tables in the supplementary material.

Table~\ref{tbl-sim-estimation-fs-h10-n1000} reports estimation and
coverage where forest search identified a subgroup in 66\% of the 500
simulations. The naive plug-in at \(\widehat H\) inherits the selection
bias of the identified subgroup, over-estimating the standard
within-subgroup analysis of the true harm subgroup (the oracle) by 31\%,
and the population effect at the realized region \(\theta(\widehat H)\)
by 85\%. Relative to the oracle, both FB and MR remove the bulk of it,
and on this fixed family they behave similarly, with MR providing a more
pronounced bias reduction. The conditional estimand is itself a
per-replicate random target, varying with the realized region
\(\widehat H\). Because the identifier's sensitivity and specificity are
modest at this sample size, \(\widehat H\) typically absorbs subjects
from the complement, for whom the effect is a substantial benefit
relative to the null marginal hazard ratio; the population hazard ratio
\(\theta(\widehat H)\) is thereby pulled below one, and in this
borderline-null configuration the region forest search labels harmful
is, in the population, mildly protective. While a detection rate of 66\%
would generally be worrisome in this scenario, the FB and MR de-biasing
procedures provide a reasonable degree of mitigation.

The oracle row carries no entry in the \(b^{\theta}\) and \(C^{\theta}\)
columns. It refits on the true region \(H\), so it estimates the effect
there rather than the conditional estimand at the realized region; those
entries would measure the mismatch between the two regions, not a bias
or a coverage rate. The mismatch is reported instead as the
identification factor \(\kappa = \beta(H)/\beta(\widehat H) = 1.33\) ---
the true harm region carries a population effect 33 per cent above the
effect at the region the search returned, the effect-scale counterpart
to the classification rates above. The complement confirms the reading:
\(\widehat H^{c}\) is 87.2\% of the trial and nearly coincides with
\(H^{c}\), and there the suppressed entry was -1.5\%.

\begin{table}

\caption{\label{tbl-sim-estimation-fs-h10-n1000}Estimation and coverage
for forest search over the 329 detected replicates (66\% of 500
simulations; n = 1,000). Scenario hazard-ratio effects
\(\theta^{\dagger}(H) = 1.00\), \(\theta^{\dagger}(H^{c}) = 0.66\),
\(\theta^{\ddagger}(H) = 1.00\), \(\theta^{\ddagger}(H^{c}) = 0.58\);
the conditional estimand \(\theta(\widehat H)\) has mean 0.77, median
0.76 (hazard-ratio scale) and SD 0.16 (\(\beta\) scale). Classification
of \(\widehat H\) vs the true harm subgroup: sens 0.44, spec 0.92, PPV
0.43, NPV 0.92. The oracle refits on the true region \(H\), so it does
not estimate the conditional estimand \(\theta(\widehat H)\); its
\(b^{\theta}\) and \(C^{\theta}\) entries are a region mismatch rather
than a bias or coverage rate and are suppressed. That mismatch is the
identification factor \(\kappa = \beta(H)/\beta(\widehat H)\) = 1.33
(median 1.32), with \(\beta(H)\) = 1.003.}

\centering{

\centering\begingroup\fontsize{10}{12}\selectfont

\begin{tabular}[t]{lcccccccccccc}
\toprule
\multicolumn{5}{c}{ } & \multicolumn{4}{c}{Relative bias (\%)} & \multicolumn{4}{c}{95\% coverage} \\
\cmidrule(l{3pt}r{3pt}){6-9} \cmidrule(l{3pt}r{3pt}){10-13}
Estimator & Avg & \(\text{SD}_{\beta}\) & \(\widehat{\text{SD}}_{\beta}\) & Length & \(b^{\text{oracle}}\) & \(b^{\theta}\) & \(b^{\ddagger}\) & \(b^{\dagger}\) & \(C^{\text{oracle}}\) & \(C^{\theta}\) & \(C^{\ddagger}\) & \(C^{\dagger}\)\\
\midrule
\addlinespace[0.3em]
\multicolumn{13}{l}{\textbf{Harm Ĥ (mean |Ĥ| = 128, 12.8\% ITT)}}\\
\hspace{1em}oracle & 1.10 & 0.22 & 0.23 & 1.01 & +0.0 & — & +10.2 & +10.2 & 1.00 & — & 0.94 & 0.94\\
\hspace{1em}naive & 1.39 & 0.13 & 0.25 & 1.45 & +30.7 & +84.8 & +38.8 & +38.8 & 0.88 & 0.30 & 0.91 & 0.91\\
\hspace{1em}FB & 1.11 & 0.15 & 0.27 & 1.27 & +3.8 & +46.9 & +10.5 & +10.5 & 0.98 & 0.82 & 0.99 & 0.99\\
\hspace{1em}MR & 0.92 & 0.18 & 0.30 & 1.16 & -14.2 & +21.7 & -8.4 & -8.5 & 0.95 & 0.99 & 0.99 & 0.99\\
\addlinespace[0.3em]
\multicolumn{13}{l}{\textbf{Complement Ĥᶜ (mean |Ĥᶜ| = 872, 87.2\% ITT)}}\\
\hspace{1em}oracle & 0.65 & 0.10 & 0.10 & 0.25 & +0.0 & — & +10.4 & -1.7 & 1.00 & — & 0.83 & 0.95\\
\hspace{1em}naive & 0.63 & 0.09 & 0.10 & 0.24 & -2.5 & -4.1 & +7.5 & -4.3 & 1.00 & 0.95 & 0.91 & 0.93\\
\hspace{1em}FB & 0.64 & 0.09 & 0.14 & 0.35 & -0.0 & -1.7 & +10.2 & -1.9 & 1.00 & 0.99 & 0.96 & 0.99\\
\hspace{1em}MR & 0.66 & 0.10 & 0.18 & 0.47 & +2.0 & +0.3 & +12.4 & +0.1 & 1.00 & 1.00 & 1.00 & 1.00\\
\bottomrule
\end{tabular}
\endgroup{}

}

\end{table}%

\subsection{Forest search under genuine harm: estimation and
coverage}\label{sec-sim-estimation-fs-h15}

This design point moves from the borderline null to a genuine harm
signal (\(\theta^{\dagger}(H) = 1.51\), \(\theta^{\ddagger}(H) = 1.67\))
at \(n = 1{,}000\), with forest search on its fixed enumerated family,
so the FB and MR de-biased procedures target the \emph{same} estimand.
Table~\ref{tbl-sim-estimation-fs-h15} reports estimation and coverage
where forest search identified a subgroup in 96\% of the 400
simulations; with a real effect present the identifier now recovers the
true membership far more reliably than at the null (see the
classification summary in the caption). The winner's curse weakens as
the signal strengthens: the naive plug-in overstates the marginal hazard
ratio by only 10\% here, against the much larger overstatement at the
null (Table~\ref{tbl-sim-estimation-fs-h10}). Scored against the
marginal and controlled-direct-effect targets, multiplier resampling
tips mildly negative (MR bias -20.7\% against the marginal and -28.4\%
against the CDE) -- an artifact of scoring a selected region against a
fixed \emph{true}-subgroup target. Against the conditional estimand
\(\theta(\widehat H)\) that Theorem 2 of the main paper characterizes,
however, the same intervals are near-nominal: MR point bias is +1.8\%
and its coverage of \(\theta(\widehat H)\) is 0.94. Switching the
scoring target to \(\theta(\widehat H)\) thus resolves the apparent
over-correction seen against true-subgroup targets, and is the cleanest
illustration that the conditional estimand is the parameter the
correction is built to cover; its stability across sample size is traced
in Figure~\ref{fig-mr-coverage-h15c}.

\begin{table}

\caption{\label{tbl-sim-estimation-fs-h15}Estimation and coverage for
forest search over the 384 detected replicates (96\% of 400 simulations;
n = 1,000). Scenario hazard-ratio effects
\(\theta^{\dagger}(H) = 1.51\), \(\theta^{\dagger}(H^{c}) = 0.66\),
\(\theta^{\ddagger}(H) = 1.67\), \(\theta^{\ddagger}(H^{c}) = 0.58\);
the conditional estimand \(\theta(\widehat H)\) has mean 1.24, median
1.31 (hazard-ratio scale) and SD 0.28 (\(\beta\) scale). Classification
of \(\widehat H\) vs the true harm subgroup: sens 0.70, spec 0.96, PPV
0.71, NPV 0.96. Estimators, targets, and SD scales are defined with the
main-text estimation tables (Section 7). The oracle refits on the true
region \(H\), so it does not estimate the conditional estimand
\(\theta(\widehat H)\); its \(b^{\theta}\) and \(C^{\theta}\) entries
are a region mismatch rather than a bias or coverage rate and are
suppressed. That mismatch is the identification factor
\(\kappa = \beta(H)/\beta(\widehat H)\) = 1.37 (median 1.20), with
\(\beta(H)\) = 1.566.}

\centering{

\centering\begingroup\fontsize{10}{12}\selectfont

\begin{tabular}[t]{lcccccccccccc}
\toprule
\multicolumn{5}{c}{ } & \multicolumn{4}{c}{Relative bias (\%)} & \multicolumn{4}{c}{95\% coverage} \\
\cmidrule(l{3pt}r{3pt}){6-9} \cmidrule(l{3pt}r{3pt}){10-13}
Estimator & Avg & \(\text{SD}_{\beta}\) & \(\widehat{\text{SD}}_{\beta}\) & Length & \(b^{\text{oracle}}\) & \(b^{\theta}\) & \(b^{\ddagger}\) & \(b^{\dagger}\) & \(C^{\text{oracle}}\) & \(C^{\theta}\) & \(C^{\ddagger}\) & \(C^{\dagger}\)\\
\midrule
\addlinespace[0.3em]
\multicolumn{13}{l}{\textbf{Harm Ĥ (mean |Ĥ| = 123, 12.3\% ITT)}}\\
\hspace{1em}oracle & 1.61 & 0.21 & 0.22 & 1.42 & +0.0 & — & -3.6 & +6.8 & 1.00 & — & 0.94 & 0.94\\
\hspace{1em}naive & 1.65 & 0.17 & 0.24 & 1.59 & +4.3 & +42.4 & -1.1 & +9.6 & 1.00 & 0.74 & 0.98 & 0.97\\
\hspace{1em}FB & 1.37 & 0.20 & 0.27 & 1.53 & -14.1 & +17.2 & -18.1 & -9.3 & 0.99 & 0.92 & 0.91 & 0.95\\
\hspace{1em}MR & 1.20 & 0.25 & 0.30 & 1.52 & -25.6 & +1.8 & -28.4 & -20.7 & 0.95 & 0.94 & 0.76 & 0.85\\
\addlinespace[0.3em]
\multicolumn{13}{l}{\textbf{Complement Ĥᶜ (mean |Ĥᶜ| = 877, 87.7\% ITT)}}\\
\hspace{1em}oracle & 0.63 & 0.10 & 0.10 & 0.25 & +0.0 & — & +8.3 & -3.6 & 1.00 & — & 0.88 & 0.94\\
\hspace{1em}naive & 0.64 & 0.10 & 0.10 & 0.24 & +0.6 & -3.0 & +8.9 & -3.1 & 1.00 & 0.95 & 0.87 & 0.94\\
\hspace{1em}FB & 0.65 & 0.10 & 0.14 & 0.36 & +2.9 & -0.8 & +11.4 & -0.8 & 1.00 & 0.99 & 0.95 & 0.99\\
\hspace{1em}MR & 0.66 & 0.10 & 0.18 & 0.48 & +4.5 & +0.8 & +13.2 & +0.8 & 1.00 & 1.00 & 0.99 & 1.00\\
\bottomrule
\end{tabular}
\endgroup{}

}

\end{table}%

\subsection{\texorpdfstring{Robustness to noise covariates: a matched
pair at
\(n = 1{,}500\)}{Robustness to noise covariates: a matched pair at n = 1\{,\}500}}\label{sec-sim-noise-pair-n1500}

The sample-size sweeps and the estimation tables let the identifier work
from the seven real prognostic covariates.
Table~\ref{tbl-sim-noise-pair-n1500} instead adds \textbf{three
standard-normal noise covariates} to the pool and holds \(n = 1{,}500\)
fixed, varying only the planted signal --- a weaker harm effect
(\(\theta^{\dagger}(H) = 1.5\)) and a stronger, well-separated one
(\(\theta^{\dagger}(H) = 2.0\)). The identifier is not misled by the
distractors: a noise covariate enters the selected region in only 7.8\%
of detections at the weaker signal and 0.2\% at the stronger one, the
true menopausal-status partner is recovered in 63.1\% and 77.0\%
respectively (against its \texttt{age} proxy), and positive-predictive
value rises from 0.79 to 0.91. As the signal separates the winner's
curse all but vanishes (naive marginal bias +4.5 to +0.6), and coverage
of the conditional estimand \(\theta(\widehat H)\) that Theorem 2
characterizes runs from mildly below nominal at the weaker signal to
near-nominal at the stronger (0.88 to 0.90); the added noise covariates
manufacture near-duplicate candidate cuts --- the
\(\eta_n = o(n^{-1/2})\) boundary that (A5) excludes
(Section~\ref{sec-ap-delicate}) --- where the infinitesimal-jackknife
variance can understate how much the selected region moves, so a mild
shortfall there is what the theory predicts at the boundary it does not
certify, rather than a failure within the regime it does. Scored against
the fixed marginal target the multiplier correction tips negative in
both cells --- the over-correction of an already-small curse discussed
for the genuine-harm design point --- so the conditional estimand, not
the marginal, remains the parameter the interval is built to cover.

\begin{table}

\caption{\label{tbl-sim-noise-pair-n1500}Robustness to three added
standard-normal noise covariates: forest search at \(n = 1{,}500\) (500
simulations per cell), harm subgroup, at a weaker
(\(\theta^{\dagger}(H)\approx1.5\)) and a stronger
(\(\theta^{\dagger}(H)\approx2.0\)) signal against a fixed complement
\(\theta^{\dagger}(H^{c})\approx0.66\). Detection, PPV and coverage
\(C\) are proportions; noise-covariate leakage and true-partner recovery
are percentages of detected replicates; \(b^{\theta}\) and
\(b^{\dagger}\) are relative bias (\%) against the conditional estimand
\(\theta(\widehat H)\) and the marginal hazard ratio, respectively.}

\centering{

\centering\begingroup\fontsize{10}{12}\selectfont

\begin{tabular}[t]{lcc}
\toprule
  & $\theta^{\dagger}(H)=1.5$ & $\theta^{\dagger}(H)=2.0$\\
\midrule
Detection rate & 1.00 & 1.00\\
PPV & 0.79 & 0.91\\
Noise-covariate leakage & 7.8 & 0.2\\
True meno partner recovered & 63.1 & 77.0\\
$\theta(\widehat H)$ mean & 1.32 & 1.89\\
Naive winner's curse $b^{\dagger}$ & +4.5 & +0.6\\
MR $b^{\theta}$ & -5.3 & -6.5\\
MR $C^{\theta}$ & 0.88 & 0.90\\
MR $b^{\dagger}$ & -19.5 & -13.3\\
MR $C^{\dagger}$ & 0.78 & 0.84\\
\bottomrule
\end{tabular}
\endgroup{}

}

\end{table}%

\subsection{Noise covariates at the borderline null: coverage robustness
under selection leakage}\label{sec-noise-null}

Section S8.6 added noise covariates at a separated signal, where the
strong effect leaves them largely unused. This section reverses that. It
holds the borderline null \(\theta^{\dagger}(H)=1.00\) --- where no
genuine effect disciplines the search --- and sweeps the number of
standard-normal noise covariates (\(k\in\{0,3,6\}\)) against per-trial
size \(n=\) 500--2,000, with forest search on its enumerated family. It
is the regime in which noise is most able to mislead selection, and it
separates cleanly what noise does and does not damage.

\textbf{Selection is heavily contaminated.} With six noise covariates at
\(n=500\), a noise cut enters the realized region in 71\% of detections,
and the true oestrogen-receptor anchor is displaced from 49\% of rules
with no noise to 28\% with six; the false-harm detection rate itself
rises from 69\% to 88\%. The contamination recedes with \(n\) as the
real signal outcompetes the distractors --- noise leakage falls to 54\%
by \(n=\) 2,000 --- but at small \(n\) the selected region is, more
often than not, partly noise-defined.

\begin{figure}

\centering{

\pandocbounded{\includegraphics[keepaspectratio]{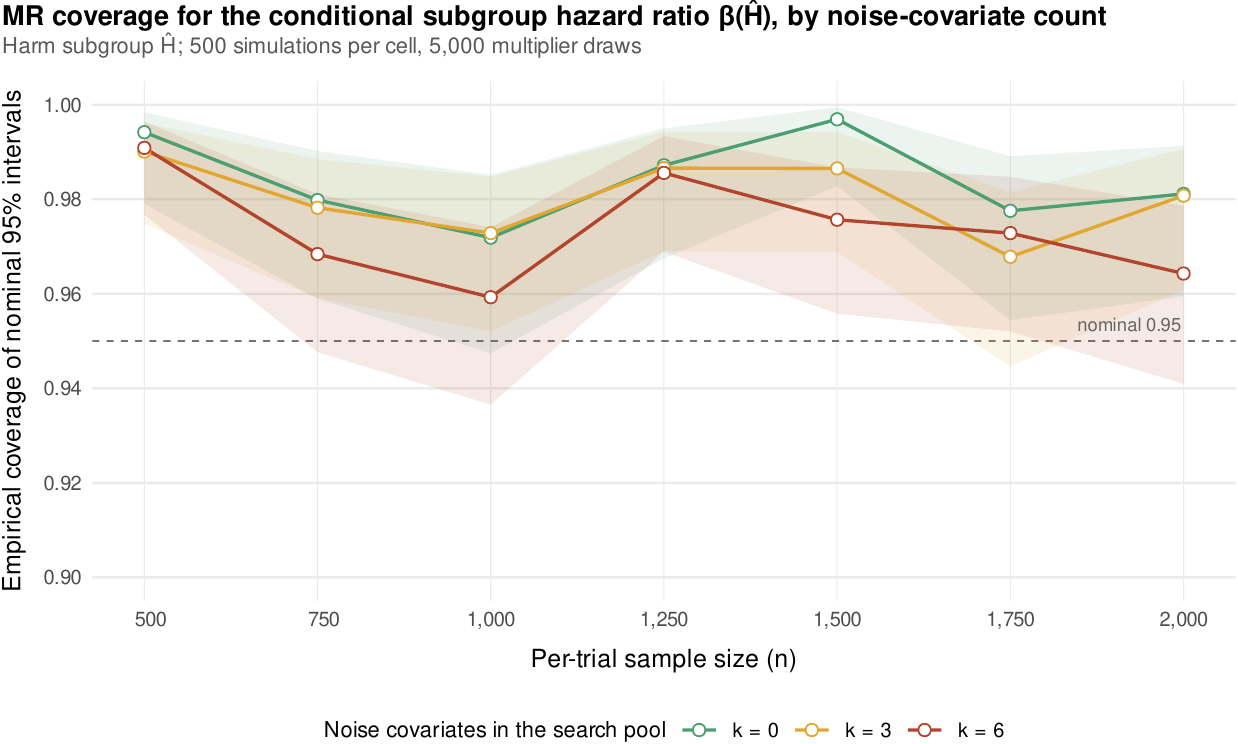}}

}

\caption{\label{fig-mr-coverage-knoise-h10}\textbf{Coverage of MR
intervals for the conditional subgroup hazard ratio
\(\beta(\widehat{H})\) on the harm subgroup \(\widehat{H}\), by sample
size and noise-covariate count.} Empirical coverage of nominal 95\%
intervals against per-trial sample size \(n\), with one curve per
noise-covariate count \(k \in \{0, 3, 6\}\). Each curve is a separate
sweep of the same data-generating mechanism, differing only in the
number of inert N(0,1) covariates added to the search pool; the outcome
model is identical throughout, so separation between curves reflects
selection noise rather than a change in the estimand. The mechanism sets
\(\theta^{\dagger}(H) = 1.00\) in the harm subgroup and
\(\theta^{\dagger}(H^{c}) = 0.66\) in its complement; 500 simulations
per cell, 5,000 multiplier draws; detector: consistency. The dashed line
marks the 0.95 nominal level. Shaded bands are Wilson 95\% intervals on
each rate. Rates are conditional on a subgroup being identified.}

\end{figure}

\textbf{Inference is not.} Across the entire grid the conditional
coverage of \(\theta(\widehat H)\) stays between 0.96 and 1.00
(Figure~\ref{fig-mr-coverage-knoise-h10}), the conditional bias shifts
only slightly with noise at each \(n\), and the median interval width
does not move --- the minimum-size and minimum-event guards keep every
realized region, noise-defined or not, well-powered enough for a stable
fit (Table~\ref{tbl-noise-null}). What noise costs is paid entirely in
identification: sensitivity and positive predictive value against the
true harm subgroup fall by roughly a third --- from 0.57 and 0.58 with
no noise to 0.38 and 0.39 with six noise covariates at \(n=\) 2,000 ---
while the effect at whatever region was drawn continues to be covered at
nominal. With no noise covariates the identification gain shows up on
the effect scale as well: the identification factor
\(\kappa = \beta(H)/\beta(\widehat H)\) falls from \(1.40\) at \(n=\)
500 to \(1.34\), \(1.25\), and \(1.24\) at \(n=\) 1,000, 1,500, and
2,000, the counterpart to positive predictive value rising from \(0.32\)
to \(0.58\) across the same range. The data-generating mechanism is
fixed across the sweep, so \(\beta(H)\) does not move and the whole of
that decline is \(\beta(\widehat H)\) approaching the effect at the true
region. It is quoted at \(k=0\) only: \(\beta(\widehat H)\) is the
effect at the realized region evaluated in the super-population, and a
region defined by a noise covariate has no counterpart there, so at
\(k>0\) the quantity would be conditional on the region having escaped
contamination.

This is the conditional estimand made concrete. Coverage is of
\(\beta(\widehat H)\), the effect in the region the search actually
returned; noise changes which region that is, and the interval follows
it, so identification quality and inferential coverage move
independently. The correction certifies the effect at the region one
found, not that one found the right region --- noise buys a worse
subgroup, correctly estimated. The one residual is the mild shortfall at
the smallest \(n\) and heaviest noise (conditional coverage 0.991 at
\(k=6\), \(n=500\), recovering by \(n=750\)): the near-duplicate cuts
the noise covariates manufacture are exactly the \(\eta_n=o(n^{-1/2})\)
boundary that (A5) excludes (Section~\ref{sec-ap-delicate}), where the
infinitesimal-jackknife variance can understate movement --- the same
boundary, and the same recovery in \(n\), as Section S8.6. The marginal
target is not shown here: at the null it over-covers trivially and would
obscure rather than reveal the misclassification, whose true-subgroup
cost is read directly from the classification rates above; its
separated-signal behaviour is in Section S8.6.

{

\begin{longtable}[]{@{}
  >{\raggedleft\arraybackslash}p{(\linewidth - 14\tabcolsep) * \real{0.0714}}
  >{\raggedleft\arraybackslash}p{(\linewidth - 14\tabcolsep) * \real{0.0571}}
  >{\raggedleft\arraybackslash}p{(\linewidth - 14\tabcolsep) * \real{0.1571}}
  >{\raggedleft\arraybackslash}p{(\linewidth - 14\tabcolsep) * \real{0.2714}}
  >{\raggedleft\arraybackslash}p{(\linewidth - 14\tabcolsep) * \real{0.1286}}
  >{\raggedleft\arraybackslash}p{(\linewidth - 14\tabcolsep) * \real{0.1429}}
  >{\raggedleft\arraybackslash}p{(\linewidth - 14\tabcolsep) * \real{0.0857}}
  >{\raggedleft\arraybackslash}p{(\linewidth - 14\tabcolsep) * \real{0.0857}}@{}}

\caption{\label{tbl-noise-null}Forest search at the borderline null
\(\theta^{\dagger}(H)=1.00\), by sample size \(n\) and number of noise
covariates \(k\). \textbf{Inference block} (\(C_\theta\) = coverage of
the conditional estimand \(\theta(\widehat H)\); bias\(_\theta\) = MR
relative bias against it; CI width = median MR interval width,
hazard-ratio scale) is stable across \(k\). \textbf{Identification
block} (leak = share of realized rules containing a noise cut; sens, PPV
of \(\widehat H\) against the true harm subgroup) degrades sharply in
\(k\). All quantities conditional on detection; read live from the
committed sweep payloads.}

\tabularnewline

\toprule\noalign{}
\begin{minipage}[b]{\linewidth}\raggedleft
\(n\)
\end{minipage} & \begin{minipage}[b]{\linewidth}\raggedleft
\(k\)
\end{minipage} & \begin{minipage}[b]{\linewidth}\raggedleft
\(C_\theta\)
\end{minipage} & \begin{minipage}[b]{\linewidth}\raggedleft
bias\(_\theta\) (\%)
\end{minipage} & \begin{minipage}[b]{\linewidth}\raggedleft
CI width
\end{minipage} & \begin{minipage}[b]{\linewidth}\raggedleft
leak (\%)
\end{minipage} & \begin{minipage}[b]{\linewidth}\raggedleft
sens
\end{minipage} & \begin{minipage}[b]{\linewidth}\raggedleft
PPV
\end{minipage} \\
\midrule\noalign{}
\endhead
\bottomrule\noalign{}
\endlastfoot
500 & 0 & 0.994 & +30.9 & 1.47 & 0 & 0.373 & 0.322 \\
500 & 3 & 0.990 & +30.8 & 1.55 & 52 & 0.305 & 0.265 \\
500 & 6 & 0.991 & +30.9 & 1.58 & 71 & 0.267 & 0.233 \\
1000 & 0 & 0.972 & +22.9 & 1.04 & 0 & 0.419 & 0.423 \\
1000 & 3 & 0.973 & +24.9 & 1.05 & 52 & 0.306 & 0.311 \\
1000 & 6 & 0.959 & +28.4 & 1.08 & 71 & 0.251 & 0.256 \\
1500 & 0 & 0.997 & +12.0 & 0.80 & 0 & 0.559 & 0.560 \\
1500 & 3 & 0.987 & +15.7 & 0.83 & 40 & 0.423 & 0.434 \\
1500 & 6 & 0.976 & +17.8 & 0.82 & 60 & 0.345 & 0.357 \\
2000 & 0 & 0.981 & +9.9 & 0.67 & 0 & 0.574 & 0.576 \\
2000 & 3 & 0.981 & +12.7 & 0.68 & 37 & 0.452 & 0.457 \\
2000 & 6 & 0.964 & +14.7 & 0.68 & 54 & 0.379 & 0.387 \\

\end{longtable}

}

\begin{figure}

\centering{

\pandocbounded{\includegraphics[keepaspectratio]{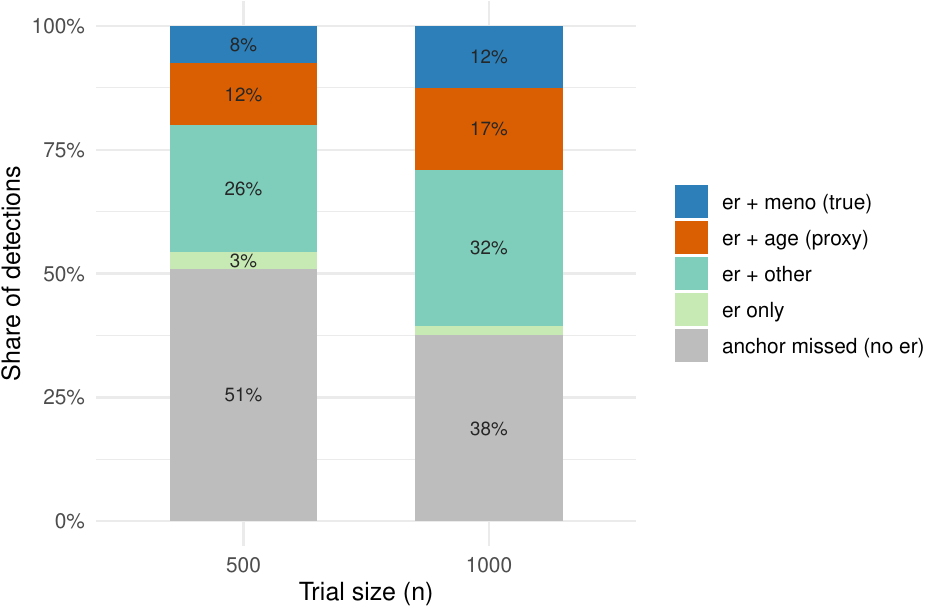}}

}

\caption{\label{fig-fs-id-structure}Composition of the FS-identified
subgroup by structure, at \(n=500\) and \(n=1000\) (500 replicates
each), at the borderline null (\(\theta^{\dagger}(H) = 1.00\)). As \(n\)
grows the anchor is missed less often and spurious second factors
recede, but the share in which \(\mathrm{age}\) substitutes for the true
\(\mathrm{meno}\) partner does not---reflecting the persistent near-tie
between \(\{\mathrm{er} \cap \mathrm{meno}\}\) and its \(\mathrm{age}\)
proxy.}

\end{figure}%

\begin{table}

\caption{\label{tbl-fs-id-structure}FS identification structure across
trial size (500 replicates per size), at the borderline null
(\(\theta^{\dagger}(H) = 1.00\)). The anchor-recovery rate rises with
\(n\) while the conditional meno:age partner split is essentially
invariant---the empirical signature of the near-tie regime.
Classification accuracy of the identified subgroup is reported in the
footnote.}

\centering{

\centering
\begin{threeparttable}
\begin{tabular}[t]{lcc}
\toprule
Quantity & 500 & 1000\\
\midrule
Detection rate (all replicates) & 68.8\% & 64.0\%\\
\addlinespace[0.3em]
\multicolumn{3}{l}{\textbf{Among detections}}\\
\hspace{1em}er + meno (true partner) & 7.6\% & 12.5\%\\
\hspace{1em}er + age (proxy partner) & 12.5\% & 16.6\%\\
\hspace{1em}er + other factor & 25.6\% & 31.6\%\\
\hspace{1em}er only & 3.5\% & 1.9\%\\
\hspace{1em}anchor missed (no er) & 50.9\% & 37.5\%\\
\hspace{1em}Anchor recovered (er, any pairing) & 49.1\% & 62.5\%\\
\hspace{1em}Conditional partner split, meno : age & 38 : 62 & 43 : 57\\
\bottomrule
\end{tabular}
\begin{tablenotes}
\item Classification of the identified subgroup against the planted membership, averaged over detections (500 / 1000): sensitivity 0.37 / 0.42, specificity 0.88 / 0.92, PPV 0.32 / 0.42, NPV 0.91 / 0.92.
\end{tablenotes}
\end{threeparttable}

}

\end{table}%

\begin{table}

\caption{\label{tbl-fs-id-structure-h15}FS identification structure
across trial size (500 replicates per size) under a genuine harm signal
(\(\theta^{\dagger}(H) = 1.51\)). In contrast to the borderline-null
case, the anchor-recovery rate rises with \(n\) and the conditional
meno:age partner split shifts toward the true partner---the near-tie
resolves as the effect separates. Classification accuracy of the
identified subgroup is reported in the footnote.}

\centering{

\centering
\begin{threeparttable}
\begin{tabular}[t]{lcc}
\toprule
Quantity & 500 & 1000\\
\midrule
Detection rate (all replicates) & 88.2\% & 98.2\%\\
\addlinespace[0.3em]
\multicolumn{3}{l}{\textbf{Among detections}}\\
\hspace{1em}er + meno (true partner) & 23.8\% & 41.3\%\\
\hspace{1em}er + age (proxy partner) & 17.0\% & 24.6\%\\
\hspace{1em}er + other factor & 22.4\% & 14.7\%\\
\hspace{1em}er only & 1.8\% & 0.6\%\\
\hspace{1em}anchor missed (no er) & 34.9\% & 18.7\%\\
\hspace{1em}Anchor recovered (er, any pairing) & 65.1\% & 81.3\%\\
\hspace{1em}Conditional partner split, meno : age & 58 : 42 & 63 : 37\\
\bottomrule
\end{tabular}
\begin{tablenotes}
\item Classification of the identified subgroup against the planted membership, averaged over detections (500 / 1000): sensitivity 0.58 / 0.70, specificity 0.92 / 0.96, PPV 0.52 / 0.70, NPV 0.94 / 0.96.
\end{tablenotes}
\end{threeparttable}

}

\end{table}%

\section{Computational details and
reproducibility}\label{computational-details-and-reproducibility}

All analyses use \texttt{forestsearch} development version 0.2.0, which
extends the CRAN release with the GLM pathway and the
multiplier-resampling inference used here; the package source
accompanies the submission. The simulation study's data-generating
mechanisms are \texttt{create\_gbsg\_dgm()} and
\texttt{generate\_aft\_dgm\_flex()}. Scripts reproducing each table and
figure, keyed by label, are listed in the README of the accompanying
reproducibility archive.

\protect\phantomsection\label{refs}

\end{document}